\documentclass[a4paper,twocolumn,11pt,unpublished]{quantumarticle}
\pdfoutput=1
\usepackage[utf8]{inputenc}
\usepackage[english]{babel}
\usepackage[T1]{fontenc}
\usepackage{amsmath}
\usepackage{hyperref}
\hypersetup{colorlinks=false, linkcolor=black}
\usepackage{qcircuit}
\usepackage{mathtools}
\usepackage{caption}
\usepackage{subcaption}
\usepackage{algorithm}
\usepackage{algpseudocode}

\usepackage{tikz}
\usepackage{lipsum}
\usepackage[numbers]{natbib}

\usepackage{amsthm}
\usepackage[acronym]{glossaries}
\glsdisablehyper  

\newacronym{vqe}{VQE}{variational quantum eigensolver}
\newacronym{nisq}{NISQ}{noisy intermediate-scale quantum}
\newacronym{bois}{BOIS}{Bayesian optimisation with information sharing}
\newacronym{bo}{BO}{Bayesian optimisation}
\newacronym{iid}{iid}{independent and identically distributed}
\newacronym{rbf}{RBF}{radial basis function}
\newacronym{lcb}{LCB}{lower confidence bound}
\newacronym{qngd}{QNGD}{quantum natural gradient descent}
\newacronym{qaoa}{QAOA}{quantum approximate optimisation algorithm}
\newacronym{nft}{NFT}{Nakanishi-Fujii-Todo algorithm}
\newacronym{emicore}{EMICoRe}{Expected Maximum Improvement
over Confident Regions}
\newacronym{bfgs}{BFGS}{Broyden–Fletcher–Goldfarb–Shanno}
\newacronym{lbfgs}{L-BFGS}{Limited-memory Broyden–Fletcher–\allowbreak Goldfarb–\allowbreak Shanno}
\newacronym[shortplural={PES}, firstplural={potential energy surfaces}]{pes}{PES}{potential energy surface}
\newacronym{onb}{ONB}{orthonormal basis}
\newacronym{stso}{STOs}{Slater-type orbitals}
\newacronym{hf}{HF}{Hartree-Fock}
\newacronym{qubit}{qubit}{quantum bit}
\newacronym{qc}{QC}{quantum computation}

\newacronym{spsa}{SPSA}{simultaneous perturbation stochastic approximation}
\newacronym{iff}{iff}{if and only if}

\algnewcommand\algorithmicforeach{\textbf{for each}}
\algdef{S}[FOR]{ForEach}[1]{\algorithmicforeach\ #1\ \algorithmicdo}

\begin{document}

\title{Bayesian optimisation with improved information sharing for the variational quantum eigensolver}
\onecolumn
\author{Milena Röhrs}
\affiliation{Department of High Performance Computing,
	Fraunhofer ITWM,
	Fraunhofer-Platz 1,
        67663 Kaiserslautern, Germany}
\author{Alexey Bochkarev}
\affiliation{Department of Mathematics, RPTU Kaiserslautern-Landau,
Gottlieb-Daimler-Straße,
67663 Kaiserslautern, Germany}
\author{Arcesio C. Medina}
\affiliation{Department of High Performance Computing,
	Fraunhofer ITWM,
	Fraunhofer-Platz 1,
        67663 Kaiserslautern, Germany}
\maketitle
\begin{abstract}
\onecolumn
This work presents a detailed empirical analysis of Bayesian optimisation with information sharing (BOIS) for the \gls{vqe}~\cite{self_variational_2021}. The method is applied to computing the potential energy surfaces
(PES) of the hydrogen and water molecules. We performed noise-free simulations and investigated the algorithms' performance under the influence of noise for the hydrogen molecule, using both emulated and real quantum hardware (IBMQ System One). Based on the noise free simulations we compared different existing information sharing schemes and proposed a new one, which trades parallelisability of the algorithm for a significant reduction in the amount of quantum computing resources required until convergence. In particular, our numerical experiments show an improvement by a factor of 1.5 as compared to the previously reported sharing schemes in $\mathrm{H_2}$, 
and at least by a factor of 5 as compared to no sharing in $\mathrm{H_2O}$. 
Other algorithmic aspects of the Bayesian optimisation, namely, the acquisition weight decrease rate and kernel, are shown to have an influence on the \gls{qc} demand of the same order of magnitude.
\end{abstract}

\titlepage
\twocolumn


\section{Introduction}
Currently existing quantum computers still do not quite implement the vision of Feynman from 40 years ago \cite{Preskill2021-rj}. 
In the current \gls{nisq} technology era, available devices are characterised by a very limited number of qubits and the size of quantum circuits is further limited by hardware architecture and decoherence effects \cite{Preskill2018-zt, Preskill1997-wu}.
Therefore, algorithms with low circuit depths that are robust to noise are required.\par 
For this purpose, hybrid quantum algorithms such as the \gls{vqe} algorithm were introduced \cite{peruzzo2014variational}. In VQE, the Hamiltonian $H$ of a system is expanded in a basis that comprises Kronecker products of Pauli matrices $P_i\in \{I, X, Y, Z\}^{\otimes n}$ \cite{self_variational_2021}:
\begin{equation}
\label{eq:Ham-expand}
    H = \sum_{i=1}^{4^n}{h_i}P_i.
\end{equation}
Due to the variational theorem \cite{levine2014quantum} the ground state energy of such a system can be obtained as the value of a global solution of the following optimisation problem:
\begin{equation}
\label{equ:vqe-min}
    \min_{\theta} \sum_{i=1}^{4^n}{h_i}\langle U(\theta)\Psi_0|P_i|U(\theta)\Psi_0 \rangle,
\end{equation}
given a sufficiently expressive ansatz $U(\theta)$ over the parameters $\theta$ acting on an initial state $\Psi_0$. While the optimisation and summation is performed via classical hardware, the evaluation of the Pauli expectation values involves a quantum computer \cite{Bauer2020-tg}.\par
In practice, besides expressivity issues of the ansatz, VQE suffers from trainabilty problems. Furthermore, given the high costs of computing time on a quantum computer as well as the current rather long queuing times, there are compelling reasons to keep the number of iterations small. In particular, an optimisation method designed to effectively deal with objective functions that are costly to evaluate is needed.\par
This problem is well studied in machine learning, for example in the context of hyperparameter optimisation \cite{rasmussen_gaussian_2019, sigoptBayesianOptimization}. A common approach to address this issue is the application of Bayesian optimisation. 
Its key idea is to model the unknown and hard to evaluate objective function by a statistical surrogate, which is often chosen as a Gaussian process. In such a statistical model all points in the feasible set are treated as random variables whose values depend on each other in a way specified by a parameterised covariance function, often called the kernel.
In the context of VQE, each point in the feasible set represents a set of ansatz parameters and the associated values are the energies of the corresponding state.
At the beginning of each iteration, the already evaluated points are incorporated into the surrogate model by updating its distribution according to Bayes' rule. The point to be evaluated next is then determined by optimising an acquisition function. This function is responsible for balancing the exploration of unknown areas of the feasible set and exploitation of points that are promising candidates for an optimum as driving forces of the optimisation \cite{frazier_tutorial_2018,shahriari_taking_2016, rasmussen_gaussian_2019,wang_intuitive_2022}.\par
In general, using surrogate models to mimic the objective function of a variational quantum algorithm is an active and promising field of research \cite{shaffer_surrogate-based_2023, Cheng2024}.
In particular, initial attempts of applying \gls{bo} to hybrid quantum algorithms showed encouraging results. 
Otterbach \textit{et al.} \cite{otterbach_unsupervised_2017} made use of Bayesian optimisation in the \gls{qaoa} applied to the clustering problem leading to a hybrid algorithm that was robust to realistic noise. In \cite{moseley2018Bayesian}, the often used Nelder-Mead and approximate gradient methods were substituted with Bayesian optimisation as the classical optimisation routine in \gls{vqe}, resulting in a faster convergence to the ground state energy in comparison to \gls{spsa} optimisation in simulations.\par
While also identifying challenges such as the scaling behaviour of \gls{bo}, several other studies confirmed its potential for \gls{vqe} and other hybrid quantum algorithms \cite{zhu_training_2019, iannelli_noisy_2021, Rattew2020-ib}. However, the incentive to further reduce the number of iterations and maximise the knowledge gain per quantum job remained. In an attempt to address this issue, Self \textit{et al.} \cite{self_variational_2021, sauvage2021optimising} noticed that one can even further reduce the quantum computational resources required for convergence by ``recycling'' some of the quantum computation results when simultaneously solving similar problems, such as finding the ground state energies associated with different molecular geometries. 
More precisely, in the process of calculating the objective function value for the Hamiltonian of a certain geometry for a given set of ansatz parameters, the bracket terms in Equation~\eqref{equ:vqe-min} are evaluated on a quantum device.
Without any quantum computational overhead, one can then obtain the energy expectation value for another Hamiltonian (expanded in the same basis states) corresponding to a similar system (for the same ansatz parameters) by only changing the linear combination coefficients. This ``freely available'' information can be incorporated into the other geometries surrogate model, leading to a more informed decision on where to sample next.
Given the scarce quantum computer resources available, this is an interesting approach to explore further.\par
\subsection{Contributions}
This study expands on \cite{self_variational_2021, sauvage2021optimising} 
by improving the sharing algorithm, with the new ``immediate sharing'' scheme leading to an overall decrease in \gls{qc} demand. This is demonstrated based on numerical experiments counting the number of iterations required to obtain accurate \glspl{pes} of the $\mathrm{H_2}$ molecule. Results obtained using the new sharing scheme both in noise-free and noisy simulations as well as on a real quantum computer are presented.
By comparison of different sharing schemes in noise-free simulations we determined that, in the given setup, sharing beyond the nearest neighbours yields a significant advantage. Together with the results by Self et al. \cite{self_variational_2021} that presented a family of problems where such sharing yielded only moderate improvement, this illustrates the necessity to investigate more sophisticated sharing thresholds. A simple idea for an information sharing rule based on the similarity of the Pauli decompositions is suggested for further investigation, as an alternative to nearest neighbours sharing.  
A few other algorithmic aspects are investigated and identified as equally impactful on the convergence behavior of the \gls{vqe} optimisation as the sharing algorithm. This includes the covariance function used to create the statistical model, which serves as a surrogate for the \gls{vqe} objective function, and the decrease rate of the acquisition weights. 
Finally, the new sharing scheme is applied to the $\mathrm{H_2O}$ molecule allowing the assessment of the algorithms performance in a case in which the set of non-zero expansion coefficients is not the same among all geometries of the \gls{pes}. Despite the small \gls{qc} overhead per iteration, the overall costs are notably reduced due to a much faster convergence in terms of iterations. These results highlight the potential of refining information sharing approaches in the context of VQE.

\section{Methods}
\label{sec:methods}
\begin{figure}
\centering
    \includegraphics[trim = 0mm 0mm 0mm 0mm, clip, width=0.95\linewidth]{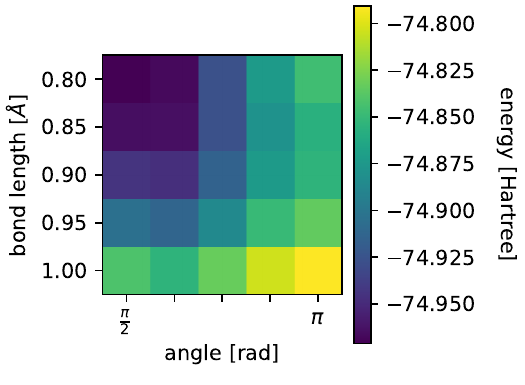}
    \captionof{figure}{$\mathrm{H_2O}$ geometries and energies.}
    \label{fig:refPES}
\end{figure}

The experiments carried out aimed to determine the electronic energy of the ground state of the hydrogen molecule $\mathrm{H_2}$ and the water molecule $\mathrm{H_2O}$ for different geometries
using VQE. The geometry of the $\mathrm{H_2}$ molecule is fully described by different H--H bond lengths. 
Eight equidistant geometries ranging from 0.3 to 1.0 \AA\ were investigated.  For the water molecule, depending on whether the symmetry is fixed to be $\mathrm{C_{2v}}$ or the nuclei are given full positional freedom, it is specified by two or three coordinates (the angle, and one or two bond lengths, respectively). The geometry parameter grid in the $\mathrm{H_2O}$ 2D case is shown in Figure~\ref{fig:refPES}. 
In the 3D case, a $3\times3\times3$ grid was chosen with an angle ranging from 1.7 to 1.9 rad and bond lengths ranging from 0.8 to 1.0 \AA. The latter case is investigated to check the method's behaviour for a \gls{pes} of larger dimension.
The accuracy of the results is quantified by comparison to a reference that was obtained by numerical diagonalisation, which can be considered as the true optimum of the VQE optimisation. The experimental details and default settings are provided below.
\subsection{Quantum Calculation}
\label{sec:def-qc-problem}
\begin{figure}[h!]
\centering
\[
\begin{array}{c}
\Qcircuit @C=1em @R=1em {
    \lstick{q_0} & 
            \gate{R_y(\theta_0)} &
            \ctrl{1} &
                \gate{R_y(\theta_2)} & 
                        \ctrl{1} &
                            \gate{R_y(\theta_4)}\\
    \lstick{q_1} & 
            \gate{R_y(\theta_1)} &
            \targ &
                \gate{R_y(\theta_3)} & 
                        \targ &
                            \gate{R_y(\theta_5)}}
\end{array}
\]
\caption{Ansatz circuit for the $\mathrm{H_2}$ molecule.} \label{RealAmpl22}
\label{fig:H2_ansatz}
\end{figure}
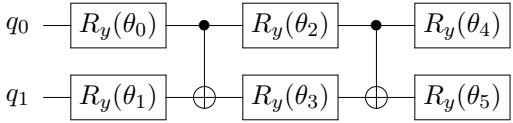

The Python packages used for the quantum calculations were Qiskit \cite{Qiskit}, qiskit\_nature \cite{QiskitNature} and qiskit\_ibm\_runtime. The second quantisation Hamiltonian for each of the molecular geometries was obtained via the PySCFDriver implemented in Qiskit Nature, choosing STO-3G as the chemical basis set. 
To further reduce the necessary number of qubits for the investigation of $\mathrm{H_2O}$, an active space consisting of four orbitals around the Fermi level occupied by in total four active electrons was chosen, similar to \cite{de_gracia_trivino_complete_2023}. For both molecules, the Pauli expansions of the resulting Hamiltonians after parity transformation can be found in the supplementary material.
For $\mathrm{H_2}$, an heuristic ansatz \cite{self_variational_2021} shown in Figure~\ref{fig:H2_ansatz} is used.\par 
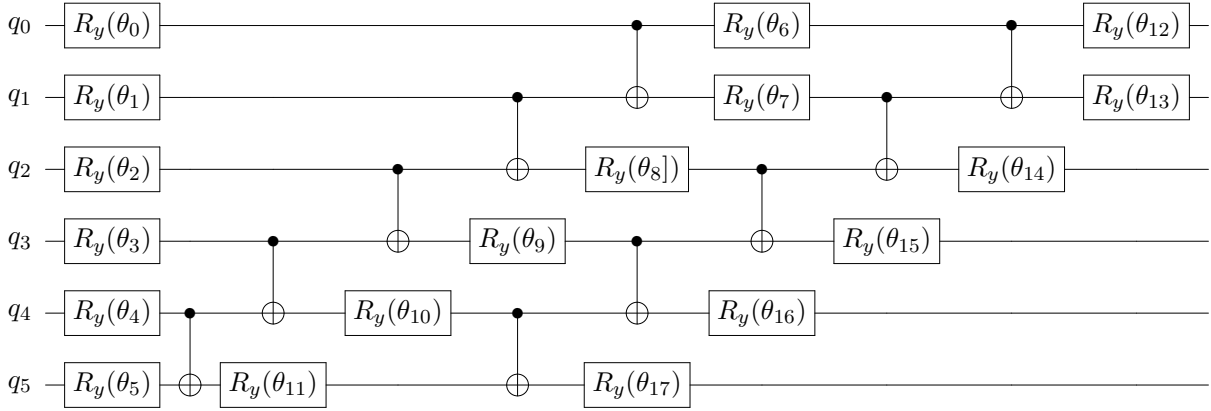
\begin{figure*}
\[
\begin{array}{c}
\Qcircuit @C=0.7em @R=1em {
\lstick{q_0} & \gate{R_y(\theta_0)} &\qw&\qw&\qw&\qw& \ctrl{1} & \gate{R_y(\theta_6)}&\qw&\ctrl{1}&\gate{R_y(\theta_{12})}&\qw \\
\lstick{q_1} & \gate{R_y(\theta_1)} &\qw&\qw&\qw&\ctrl{1}& \targ{} & \gate{R_y(\theta_7)}&\ctrl{1}&\targ{}&\gate{R_y(\theta_{13})}&\qw \\
\lstick{q_2} & \gate{R_y(\theta_2)} &\qw&\qw& \ctrl{1}& \targ{} & \gate{R_y(\theta_8])}& \ctrl{1}& \targ{} & \gate{R_y(\theta_{14})} & \qw&\qw \\
\lstick{q_3} & \gate{R_y(\theta_3)} &\qw& \ctrl{1} &\targ{} & \gate{R_y(\theta_9)} & \ctrl{1} & \targ{}&\gate{R_y(\theta_{15})} & \qw&\qw&\qw \\
\lstick{q_4} & \gate{R_y(\theta_4)} & \ctrl{1} & \targ{} & \gate{R_y(\theta_{10})}& \ctrl{1} &\targ{} & \gate{R_y(\theta_{16})} &\qw&\qw&\qw& \qw \\
\lstick{q_5} & \gate{R_y(\theta_5)} & \targ{} & \gate{R_y(\theta_{11})} &\qw& \targ{} & \gate{R_y(\theta_{17})}  &\qw&\qw&\qw&\qw& \qw \\
}
\end{array}
\]
\caption{RealAmplitudes ansatz circuit for the $\mathrm{H_2O}$ molecule.} 
\label{fig:RealAmpl_44}
\end{figure*}

For $\mathrm{H_2O}$ the RealAmplitudes ansatz with two repetitions for six qubits from the Qiskit circuit library was chosen (Figure~\ref{fig:RealAmpl_44}).

The simulations without noise were performed locally with the qiskit primitive estimator, while the noise-corrupted calculations were performed, on the one hand, using noise models extracted from qiskit fake providers and on the other hand, on a real backend (IBMQ System One) from the qiskit runtime service. Apart from the inherent noise handling strategies of Bayesian optimisation, that were turned on for the noisy calculations, no error correction or mitigation techniques were employed. The number of execution shots was set to 1000 except for the initial points and the last five iterations, for which 8192 shots were used.
\subsection{Bayesian Optimisation}
The Bayesian optimisation was realised with the Python packages GPyOpt \cite{gpyopt2016} and GPy \cite{gpy2014}. If not noted otherwise, the Gaussian process model was built with the Mat\`ern 5/2 kernel as implemented in GPy. In selected cases, the \gls{rbf} kernel was used. 
For the kernel hyperparameter optimisation, five \gls{lbfgs} optimisations (modified to deal with bounds \cite{Zhu1997-if}) of the negative log marginal likelihood were performed with different random starting points with at most 1000 iterations, from which the best set of resulting hyperparameters was selected.
\Gls{lcb} was used as the acquisition function:
\begin{equation}
    a_{LCB}(\theta)=\mu(\theta)-\kappa_n\sigma(\theta)
\end{equation}
with $\mu(\theta)$ and $\sigma^2(\theta)$ being the mean and the variance computed from the conditional probability density of the surrogate model for ansatz parameters $\theta$. 
The shift from exploration to exploitation was achieved by calculating the acquisition weight $\kappa_n$ for each iteration $n\ge 0$ according to the formula:
\begin{equation}
\label{eq:AW}
    \kappa_n \coloneqq 
    \max
    \left(0.000001,\, \kappa_0\left(1-\frac{n}{n_{max}}\right)\right)
\end{equation}
where $n_{max}=100$ is the maximal number of iterations and  $k_0$ was set to 1.
For each iteration, the optima of the acquisition functions were determined with the \gls{lbfgs} algorithm with at most 100 iterations.\par
Choosing a fixed number of iterations $n_{max}$ for the Bayesian optimisation is the default stopping criterion in standard implementations such as GPyOpt \cite{gpyopt2016}. More elaborate stopping criteria exist \cite{Bolton2004-zl, makarova2022automatic}, but are not the focus of this work. Instead, the objective was to determine the QC demand that is needed to get sufficiently close to the exact VQE solution. For this purpose,  
in the noise-free experiments, the calculation was stopped when the relative deviation from the reference was smaller than a predefined value (details on this can be found in the \href{https://github.com/milena-roehrs/BOIS-with-immediate-sharing/blob/main/README.md}{code documentation}). Further research is needed to develop an efficient convergence criterion for BOIS with unknown reference values.
For both molecules, each 100 initial ansatz parameters $X_{init}\in[0, 2\pi]^{30 \times r}$ (with r being the number of rotational gates in the respective ansatz and 30 the number of initial points of the feasible set) were generated randomly and used for repeated experiments of the various sharing schemes (except ``no sharing'') to ensure comparability.\par

\subsection{Information Sharing Schemes}
The Pauli-expansion of the Hamiltonians of the different problems defining the VQE objective functions for the different geometries according to Equation~\eqref{equ:vqe-min} can be found in the supplementary material. For simplicity, the expectation values of a Kronecker product of Pauli matrices for a certain set of ansatz parameters, which are represented as the brackets in Equation~\eqref{equ:vqe-min} and correspond to the circuits evaluated on the quantum device, are further referred to as ``Paulis''.
\begin{figure}[ht]
\begin{algorithm}[H]
\caption{Non-Immediate Sharing \cite{self_variational_2021}\label{alg:non-immediate}}
\begin{algorithmic}
    \State Generate starting points: $\theta_{1}^{0}, \ldots, \theta_{K}^{0}$
    \State Compute all Paulis $\mathcal{P}_{i}^{0}$ relevant for any $f^d(\theta_{i}^{0})$
    \ForEach{geometry $d$}
        \State $\mathcal{P}_{i}^{d,0} \gets \mathcal{P}_{i}^{0}$
        \State Compute $f^d(\theta_{i}^{0})$
        \State Set $F^d := \mathrm{min} \left\{f^d(\theta_{i}^{0})\mid i \in \left\{1,...,K\right\}\right\}$
        \State Set information $I^d := \cup_{i}\left\{(\theta_{i}^{0}, f^d(\theta_{i}^{0})\right\}$
    \EndFor
    \ForEach{iteration $n$}
        \ForEach{geometry $d$}
            \State Update $d$-model using $I_d$
            \State Pick $\theta^{d, n}$ as per updated $d$-model
            \State Compute Paulis $\mathcal{P}^{d,n}$ for $f^d(\theta^{d, n})$
            \State Compute $f^d := f^d(\theta^{d,n})$
            \State \algorithmicif{} $F^d > f^d$: update $F^d:=f^{d}$
            \State Update $I^d := I^{d}\cup\{(\theta^{d,n}, f^d(\theta^{d, n}))\}$
        \EndFor
        \ForEach{geometry $d$}
            \ForEach{sharing partner $d^{\prime}$}
                \State Compute $f^{d^{\prime}}(\theta^{d, n})$ from $\mathcal{P}^{d,n}$
                \State $I^{d^{\prime}} := I^{d^\prime} \cup \{(\theta^{d,n}, f^{d^{\prime}}(\theta^{d, n})\}$
            \EndFor
        \EndFor
    \EndFor
    \State \textbf{return} $F^{d}$ for all geometries $d$.
\end{algorithmic}
\end{algorithm}
\end{figure}

This work differentiates between four sharing schemes for exchanging evaluated Paulis among the different optimisation problems. First, they differ by the set of sharing partners which includes either only the nearest neighbours (``nearest-neighbour sharing'') or also all other geometries (``all-to-all sharing''). Second, a distinction is made between the algorithm from reference \cite{self_variational_2021}, where there is a sharing phase after each iteration (Algorithm \ref{alg:non-immediate}) and the newly introduced ``immediate sharing'' approach (Algorithm \ref{alg:immediate})

As in the usual \gls{bo} algorithm, both algorithms begin with an initialisation loop, which computes the initial $K$ points and prepares the sets of available information $I^d$ for each geometry $d$. Then, in the original, ``non-immediate sharing'' scheme (Algorithm~\ref{alg:non-immediate}), the main iteration loop comprises two inner loops. First, we iterate over geometries, pick the next points to investigate ($\theta^{d,n}$) and, effectively, calculate all the necessary Pauli terms to evaluate the objective function values $f^d$. In the second inner loop, we iterate over all geometries again and over the respective sharing partners for each geometry, in order to update the information sets with the shared information.

In contrast, the ``immediate sharing'' scheme we propose (Algorithm~\ref{alg:immediate}) implies only one inner loop per iteration, which enumerates the geometries.
The key idea is to share the information immediately after it is gained (Pauli terms computation step) instead of waiting until the end of the iteration. As a result, the remaining optimisers can make a more informed choice about which point $\theta^{d,n}$ to evaluate next already within the same iteration. Note that the ability of Algorithm~\ref{alg:non-immediate} to run on several quantum devices in parallel is traded for an overall smaller \gls{qc} demand in Algorithm~\ref{alg:immediate}.
It is also important to recognise that in the case of the $\mathrm{H_2}$ molecule sharing is much more straightforward than for $\mathrm{H_2O}$, as the set of relevant Paulis is the same among all geometries investigated. For water, in the initialisation phase of the algorithm, the optimiser of a geometry first calculates only the Paulis relevant for this configuration. Once it shares with another geometry, the missing Paulis are calculated and stored, in case another sharing partner needs them, too. In the iteration phase, all Paulis requested by the sharing partners are directly included in a geometries function evaluation.  
\begin{figure}[ht]
\begin{algorithm}[H]
\caption{Immediate Sharing}
\label{alg:immediate}
\begin{algorithmic}
    \State Generate starting points: $\theta_{1}^{0}, \ldots, \theta_{K}^{0}$
    \State Compute all Paulis $\mathcal{P}_{i}^{0}$ relevant for any $f^d(\theta_{i}^{0})$
    \ForEach{geometry $d$}
        \State $\mathcal{P}_{i}^{d,0} \gets \mathcal{P}_{i}^{0}$
        \State Compute $f^d(\theta_{i}^{0})$
        \State Set $F^d := \mathrm{min} \left\{f^d(\theta_{i}^{0})\mid i \in \left\{1,...,K\right\}\right\}$
        \State Set information $I^d := \cup_{i}\left\{(\theta_{i}^{0}, f^d(\theta_{i}^{0})\right\}$
    \EndFor
    \ForEach{iteration $n$}
        \ForEach{geometry $d$}
            \State Update $d$-model using $I_d$
            \State Pick $\theta^{d, n}$ as per updated $d$-model
            \State Compute Paulis $\mathcal{P}^{d,n}$ for $f^d(\theta^{d, n})$
            \State Compute $f^d := f^d(\theta^{d,n})$
            \State \algorithmicif{} $F^d > f^d$: update $F^d:=f^{d}$
            \State Update $I^d := I^{d}\cup\{(\theta^{d,n}, f^d(\theta^{d, n}))\}$
            \ForEach{sharing partner $d^{\prime}$}
                \State Compute $f^{d^{\prime}}(\theta^{d, n})$ from $\mathcal{P}^{d,n}$
                \State $I^{d^{\prime}} := I^{d^\prime} \cup \{(\theta^{d,n}, f^{d^{\prime}}(\theta^{d, n})\}$
            \EndFor
        \EndFor
    \EndFor
    \State \textbf{return} $F^{d}$ for all geometries $d$.
\end{algorithmic}
\end{algorithm}
\end{figure}

\section{Results and Discussion}
\label{ch:results}
 First, the different sharing schemes are compared in noiseless simulations for $\mathrm{H_2}$. For this purpose, histograms of the iterations of convergence are provided. The iteration of convergence is defined here as the first iteration in which the last of the eight optimisations corresponding to the different bond lengths reaches convergence. This is followed by an analysis of an illustrative example calculation highlighting the strengths and weaknesses of the individual sharing schemes. Subsequently, the performance in the presence of noise is investigated. This is carried out, on the one hand,  via noise models that simulate quantum computers of different sizes and, on the other hand, on the real quantum computer ``ibmq\_ehningen'', also known as IBM Q System One.
 In addition, the influence of two characteristics of the Bayesian optimisation and its surrogate model, namely, the use of different kernels and acquisition weight decrease rates, is examined. Finally, a modified immediate all-to-all sharing method is applied to calculate part of the $\mathrm{H_2O}$ \gls{pes}, with and without fixed symmetry.
\subsection{Comparison of Sharing Schemes}
For each of the sharing schemes ``immediate all-to-all sharing'', ``all-to-all sharing'' and ``no sharing'' 100 repetitions were performed.
\begin{figure}
\centering
    \includegraphics[width=0.9\linewidth]{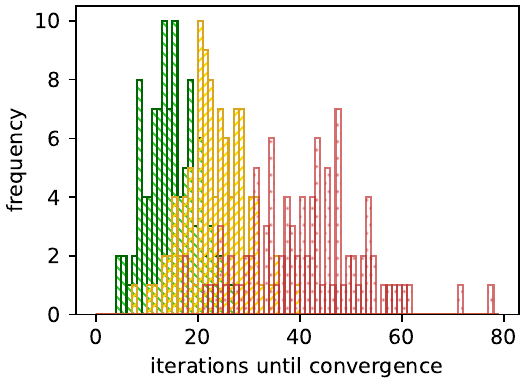}
    \captionof{figure}{Iteration of convergence when using \gls{bo} with no sharing (red - dots), ``non-immediate all-to-all sharing'' (yellow - ascending stripes) and ``immediate all-to-all sharing'' (green - descending stripes) for calculating the $\mathrm{H_2}$-\gls{pes}.}
    \label{fig:effect-of-sharing}
\end{figure}
A histogram of the resulting iterations of convergence is shown in Figure~\ref{fig:effect-of-sharing}. As one can see, the ``immediate all-to-all sharing scheme'' leads to the fastest convergence requiring on average 15 iterations, followed by the ``all-to-all sharing'' scheme with 23. The ``no-sharing'' case reaches convergence the latest, requiring on average 41 iterations.\par
\begin{figure*}
  \centering
  \begin{subfigure}{0.45\textwidth}
    \centering
    \includegraphics[width=\linewidth]{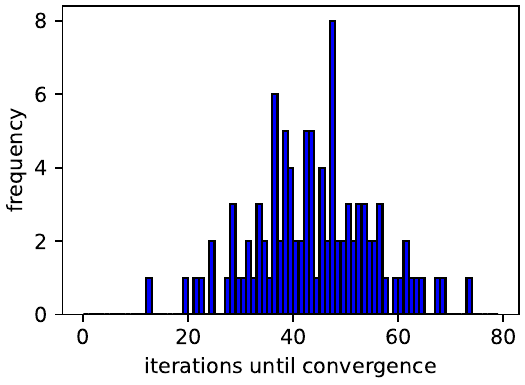}
    \caption[Convergence iteration with ``nearest-neighbour sharing'']{Nearest-neighbour sharing.}
    \label{fig:NNhisto}
  \end{subfigure}
  \begin{subfigure}{0.45\textwidth}
    \centering
    \includegraphics[width=\linewidth]{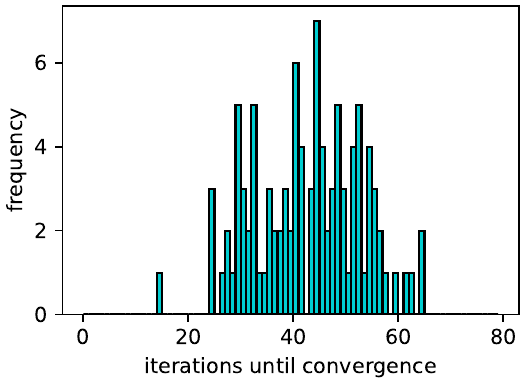}
    \caption[Convergence iteration with ``immediate nearest-neighbour sharing'']{Immediate nearest-neighbour sharing.}
    \label{fig:iNNhisto}
  \end{subfigure}
  \caption[Convergence histograms of ``nearest-neighbour sharing'' schemes]{Convergence iteration of the different ``nearest-neighbour sharing'' schemes.}
  \label{fig:nn-effect}
\end{figure*}

For $\mathrm{H_2}$, the number of iterations is a direct measure of the number of quantum calculations performed, as each iteration, independent of the sharing scheme, corresponds to eight energy evaluations. Each of them consists of five Pauli string expectation values that are calculated as the mean of a defined number of executions (shots) of the quantum circuit.
Thus, the results of the experiment confirm that the sharing of information reduces the number of quantum jobs needed to obtain an accurate \gls{pes}. In addition, the newly introduced ``immediate sharing'' scheme offers a notable improvement over the other sharing schemes.
The histograms obtained in a similar way when restricting information sharing to nearest neighbours can be found in Figure~\ref{fig:nn-effect}. 
As one can see, both ``nearest-neighbour sharing'' and ``immediate nearest-neighbour sharing'' lead to histograms very similar to the ``no sharing'' case displayed in Figure~\ref{fig:effect-of-sharing}. This indicates that in the given setup, on the one hand, the neighbour geometries around a specific inter-nuclear distance do not suffice to improve the convergence speed of the \gls{bo} and, on the other hand, the optimisation problems whose geometries are further away in the geometry grid are still similar enough to contribute valuable information by sharing.\par
\begin{figure*}
  \begin{subfigure}{1\textwidth}
    \centering
    \includegraphics[width=0.7\linewidth]{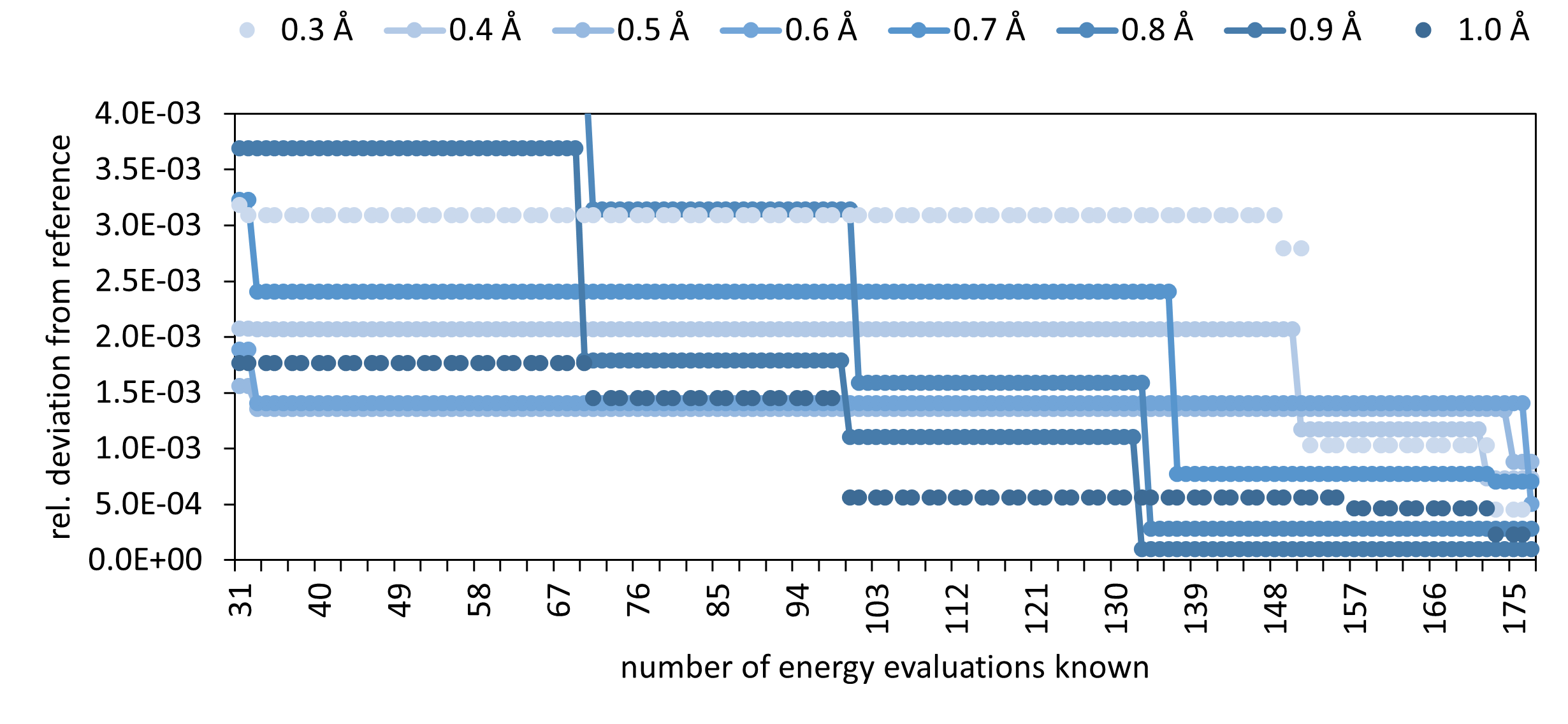}
    \caption{Nearest-neighbour sharing scheme.}
    \label{fig:NNconv}
  \end{subfigure}%
  \hfill
  \begin{subfigure}{1\textwidth}
    \centering
    \includegraphics[width=0.7\linewidth]{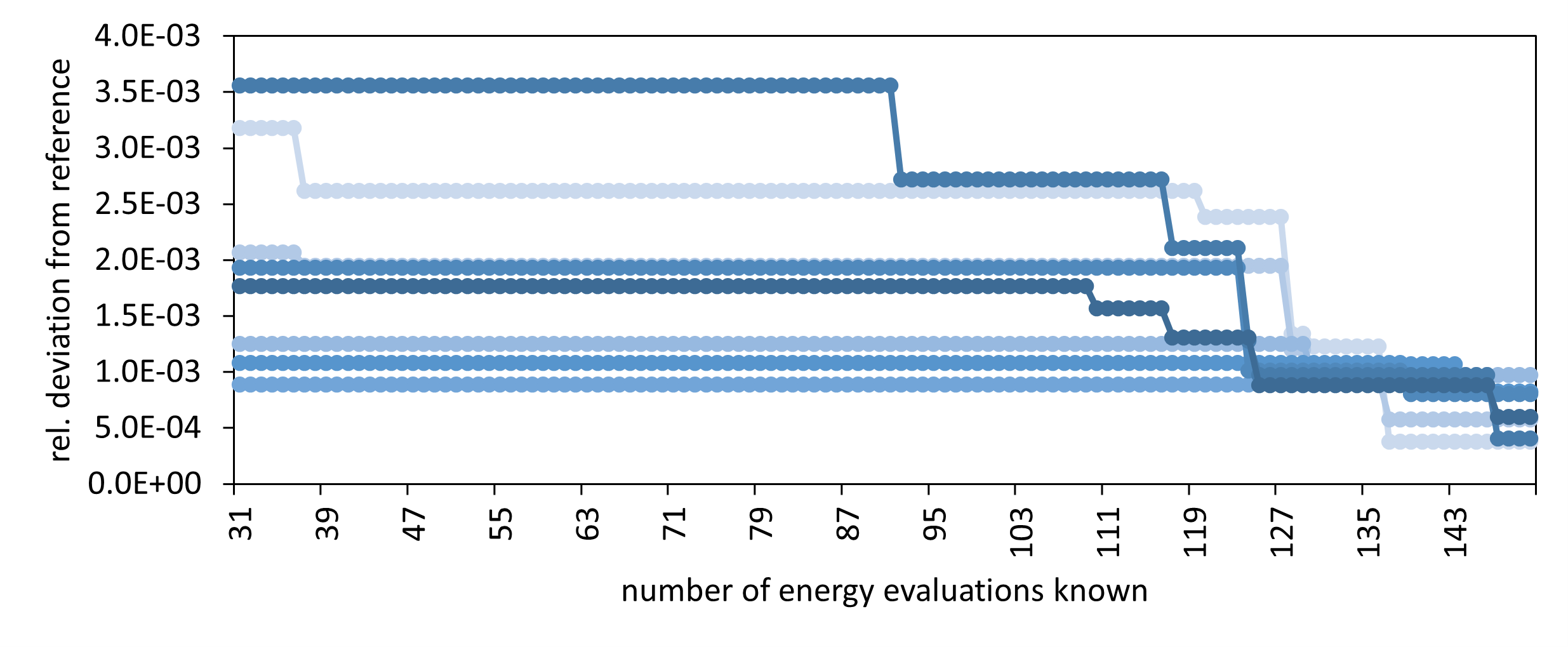}
    \caption{``Immediate all-to-all sharing'' scheme.}
    \label{fig:iaaconv}
  \end{subfigure}
  \hfill
  \begin{subfigure}{1\textwidth}
    \centering
    \includegraphics[width=0.7\linewidth]{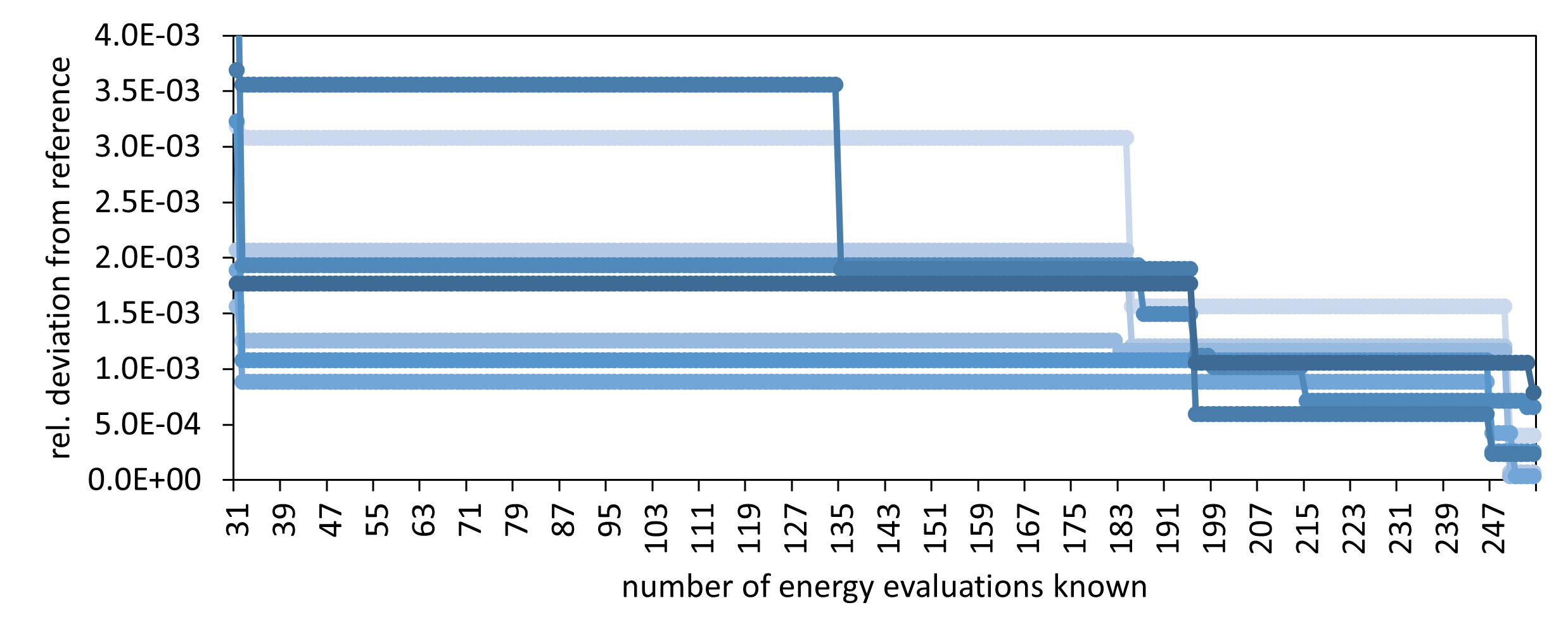}
    \caption{``All-to-all sharing'' scheme.}
    \label{fig:aaconv}
  \end{subfigure}
  \caption[Convergence of different sharing schemes without noise]{Convergence of the relative deviation of the best seen energy so far from the reference. The horizontal axis measures the number of energy evaluations known to (the majority of) the optimisations (with the exception of the 0.3 and 1.0 Å optimisations in the ``nearest-neighbour sharing'' case, which know one energy point less per iterations). The different iterations are marked by ticks. In the ``all-to-all sharing'' scheme, an optimisation model learns eight energies per iteration, and in the ``nearest-neighbour sharing'' scheme three. }
  \label{fig:conv-diff-schemes}
\end{figure*}
The above findings are explained in more detail below through an illustrative example. 
One set of input points has been chosen for analysing the corresponding ``immediate all-to-all'', ``all-to-all'' and ``nearest-neighbour'' sharing calculations. They converged after \href{https://github.com/milena-roehrs/BOIS-with-immediate-sharing/blob/main/H2_raw_data/comparison_of_sharing_schemes/immediate_aa/outdata_rep30}{15}, \href{https://github.com/milena-roehrs/BOIS-with-immediate-sharing/blob/main/H2_raw_data/comparison_of_sharing_schemes/all-to-all-sharing/outdata_rep30}{28}, and \href{https://github.com/milena-roehrs/BOIS-with-immediate-sharing/blob/main/H2_raw_data/comparison_of_sharing_schemes/NN/outdata_rep30}{49} iterations, respectively, and hence can be considered representative for the different convergence speeds.
\par For a certain set of input points and geometry, the first 30 energy evaluations, that correspond to these input points, are the same among the different sharing schemes. The first real optimisation step belonging to the 0.3 \AA\ geometry is associated with the 31st energy evaluation. Starting from this point, the relative deviation of the best value seen so far from the reference is plotted in Figure~\ref{fig:conv-diff-schemes}.\par Despite some randomness being present through the choice of the starting points of the kernel's hyperparameter optimisation, the points evaluated for distance 0.3 Å for the different methods have a pairwise Manhattan distance of less than $6\cdot10^{-6}$, hence they are essentially the same (In the ``non-immediate sharing'' schemes this is also the case for the other geometries). The point found by the optimisation of the acquisition function immediately meets the accuracy requirement when being shared with the 0.6 Å geometry. This means, one of the eight optimisations has already converged after the first iteration for the ``immediate) all-to-all sharing'' cases. 
In contrast, in the ``nearest-neighbour sharing'' scheme, in which the optimisation at 0.6 Å does not get the information from the optimisation at 0.3 Å, the former requires 49 iterations to converge. In both ``all-to-all sharing'' cases, four of the eight optimisations first meet the convergence criterion at a point that was shared from some optimisation that was not the neighbour. This is clear evidence that information sharing between longer distances directly influences the iteration in which convergence is reached.\par
Besides, the strength of the immediate sharing principle to avoid calculating redundant information can be seen from the examples output. In the non-immediate ``all-to-all sharing'' case, all but one point evaluated in the first iteration have a pairwise Manhattan distance of less than 0.41. In contrast, in the ``immediate sharing'' scheme, the Manhattan distance in the first iteration is at least 1. In other words, among different optimisations basically the same point of the feasible set is determined to be the most interesting one for the next evaluation. This makes sense because without using ``immediate sharing'' the unexplored regions of the feasible set (which dominate the acquisition function in the exploration phase via large variances) are the same for each of the eight optimisations.\par In contrast, when updating the model immediately, once a point is found, the variance at the former optimum becomes zero and very small in the surrounding such that the \gls{lcb} value is higher resulting in the optimisation of the acquisition function to end up at a different point. The influence of the different schemes is harder to analyse when the algorithm shifts to exploitation. However, it remains plausible that incorporating more information early on improves the surrogate model, which speeds-up the optimisation.\par
We saw above that the ``immediate all-to-all sharing'' scheme performed best when only taking into account the quantum computation demand, which is the focus of this work. This is why in the following, this method is predominantly used. However, before continuing with incorporating noise, there will be a brief analysis on the scaling behaviour of the classical computations and its implications for geometries composed of a large number of points. Unfortunately, updating the surrogate model scales as $\mathcal{O}(n^3)$, with $n$ being the number of function evaluations included in the model so far \cite{rasmussen_gaussian_2019}. Therefore, when investigating families of problems defined by more than one physical parameter, the calculation cost for the classical part quickly explodes. Thus, it becomes essential to implement a sharing cutoff \cite{self_variational_2021}. As the above discussed results indicate that simply cutting after the nearest neighbours does not in general solve the problem, the sharing cut-off needs to be more elaborate. One option would be to pairwise quantify the similarity/sharing potential of different problems of a given problem family by a function $p_n: G \times G \to [0,1]$ with G being the space of Hermitian $\mathrm{2^n \times 2^n}$ matrices.  
The sharing cut-off could now be defined as a value of this potential $\gamma \in (0,1)$ above which two problems are declared as ``similar enough'' to significantly benefit from sharing.
As a very simple candidate for such a measure we propose $1-\pi^{-1}\varphi$, with $\varphi$ being the angle of the of the Pauli coordinates of the Hamiltonians. This would classify correctly at least when the Hamiltonians are exactly the same (in this case the total demand is the number of parallel optimisations smaller compared to non-communicating optimisations) or if they have no common non-zero Pauli coordinates. 
It remains subject to further studies to determine if this measure is suitable for correctly classifying less clear cases, what alternative measures exist and how to balance between the cost of the classical and quantum calculations.
\subsection{Calculations with Noise}
\begin{figure*}[ht!]
  \centering
  \begin{subfigure}{0.3\textwidth}
    \centering
    \includegraphics[width=\linewidth]{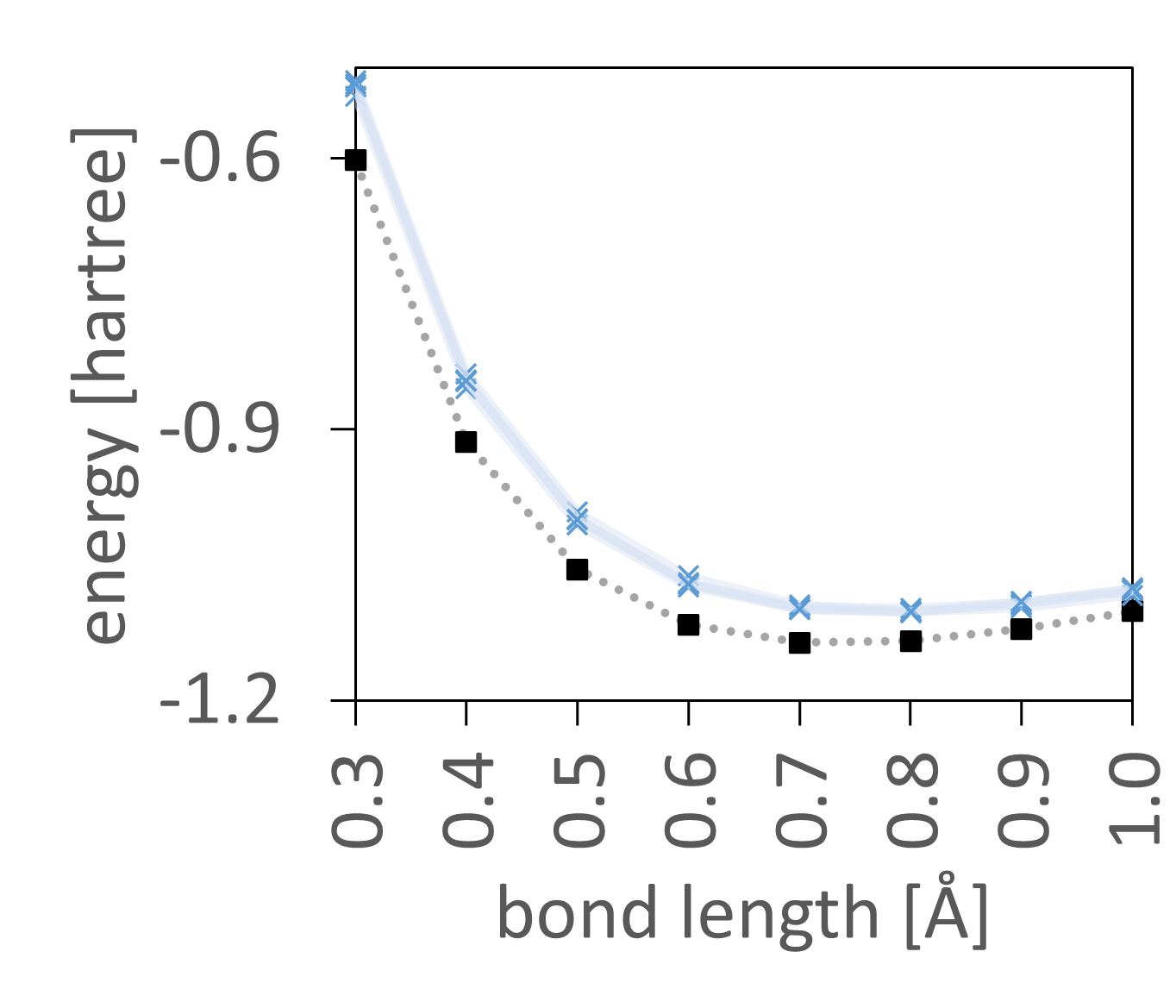}
    \caption[FakeLima PES]{5 qubit noise mode\\of ibmq\_lima.}
    \label{fig:FakeLima}
  \end{subfigure}
  \begin{subfigure}{0.3\textwidth}
    \centering
    \includegraphics[width=\linewidth]{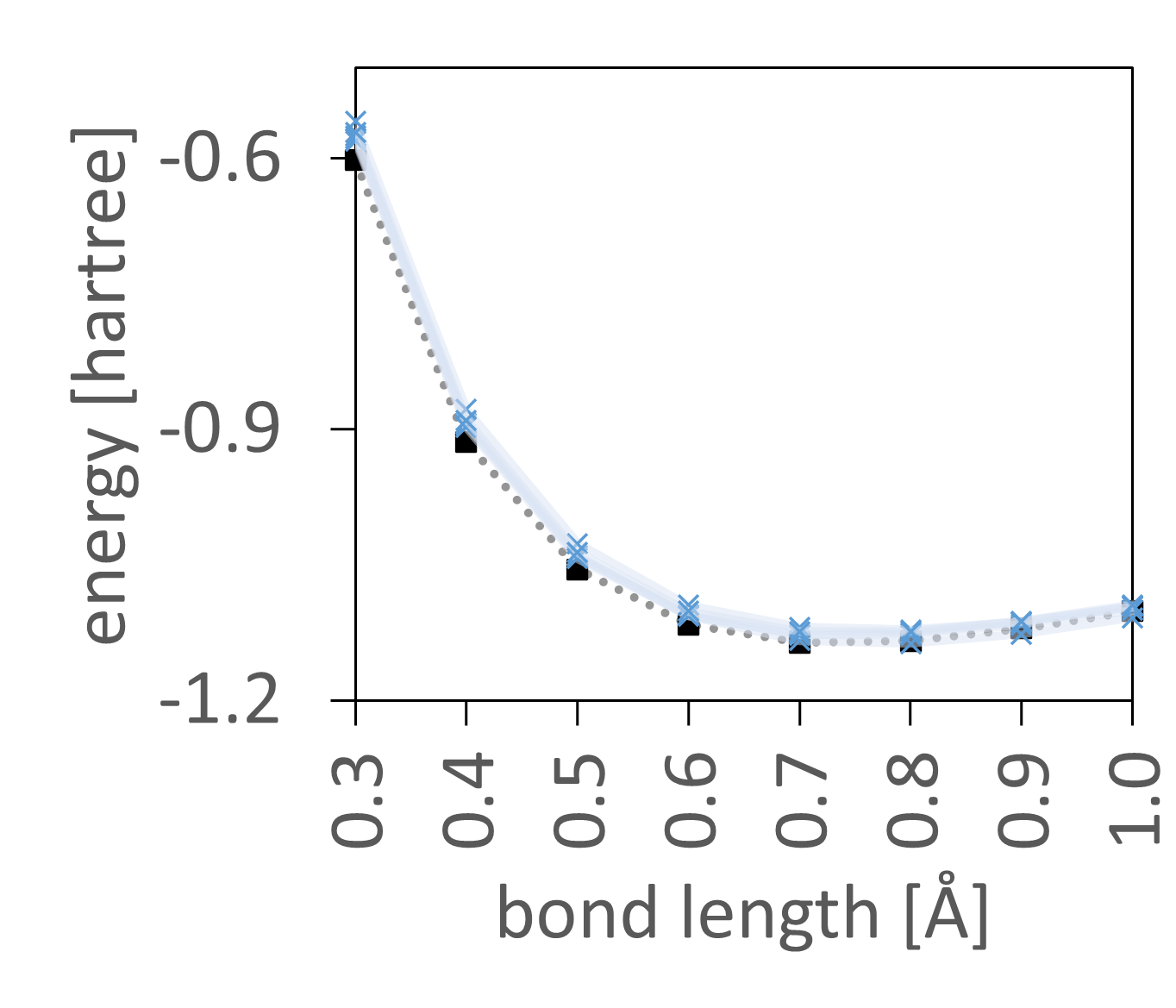}
    \caption[FakeKolkata PES]{27 qubit noise mode\\of ibmq\_kolkata.}
    \label{fig:FakeKolkata}
  \end{subfigure}
  \begin{subfigure}{0.3\textwidth}
    \centering
    \includegraphics[width=\linewidth]{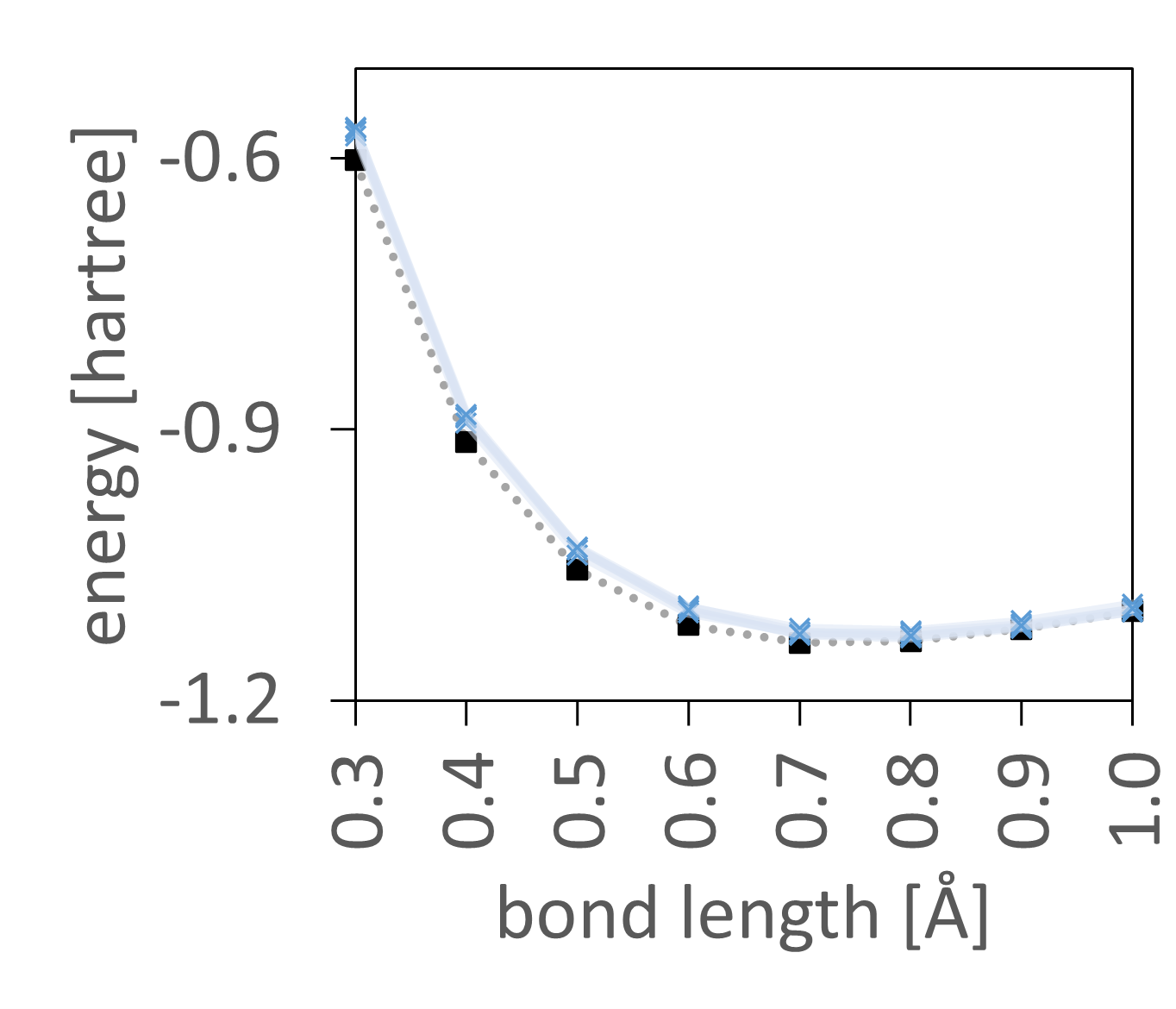}
    \caption[Real device PES]{IBM Q System One\\(ibmq\_ehningen).}
    \label{fig:EhningenPES}
  \end{subfigure}
  \caption[PES with noise]{\Gls{pes} of $\mathrm{H_2}$ calculated five times through \gls{bois} based \gls{vqe} using the ``immediate all-to-all sharing'' scheme [blue] in comparison to the exact value [black].}
  \label{fig:with_noise}
\end{figure*}

The performance of the ``immediate all-to-all sharing'' scheme with 100 iterations in the presence of noise was investigated using two noise models as well as the ibmq\_ehningen quantum computer (IBM Q System One). The experiment was repeated five times. The results are plotted in Figure~\ref{fig:with_noise}. In agreement with the results of Self \textit{et al.} \cite{self_variational_2021}, the points closely resemble the exact \gls{pes} while showing a systematic offset which can be associated with depolarisation errors. This offset is most pronounced in the calculations utilising the ibmq\_lima based noise model. This could originate from lower error rates in the FakeKolkata noise model.\par
The real device results show good agreement with the reference as after 100 iterations all optimisations reached a relative deviation of less than 0.02. 
Looking at the detailed convergence behaviour (shown in Figure~\ref{fig:ehnigen-conv-rep3} and in the supplementary material for the four other runs), this precision has already been achieved in the 32nd iteration. 
\begin{figure}[h]
\centering
    \includegraphics[trim=20 0 0 0,clip,width=0.9\linewidth]{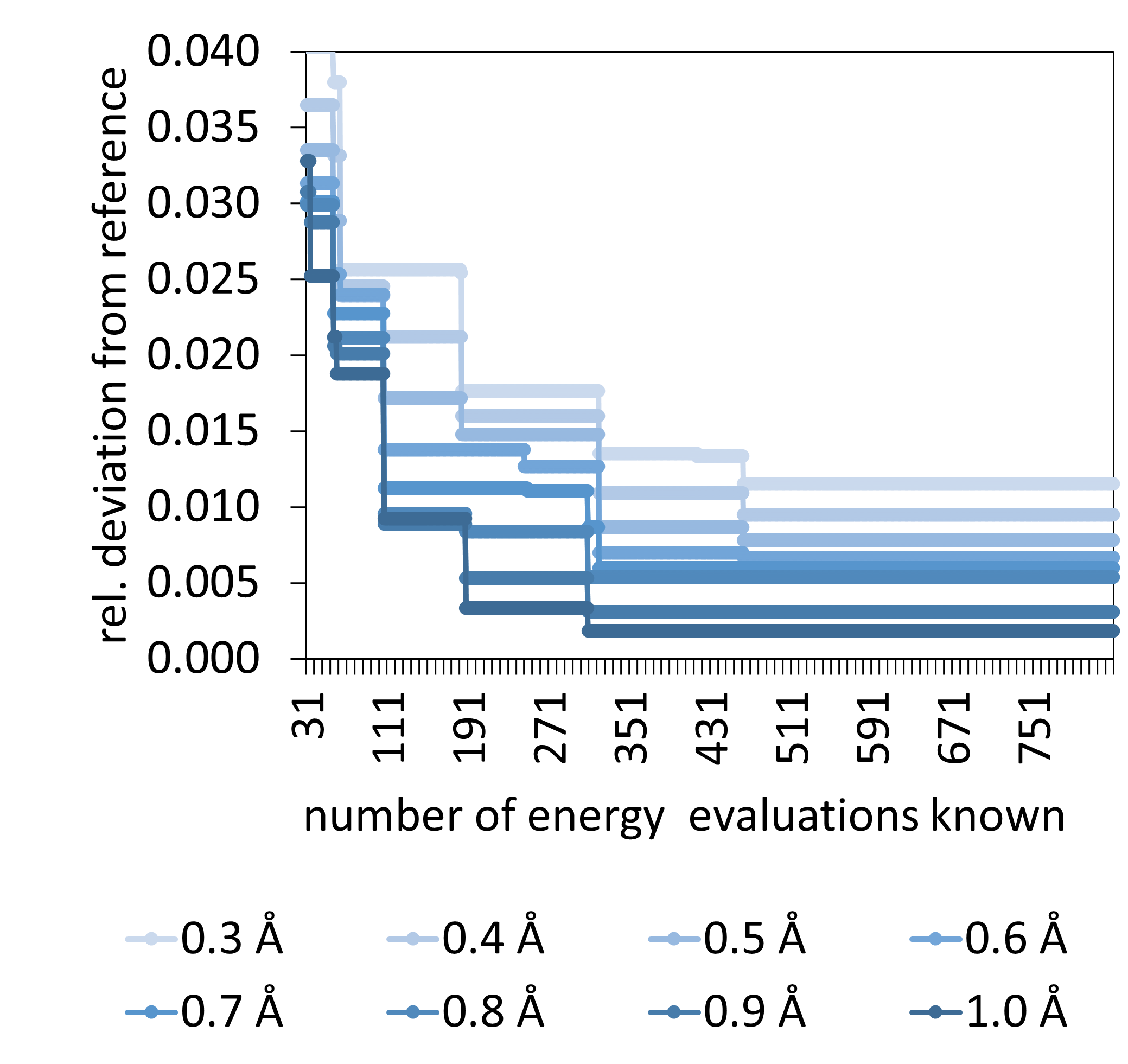}
    \captionof{figure}{Convergence of the relative deviation of the best seen energy so far from the reference on the IBM Q System One (``Repetition 3'' initial points; for the other repetitions see the supplementary material.)}
    \label{fig:ehnigen-conv-rep3}
\end{figure}
Furthermore, Figure~\ref{fig:ehnigen-conv-rep3} illustrates clearly how the improvement of the deviation of the best seen value to the reference appears in groups. This means, a set of ansatz parameters, that was found in the course of one of the optimisations, was shared and from then on served as the best seen value for other iterations, too. This behaviour appears, for example, at energy evaluation 321, which takes place in the 37th iteration. In this instance, the acquisition function corresponding to the molecule with bond length 0.5 Å shares its energy evaluation with all other optimisations leading to an improved best seen energy for the optimisations at 0.3 Å, 0.4 Å, 0.5 Å, 0.6 Å and 0.7 Å. This further supports the claim that sharing beyond nearest neighbours is important in this set-up and a suitable similarity cut-off would need to include at least neighbours of neighbours.\par One should note here that this occurrence of different optimisations reaching a new ``best seen value'' in groups is much less pronounced in the noise-free simulations in comparison to the real, noise-corrupted device. This might originate from the acquisition functions in the noise-corrupted case being a less accurate description of the gain in insight a new point could bring. As a result, many function evaluations are unsuccessful in achieving improvement. If one of the optimisations finds a set of parameters that is indeed valuable for several of the optimisations, the best seen values improve as the optimisation does not know many other points with low energies with which the new point has to compete. In contrast, in the simulation, the model is expected to be more accurate and the optimisation of the acquisition function hence might be more successful in finding valuable points. As a result, the competition to be the next ``best seen value'' is much tougher.
Another observation that further supports this hypothesis is the difference in time between subsequent improvements and their order of magnitude when appearing. In the simulation (Figure~\ref{fig:conv-diff-schemes} (b)) there is first a phase of few improvements, in agreement with the algorithm starting with exploring the feasible set, which is followed by a phase in which there is an improvement of the best seen value in nearly every iteration until convergence is reached. However, the improvement has only a order of magnitude of roughly 0.0005.
In contrast, on the real device (Figure~\ref{fig:ehnigen-conv-rep3}) throughout the whole calculations there are repeatedly long periods without any improvement (e.g. from energy evaluation 65 to 107) followed by several optimisations improving their relative deviation from reference by around 0.001--0.01 through the same shared point. This could indicate that information sharing in part compensates for unreliability in finding new valuable points, which is interesting to keep in mind, also for other types of heuristic-based optimisations. 
\subsection{Influence of the Acquisition Weight Decrease Rate}
\begin{figure}
\centering
    \includegraphics[width=0.9\linewidth]{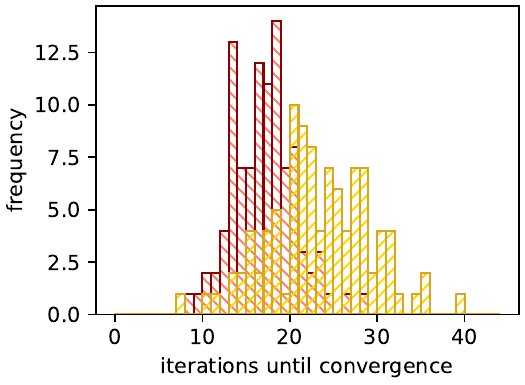}
    \captionof{figure}{Influence of the decrease rate of the acquisition weight on the iteration of convergence comparing a decrease rate of 1/30 (red - descending stripes) with 1/100 (yellow - ascending stripes).}
    \label{fig:different_AW}
\end{figure}
The decrease of the acquisition weight regulates the shift of the driving force of the optimisation from exploration to exploitation. To ensure that the optimisation reaches the exploitation phase in the final iterations, the decrease rate depends on the maximum number of iterations chosen for the calculation. This dependency should be kept in mind to avoid exaggerating the increase of convergence speed in a self-fulfilling prophecy manner. When discovering that information sharing leads to faster convergence in the simulations, one might be tempted to set the maximum number of iterations on the real device for the information sharing case to a smaller number compared to the ``no-sharing'' case.
For examining the resulting effect on the convergence, each 100 repetitions of the BO based VQE utilising the ``non-immediate all-to-all sharing'' scheme were conducted, on the one hand with 30 iterations and on the other hand with 100 chosen as the maximum number of iterations $n_{max}$. As one can see in Equation~\eqref{eq:AW}, this results in a linear decrease rate of the acquisition weight of 1/30 and 1/100, respectively.\par
Upon conducting the experiment, it became evident that a higher decrease rate leads to a faster convergence in the described set-up, as visualised in Figure~\ref{fig:different_AW}. 
When specifying a decrease rate of 1/30 the optimisation reached convergence on average within 17 iterations. In contrast, the optimisations with decrease rate 1/100 needed on average 23 iterations. This strong dependence of the convergence speed from $n_{max}$ suggests that, in this setting, a short exploration time suffices and focusing on exploitation early on is more beneficial. Still, the larger of both $n_{max}$ values was chosen in the remaining experiments to reduce the risk of converging to a local minimum.

\subsection{Influence of the Covariance Function}
\begin{figure}
\centering
    \includegraphics[width=0.9\linewidth]{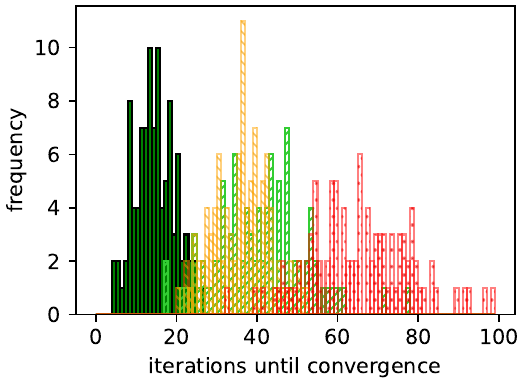}
    \captionof{figure}[Influence of the kernel]{Influence of the kernel function on the iteration of convergence for the ``immediate all-to-all sharing'' and the ``no sharing'' scheme: ``immediate all-to-all sharing'' + Mat\`ern 5/2 (dark green - fully filled), ``no sharing'' + Mat\`ern 5/2 (light green - ascending stripes), ``immediate all-to-all sharing'' + RBF (yellow - descending stripes) and ``no sharing'' + RBF (red - dots). Two of the 100 ``RBF no sharing'' calculations did not converge within the maximal number of iterations and hence do not appear in the histogram.}
    \label{fig:different_kernel}
\end{figure}
In addition, the influence of the kernel defining the surrogate model on the speed of convergence has been examined. For this purpose (in addition to the Matérn 5/2 calculations performed for the sharing scheme comparison in Figure~\ref{fig:effect-of-sharing}),
100 repetitions of ``immediate all-to-all sharing'' and ``no-sharing'' using the \gls{rbf} kernel have been executed on noiseless simulators.
For both cases, the use of the \gls{rbf} on average led to a slower convergence by a factor of 2.4 in the case with sharing and a factor of roughly 1.5 without.
However, the benefit of sharing is still similarly pronounced for the \gls{rbf} kernel as for the Matérn 5/2 kernel. 
This suggests that combining information sharing with special kernels that are particularly suited to mimic the characteristics of the VQE objective function is a very promising approach for further reducing the necessary number of iterations. For example, the ``VQE-Kernel'' introduced by Nicoli \textit{et al.} \cite{nicoli2023physicsinformed} is likely to have such a beneficial effect.

\subsection{\texorpdfstring{Quantum Demand Reduction in $\mathrm{H_2O}$}{Quantum Demand Reduction in the Water Molecule}}
\begin{figure}
\centering
    \includegraphics[width=0.9\linewidth]{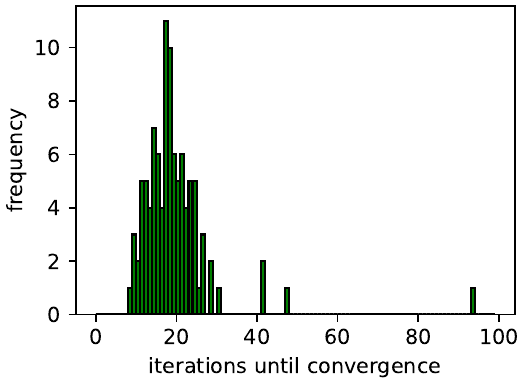}
    \captionof{figure}[Influence of the kernel]{Iteration of convergence of $\mathrm{H_2O}$ using the immediate all-to-all sharing scheme.}
    \label{fig:H2O_hist}
\end{figure}

Finally, \gls{bois} with immediate sharing has been used to investigate the $\mathrm{H_2O}$ \gls{pes}. First, the 2D \gls{pes} of $\mathrm{H_2O}$ with symmetric bonds was determined.
The historgram displaying the iterations of convergence of 100 repetitions of calculating the $\mathrm{H_2O}$ \gls{pes} is shown in Figure~\ref{fig:H2O_hist}. On average, with sharing after 20 iterations the \gls{pes} is converged. The median is 18. In contrast, without sharing, none of the repetitions resulted in a fully converged PES within 100 iterations. Only 5 \% of the individual optimisers converged, needing at least 55 iterations. In comparison, the fastest individual optimisation with sharing converged after 6 iterations.\par
It is important to note here that the quantum computational demand per iteration in the case of water is slightly higher with sharing than without. This is because of the evaluation of additional Paulis necessary for sharing with other geometries that have a slightly different set of non-zero expansion coefficients. For a function evaluation on one of the five higher-symmetric $D_{\infty h}$-geometries (optimiser 20 -- 24, as specified in the supplementary material) the Hamiltonian can be expressed as a linear combination of 55 basis matrices. If sharing the evaluation with one of the 20 $C_{2v}$ geometries, 45 of the necessary 95 Pauli expectation values are missing and need to be calculated causing the \gls{qc} overhead. An overview of the \gls{qc} demand (in the average case) caused in the different parts of the optimisation with or without sharing is provided in Table~\ref{tab:QCdemand2D}.
\begin{table}[h]
  \centering
  {\footnotesize
    \begin{tabular}{ccc}
      \hline
      & \multicolumn{2}{c}{QC Demand [Pauli Eval.]} \\
      & \multicolumn{1}{c}{with sharing} & \multicolumn{1}{c}{without sharing}\\
      \hline
      Initialisation & 2 850 & 65 250  \\
      Per Iteration & 2 375 & 2 175 \\
      All Iterations &47 500&>217 500\\
      \hline
      Total &50 350&>282 750\\
      \hline
    \end{tabular}
  }
  \caption{\footnotesize Quantum computational demand for the different parts of the optimisation algorithm with and without sharing measured in the number of Pauli-evaluations for the 2D \gls{pes}-grid.}
  \label{tab:QCdemand2D}
\end{table}
Evidently, despite the higher cost per iteration, the quantum demand for convergence with sharing is approximately five times smaller than the demand after 100 iterations without sharing, after which the \gls{pes} has not even converged. For the iterations only (when not taking into account the higher costs of initialising the surrogate model with 30 points for each geometry), the factor is approximately 4. 
\subsection{3D PES with asymmetry}
To investigate the performance of the methods for \gls{pes} with higher dimension and for Hamiltonians with a larger set of non-zero expansion coefficients, a cutout of the $\mathrm{H_2O}$ \gls{pes} including geometries with asymmetric O--H bonds has been investigated. For those cases, the Hamiltonian needs 175 Kronecker products of Pauli matrices for its expansion.
Similarly as before, the advantage of immediate sharing in comparison to no sharing is evident. Without sharing, none of the repeated calculations archived a \gls{pes} of the desired accuracy after 100 iterations.
In contrast, with sharing, on average the convergence criterion was fully met after 17 iterations. The quantum demand on the basis of this average value is shown in Table~\ref{tab:QCdemand3D}. The slightly smaller number of iterations necessary to reach convergence with respect to the 2D \gls{pes} case probably originates from the slightly larger number of simultaneously optimised geometries leading to an even larger information gain per iteration.  In total, with sharing the number of Pauli evaluations until convergence is at least a factor of 6 smaller.

\begin{table}[h]
  \centering
  {\footnotesize
    \begin{tabular}{ccc}
      \hline
      & \multicolumn{2}{c}{QC Demand [Pauli Eval.]} \\
      & \multicolumn{1}{c}{with sharing} & \multicolumn{1}{c}{without sharing}\\
      \hline
      Initialisation & 5 250 & 120 150  \\
      Per Iteration & 4 725 & 4 005 \\
      All Iterations & 80 325 &>400 500\\
      \hline
      Total &85 575&>520 650\\
      \hline
    \end{tabular}
  }
  \caption{\footnotesize Quantum computational cemand for the different parts of the optimisation algorithm with and without sharing measuered in the number of Pauli-evaluations for the 3D \gls{pes}-grid.}
  \label{tab:QCdemand3D}
\end{table}

\section{Conclusion and Outlook}

The potential energy surface of $\mathrm{H_2}$ was calculated in order to compare the convergence behaviour of different \gls{bois} based \gls{vqe} setups.
The results support that \gls{bois} based \gls{vqe} is a promising candidate to decrease the total number of quantum jobs necessary to obtain an accurate \gls{pes}, even on real devices. The novel sharing approach ``immediate sharing'' showed an additional speed-up in comparison to the methods discussed in the literature.\par 
Furthermore, we presented a computational setting for the $\mathrm{H_2}$ molecule, where exchange of information beyond nearest neighbours significantly accelerated the convergence.
In light of the fact that the cost of updating the surrogate model increases cubically with respect to the number of function evaluations, it becomes necessary to employ more sophisticated similarity heuristics that determine the pairs of problems between which exchange of information should take place. A very simple example of such a heuristic was presented, based on the angle between the expansion coefficients of the Hamiltonians. Assessment and improvement of this measure might constitute an interesting further research direction both in terms of the theoretical and experimental work.  
For example, numerical studies should be performed to determine how the the sharing speedup factor is affected by increasing dissimilarity quantified by this measure as well as the geometric similarity.\par
Alternatively, one could explore the option of using \gls{bois} with immediate all-to-all sharing as a method to find a good starting point for other optimisation algorithms, for example \gls{spsa}, to benefit from some speed-up while still keeping the number of function evaluations in the model small.\par
Moreover, it was shown that the convergence speed is sensitive to both the decrease rate of the acquisition function and the surrogate model used. On the one hand, this highlights the importance of accurately describing the exact optimisation setup when reporting benchmark studies. On the other hand, it indicates the potential of the method improving even further, for example, when combining it with covariance and acquisition functions that have been shown to be explicitly well suited for application in VQE such as the ``VQE-kernel'' presented by Nicoli \textit{et al.} \cite{nicoli2023physicsinformed} or the ``quantum kernel'' reported by Smith \textit{et al.} \cite{smith_faster_2023}.\par
Moreover, it would be interesting to conduct further studies that explore how other characteristics of the \gls{bois} setup, like the details of the hyperparameter optimisation of the surrogate model, the chemical basis set used or the choice of ansatz circuit, influence the convergence speed. In the immediate sharing setup, also the order in which the different geometries are evaluated might play a role.\par
Another approach to further improve the method could be the development of a new type of acquisition function that takes the sharing aspect into account and does not find the best point for one specific optimisation, but rather promises the largest overall information gain (a kind of utilitarian acquisition function).\par
Besides this in-depth investigations of single aspects of the algorithm, it was demonstrated that information sharing can also be beneficial in molecules with more complex geometries. It was shown for the investigated example of $\mathrm{H_2O}$ that the quantum computational overhead per iteration introduced by sharing between non-identical sets of relevant Paulis is compensated by needing much fewer iterations until convergence leading to a, in total, significantly reduced \gls{qc} demand compared to the case with no sharing.
\par
In summary, the methodology is promising and should be further explored in future research.

\section{Data and Code Availability}
A final version of the code can be accessed at \href{https://github.com/milena-roehrs/BOIS-with-immediate-sharing}{https://github.com/milena-roehrs/BOIS-with-immediate-sharing}.
The raw data supporting the findings of this study are available in the same GitHub repository.

\section{Acknowledgments}
The authors thank Professor Anita Schöbel of the University of Kaiserslautern-Landau for helpful discussions during the initial to intermediate stages of this work.

\bibliographystyle{quantum}
\bibliography{mybibliography}

\end{document}


\title{Improved BOIS for VQE:\\Supplementary Material}
\date{\today}

\maketitle

\section*{Details of individual optimisation problems}
\subsection*{$\mathrm{H_2}$}
\begin{table}[h!]
    \centering
    {\small
    \begin{tabular}{ccccccc}
      \toprule
      && \multicolumn{5}{c}{Pauli expansion coefficients} \\
      \# & bl [\AA] & II & IZ & ZI & ZZ & XX\\
      \midrule
      0 & 0.3 & -0.75374195 & 0.80864891 & -0.80864891 & -0.01328798&0.16081852\\
      1 & 0.4 & -0.86257953 & 0.68881943 & -0.68881943 & -0.01291397&0.16451542\\
      2 & 0.5 & -0.94770788&  0.58307963& -0.58307963& -0.01251643&0.16887023\\
      3 & 0.6 & -1.00712708&  0.49401379& -0.49401379& -0.01206439&0.17373064\\
      4 & 0.7 & -1.04391252&  0.42045568& -0.42045568& -0.0115074 &0.17900058\\
      5 & 0.8 & -1.0632128 &  0.35995942& -0.35995942& -0.01080973&0.18462678\\
      6 & 0.9 & -1.07028327&  0.30978728& -0.30978728& -0.00996911&0.19057169\\
      7 & 1.0 & -1.06924349&  0.26752865& -0.26752865& -0.00901493&
  0.19679058\\
      \bottomrule
    \end{tabular}
    }
    \caption{Pauli expansion coefficients corresponding to a certain optimisation number (\#) and bond length (bl) of the $\mathrm{H_2}$ molecule.}
\end{table}

\subsection*{$\mathrm{H_2O}$ symmetric}

\begin{description}
   \item[\#0] 
   \begin{itemize}
    \item   
   \textbf{Geometry:}\\
    O 0.0 0.0 0.0\\
    H 0.0 0.0 0.8\\
    H 0.0 0.8 4.898587196589413e-17
    \item
    \textbf{Pauli strings:} 'IIIIII', 'IIIIIZ', 'IIIZXX', 'IIIIYY', 'IIIIZZ', 'IIIIZI', 'IIIZZI', 'IIIZZZ', 'IIIZII', 'IIIZIZ', 'IIIZYY', 'IIIIXX', 'IIZIII', 'IIZIIZ', 'IIZZXX', 'IIZIYY', 'IZXIZX', 'IIXIZX', 'IZXIIX', 'IIXIIX', 'ZXXIII', 'ZXXIIZ', 'IYYIII', 'IYYIIZ', 'ZXXZXX', 'IYYZXX', 'ZXXIYY', 'IYYIYY', 'XXXXXX', 'YXYXXX', 'XXXYXY', 'YXYYXY', 'IZZIII', 'IZZIIZ', 'IZZZXX', 'IZZIYY', 'ZXZIZX', 'IXIIZX', 'ZXZIIX', 'IXIIIX', 'ZZIIII', 'ZZIIIZ', 'ZZIZXX', 'ZZIIYY', 'XZIXXX', 'XIIXXX', 'XZIYXY', 'XIIYXY', 'ZIIIII', 'ZIIIIZ', 'ZIIZXX', 'ZIIIYY', 'IIZIZZ', 'IZXZXZ', 'IIXZXZ', 'IZXIXI', 'IIXIXI', 'ZXXIZZ', 'IYYIZZ', 'IZZIZZ', 'ZXZZXZ', 'IXIZXZ', 'ZXZIXI', 'IXIIXI', 'XXZXXZ', 'YYIXXZ', 'XXZYYI', 'YYIYYI', 'ZZIIZZ', 'ZIIIZZ', 'IIZZZI', 'ZXXZZI', 'IYYZZI', 'XXXXZI', 'YXYXZI', 'XXXXII', 'YXYXII', 'IZZZZI', 'ZZIZZI', 'XZIXZI', 'XIIXZI', 'XZIXII', 'XIIXII', 'ZIIZZI', 'IIZZII', 'ZXXZII', 'IYYZII', 'IZZZII', 'ZZIZII', 'ZIIZII', 'IZIIII', 'ZZZIII', 'ZIZIII', 'ZYYIII', 'IXXIII'
    \item
    \textbf{Coefficients:} $-4.17946138e+00,  2.65507088e-01, -3.80039358e-02,
     -3.80039358e-02,  1.85644260e-01,  2.92418296e-01,
     -2.51893832e-01,  2.79629477e-01, -2.78214345e-01,
      2.80428143e-01,  3.69056167e-03,  3.69056167e-03,
      2.65507088e-01,  2.12870927e-01, -2.77588637e-02,
     -2.77588637e-02,  1.65321512e-02, -1.65321512e-02,
     -1.65321512e-02,  1.65321512e-02, -3.80039358e-02,
     -2.77588637e-02, -3.80039358e-02, -2.77588637e-02,
      1.28586257e-02,  1.28586257e-02,  1.28586257e-02,
      1.28586257e-02,  1.81017786e-02,  1.81017786e-02,
      1.81017786e-02,  1.81017786e-02,  1.85644260e-01,
      1.88779899e-01, -2.18648285e-02, -2.18648285e-02,
      2.44049429e-03, -2.44049429e-03, -2.44049429e-03,
      2.44049429e-03, -2.51893832e-01,  1.39471597e-01,
     -8.39966685e-03, -8.39966685e-03,  1.37041099e-02,
     -1.37041099e-02,  1.37041099e-02, -1.37041099e-02,
     -2.78214345e-01,  1.56924558e-01, -6.91065124e-03,
     -6.91065124e-03,  1.88779899e-01,  2.44049429e-03,
     -2.44049429e-03, -2.44049429e-03,  2.44049429e-03,
     -2.18648285e-02, -2.18648285e-02,  2.20039773e-01,
      1.05201448e-02, -1.05201448e-02, -1.05201448e-02,
      1.05201448e-02,  5.52352242e-03,  5.52352242e-03,
      5.52352242e-03,  5.52352242e-03,  1.52125508e-01,
      1.58540029e-01,  1.39471597e-01, -8.39966685e-03,
     -8.39966685e-03,  1.37041099e-02,  1.37041099e-02,
     -1.37041099e-02, -1.37041099e-02,  1.52125508e-01,
      1.53808036e-01,  2.65527980e-02, -2.65527980e-02,
     -2.65527980e-02,  2.65527980e-02,  1.46723346e-01,
      1.56924558e-01, -6.91065124e-03, -6.91065124e-03,
      1.58540029e-01,  1.46723346e-01,  1.60574596e-01,
      2.92418296e-01,  2.79629477e-01,  2.80428143e-01,
      3.69056167e-03,  3.69056167e-03$
      \end{itemize}
   \item[\#1] 
   \begin{itemize}
   \item
    \textbf{Geometry:}\\
    O 0.0 0.0 0.0\\
    H 0.0 0.0 0.8500000000000001\\
    H 0.0 0.8500000000000001 5.204748896376252e-17
    \item 
    \textbf{Pauli strings:}
    'IIIIII', 'IIIIIZ', 'IIIZXX', 'IIIIYY', 'IIIIZZ', 'IIIIZI', 'IIIZZI', 'IIIZZZ', 'IIIZII', 'IIIZIZ', 'IIIZYY', 'IIIIXX', 'IIZIII', 'IIZIIZ', 'IIZZXX', 'IIZIYY', 'IZXIZX', 'IIXIZX', 'IZXIIX', 'IIXIIX', 'ZXXIII', 'ZXXIIZ', 'IYYIII', 'IYYIIZ', 'ZXXZXX', 'IYYZXX', 'ZXXIYY', 'IYYIYY', 'XXXXXX', 'YXYXXX', 'XXXYXY', 'YXYYXY', 'IZZIII', 'IZZIIZ', 'IZZZXX', 'IZZIYY', 'ZXZIZX', 'IXIIZX', 'ZXZIIX', 'IXIIIX', 'ZZIIII', 'ZZIIIZ', 'ZZIZXX', 'ZZIIYY', 'XZIXXX', 'XIIXXX', 'XZIYXY', 'XIIYXY', 'ZIIIII', 'ZIIIIZ', 'ZIIZXX', 'ZIIIYY', 'IIZIZZ', 'IZXZXZ', 'IIXZXZ', 'IZXIXI', 'IIXIXI', 'ZXXIZZ', 'IYYIZZ', 'IZZIZZ', 'ZXZZXZ', 'IXIZXZ', 'ZXZIXI', 'IXIIXI', 'XXZXXZ', 'YYIXXZ', 'XXZYYI', 'YYIYYI', 'ZZIIZZ', 'ZIIIZZ', 'IIZZZI', 'ZXXZZI', 'IYYZZI', 'XXXXZI', 'YXYXZI', 'XXXXII', 'YXYXII', 'IZZZZI', 'ZZIZZI', 'XZIXZI', 'XIIXZI', 'XZIXII', 'XIIXII', 'ZIIZZI', 'IIZZII', 'ZXXZII', 'IYYZII', 'IZZZII', 'ZZIZII', 'ZIIZII', 'IZIIII', 'ZZZIII', 'ZIZIII', 'ZYYIII', 'IXXIII'
    \item 
    \textbf{Coefficients:} -4.22729389e+00,  2.56819280e-01, -4.21555207e-02,
     -4.21555207e-02,  1.80356781e-01,  2.89369626e-01,
     -2.25640195e-01,  2.76264378e-01, -2.46087190e-01,
      2.77939080e-01,  4.73282050e-03,  4.73282050e-03,
      2.56819280e-01,  2.06072769e-01, -2.96625988e-02,
     -2.96625988e-02,  1.59281505e-02, -1.59281505e-02,
     -1.59281505e-02,  1.59281505e-02, -4.21555207e-02,
     -2.96625988e-02, -4.21555207e-02, -2.96625988e-02,
      1.49782726e-02,  1.49782726e-02,  1.49782726e-02,
      1.49782726e-02,  1.82391892e-02,  1.82391892e-02,
      1.82391892e-02,  1.82391892e-02,  1.80356781e-01,
      1.86209457e-01, -2.47650066e-02, -2.47650066e-02,
      1.76738076e-03, -1.76738076e-03, -1.76738076e-03,
      1.76738076e-03, -2.25640195e-01,  1.39929485e-01,
     -9.32938431e-03, -9.32938431e-03,  1.41240460e-02,
     -1.41240460e-02,  1.41240460e-02, -1.41240460e-02,
     -2.46087190e-01,  1.54540476e-01, -7.67552087e-03,
     -7.67552087e-03,  1.86209457e-01,  1.76738076e-03,
     -1.76738076e-03, -1.76738076e-03,  1.76738076e-03,
     -2.47650066e-02, -2.47650066e-02,  2.20039773e-01,
      1.02158252e-02, -1.02158252e-02, -1.02158252e-02,
      1.02158252e-02,  5.60218036e-03,  5.60218036e-03,
      5.60218036e-03,  5.60218036e-03,  1.51853619e-01,
      1.56915346e-01,  1.39929485e-01, -9.32938431e-03,
     -9.32938431e-03,  1.41240460e-02,  1.41240460e-02,
     -1.41240460e-02, -1.41240460e-02,  1.51853619e-01,
      1.54133960e-01,  2.68210226e-02, -2.68210226e-02,
     -2.68210226e-02,  2.68210226e-02,  1.45909342e-01,
      1.54540476e-01, -7.67552087e-03, -7.67552087e-03,
      1.56915346e-01,  1.45909342e-01,  1.58288417e-01,
      2.89369626e-01,  2.76264378e-01,  2.77939080e-01,
      4.73282050e-03,  4.73282050e-03
   \end{itemize}
   \item[\#2] 
   \begin{itemize}
       \item \textbf{Geometry:}\\
        O 0.0 0.0 0.0\\
        H 0.0 0.0 0.9\\
        H 0.0 0.9 5.5109105961630896e-17
        \item
        \textbf{Pauli strings:}\\
        'IIIIII', 'IIIIIZ', 'IIIZXX', 'IIIIYY', 'IIIIZZ', 'IIIIZI', 'IIIZZI', 'IIIZZZ', 'IIIZII', 'IIIZIZ', 'IIIZYY', 'IIIIXX', 'IIZIII', 'IIZIIZ', 'IIZZXX', 'IIZIYY', 'IZXIZX', 'IIXIZX', 'IZXIIX', 'IIXIIX', 'ZXXIII', 'ZXXIIZ', 'IYYIII', 'IYYIIZ', 'ZXXZXX', 'IYYZXX', 'ZXXIYY', 'IYYIYY', 'XXXXXX', 'YXYXXX', 'XXXYXY', 'YXYYXY', 'IZZIII', 'IZZIIZ', 'IZZZXX', 'IZZIYY', 'ZXZIZX', 'IXIIZX', 'ZXZIIX', 'IXIIIX', 'ZZIIII', 'ZZIIIZ', 'ZZIZXX', 'ZZIIYY', 'XZIXXX', 'XIIXXX', 'XZIYXY', 'XIIYXY', 'ZIIIII', 'ZIIIIZ', 'ZIIZXX', 'ZIIIYY', 'IIZIZZ', 'IZXZXZ', 'IIXZXZ', 'IZXIXI', 'IIXIXI', 'ZXXIZZ', 'IYYIZZ', 'IZZIZZ', 'ZXZZXZ', 'IXIZXZ', 'ZXZIXI', 'IXIIXI', 'XXZXXZ', 'YYIXXZ', 'XXZYYI', 'YYIYYI', 'ZZIIZZ', 'ZIIIZZ', 'IIZZZI', 'ZXXZZI', 'IYYZZI', 'XXXXZI', 'YXYXZI', 'XXXXII', 'YXYXII', 'IZZZZI', 'ZZIZZI', 'XZIXZI', 'XIIXZI', 'XZIXII', 'XIIXII', 'ZIIZZI', 'IIZZII', 'ZXXZII', 'IYYZII', 'IZZZII', 'ZZIZII', 'ZIIZII', 'IZIIII', 'ZZZIII', 'ZIZIII', 'ZYYIII', 'IXXIII'
        \item \textbf{Coefficients:}\\
        -4.26701411e+00,  2.48356106e-01, -4.58833280e-02,
     -4.58833280e-02,  1.76088317e-01,  2.85974297e-01,
     -1.99943317e-01,  2.72568845e-01, -2.16337164e-01,
      2.75202431e-01,  5.81794202e-03,  5.81794202e-03,
      2.48356106e-01,  1.99024780e-01, -3.11125998e-02,
     -3.11125998e-02,  1.52610000e-02, -1.52610000e-02,
     -1.52610000e-02,  1.52610000e-02, -4.58833280e-02,
     -3.11125998e-02, -4.58833280e-02, -3.11125998e-02,
      1.72127944e-02,  1.72127944e-02,  1.72127944e-02,
      1.72127944e-02,  1.83814266e-02,  1.83814266e-02,
      1.83814266e-02,  1.83814266e-02,  1.76088317e-01,
      1.83357160e-01, -2.76338430e-02, -2.76338430e-02,
      1.09175308e-03, -1.09175308e-03, -1.09175308e-03,
      1.09175308e-03, -1.99943317e-01,  1.40136149e-01,
     -1.02748019e-02, -1.02748019e-02,  1.45012778e-02,
     -1.45012778e-02,  1.45012778e-02, -1.45012778e-02,
     -2.16337164e-01,  1.52121965e-01, -8.40637631e-03,
     -8.40637631e-03,  1.83357160e-01,  1.09175308e-03,
     -1.09175308e-03, -1.09175308e-03,  1.09175308e-03,
     -2.76338430e-02, -2.76338430e-02,  2.20039773e-01,
      9.90635390e-03, -9.90635390e-03, -9.90635390e-03,
      9.90635390e-03,  5.67478748e-03,  5.67478748e-03,
      5.67478748e-03,  5.67478748e-03,  1.51368247e-01,
      1.55320278e-01,  1.40136149e-01, -1.02748019e-02,
     -1.02748019e-02,  1.45012778e-02,  1.45012778e-02,
     -1.45012778e-02, -1.45012778e-02,  1.51368247e-01,
      1.53867131e-01,  2.69429053e-02, -2.69429053e-02,
     -2.69429053e-02,  2.69429053e-02,  1.44821043e-01,
      1.52121965e-01, -8.40637631e-03, -8.40637631e-03,
      1.55320278e-01,  1.44821043e-01,  1.56013718e-01,
      2.85974297e-01,  2.72568845e-01,  2.75202431e-01,
      5.81794202e-03,  5.81794202e-03
   \end{itemize}

   \item[\#3] 
   \begin{itemize}
       \item \textbf{Geometry:}\\
        O 0.0 0.0 0.0\\
        H 0.0 0.0 0.95\\
        H 0.0 0.95 5.817072295949927e-17
        \item \textbf{Pauli strings:}\\
        'IIIIII', 'IIIIIZ', 'IIIZXX', 'IIIIYY', 'IIIIZZ', 'IIIIZI', 'IIIZZI', 'IIIZZZ', 'IIIZII', 'IIIZIZ', 'IIIZYY', 'IIIIXX', 'IIZIII', 'IIZIIZ', 'IIZZXX', 'IIZIYY', 'IZXIZX', 'IIXIZX', 'IZXIIX', 'IIXIIX', 'ZXXIII', 'ZXXIIZ', 'IYYIII', 'IYYIIZ', 'ZXXZXX', 'IYYZXX', 'ZXXIYY', 'IYYIYY', 'XXXXXX', 'YXYXXX', 'XXXYXY', 'YXYYXY', 'IZZIII', 'IZZIIZ', 'IZZZXX', 'IZZIYY', 'ZXZIZX', 'IXIIZX', 'ZXZIIX', 'IXIIIX', 'ZZIIII', 'ZZIIIZ', 'ZZIZXX', 'ZZIIYY', 'XZIXXX', 'XIIXXX', 'XZIYXY', 'XIIYXY', 'ZIIIII', 'ZIIIIZ', 'ZIIZXX', 'ZIIIYY', 'IIZIZZ', 'IZXZXZ', 'IIXZXZ', 'IZXIXI', 'IIXIXI', 'ZXXIZZ', 'IYYIZZ', 'IZZIZZ', 'ZXZZXZ', 'IXIZXZ', 'ZXZIXI', 'IXIIXI', 'XXZXXZ', 'YYIXXZ', 'XXZYYI', 'YYIYYI', 'ZZIIZZ', 'ZIIIZZ', 'IIZZZI', 'ZXXZZI', 'IYYZZI', 'XXXXZI', 'YXYXZI', 'XXXXII', 'YXYXII', 'IZZZZI', 'ZZIZZI', 'XZIXZI', 'XIIXZI', 'XZIXII', 'XIIXII', 'ZIIZZI', 'IIZZII', 'ZXXZII', 'IYYZII', 'IZZZII', 'ZZIZII', 'ZIIZII', 'IZIIII', 'ZZZIII', 'ZIZIII', 'ZYYIII', 'IXXIII'
        \item \textbf{Coefficients:}\\
        -4.29828906e+00,  2.40013873e-01, -4.91456916e-02,
         -4.91456916e-02,  1.72583530e-01,  2.82306710e-01,
         -1.75312850e-01,  2.68611998e-01, -1.88864344e-01,
          2.72289679e-01,  6.93379425e-03,  6.93379425e-03,
          2.40013873e-01,  1.91901878e-01, -3.21168595e-02,
         -3.21168595e-02,  1.45451376e-02, -1.45451376e-02,
         -1.45451376e-02,  1.45451376e-02, -4.91456916e-02,
         -3.21168595e-02, -4.91456916e-02, -3.21168595e-02,
          1.95181640e-02,  1.95181640e-02,  1.95181640e-02,
          1.95181640e-02,  1.85210647e-02,  1.85210647e-02,
          1.85210647e-02,  1.85210647e-02,  1.72583530e-01,
          1.80269791e-01, -3.04510894e-02, -3.04510894e-02,
          4.24943569e-04, -4.24943569e-04, -4.24943569e-04,
          4.24943569e-04, -1.75312850e-01,  1.40119669e-01,
         -1.12509265e-02, -1.12509265e-02,  1.48371730e-02,
         -1.48371730e-02,  1.48371730e-02, -1.48371730e-02,
         -1.88864344e-01,  1.49673004e-01, -9.10506564e-03,
         -9.10506564e-03,  1.80269791e-01,  4.24943569e-04,
         -4.24943569e-04, -4.24943569e-04,  4.24943569e-04,
         -3.04510894e-02, -3.04510894e-02,  2.20039773e-01,
          9.60018544e-03, -9.60018544e-03, -9.60018544e-03,
          9.60018544e-03,  5.73945933e-03,  5.73945933e-03,
          5.73945933e-03,  5.73945933e-03,  1.50737925e-01,
          1.53749953e-01,  1.40119669e-01, -1.12509265e-02,
         -1.12509265e-02,  1.48371730e-02,  1.48371730e-02,
         -1.48371730e-02, -1.48371730e-02,  1.50737925e-01,
          1.53153401e-01,  2.69561931e-02, -2.69561931e-02,
         -2.69561931e-02,  2.69561931e-02,  1.43538249e-01,
          1.49673004e-01, -9.10506564e-03, -9.10506564e-03,
          1.53749953e-01,  1.43538249e-01,  1.53759630e-01,
          2.82306710e-01,  2.68611998e-01,  2.72289679e-01,
          6.93379425e-03,  6.93379425e-03
    \end{itemize}

   \item[\#4]
   \begin{itemize}
       \item \textbf{Geometry:}\\
        O 0.0 0.0 0.0\\
        H 0.0 0.0 1.0\\
        H 0.0 1.0 6.123233995736766e-17
        \item \textbf{Pauli strings:}\\
        'IIIIII', 'IIIIIZ', 'IIIZXX', 'IIIIYY', 'IIIIZZ', 'IIIIZI', 'IIIZZI', 'IIIZZZ', 'IIIZII', 'IIIZIZ', 'IIIZYY', 'IIIIXX', 'IIZIII', 'IIZIIZ', 'IIZZXX', 'IIZIYY', 'IZXIZX', 'IIXIZX', 'IZXIIX', 'IIXIIX', 'ZXXIII', 'ZXXIIZ', 'IYYIII', 'IYYIIZ', 'ZXXZXX', 'IYYZXX', 'ZXXIYY', 'IYYIYY', 'XXXXXX', 'YXYXXX', 'XXXYXY', 'YXYYXY', 'IZZIII', 'IZZIIZ', 'IZZZXX', 'IZZIYY', 'ZXZIZX', 'IXIIZX', 'ZXZIIX', 'IXIIIX', 'ZZIIII', 'ZZIIIZ', 'ZZIZXX', 'ZZIIYY', 'XZIXXX', 'XIIXXX', 'XZIYXY', 'XIIYXY', 'ZIIIII', 'ZIIIIZ', 'ZIIZXX', 'ZIIIYY', 'IIZIZZ', 'IZXZXZ', 'IIXZXZ', 'IZXIXI', 'IIXIXI', 'ZXXIZZ', 'IYYIZZ', 'IZZIZZ', 'ZXZZXZ', 'IXIZXZ', 'ZXZIXI', 'IXIIXI', 'XXZXXZ', 'YYIXXZ', 'XXZYYI', 'YYIYYI', 'ZZIIZZ', 'ZIIIZZ', 'IIZZZI', 'ZXXZZI', 'IYYZZI', 'XXXXZI', 'YXYXZI', 'XXXXII', 'YXYXII', 'IZZZZI', 'ZZIZZI', 'XZIXZI', 'XIIXZI', 'XZIXII', 'XIIXII', 'ZIIZZI', 'IIZZII', 'ZXXZII', 'IYYZII', 'IZZZII', 'ZZIZII', 'ZIIZII', 'IZIIII', 'ZZZIII', 'ZIZIII', 'ZYYIII', 'IXXIII'
        \item \textbf{Coefficients:}\\
        -4.32132886e+00,  2.31767040e-01, -5.19388670e-02,
         -5.19388670e-02,  1.69667753e-01,  2.78432259e-01,
         -1.52055301e-01,  2.64452527e-01, -1.63564802e-01,
          2.69255875e-01,  8.06903827e-03,  8.06903827e-03,
          2.31767040e-01,  1.84867190e-01, -3.27042652e-02,
         -3.27042652e-02,  1.37981811e-02, -1.37981811e-02,
         -1.37981811e-02,  1.37981811e-02, -5.19388670e-02,
         -3.27042652e-02, -5.19388670e-02, -3.27042652e-02,
          2.18493547e-02,  2.18493547e-02,  2.18493547e-02,
          2.18493547e-02,  1.86559127e-02,  1.86559127e-02,
          1.86559127e-02,  1.86559127e-02,  1.69667753e-01,
          1.77002028e-01, -3.31948825e-02, -3.31948825e-02,
         -2.22297466e-04,  2.22297466e-04,  2.22297466e-04,
         -2.22297466e-04, -1.52055301e-01,  1.39897517e-01,
         -1.22603108e-02, -1.22603108e-02,  1.51345388e-02,
         -1.51345388e-02,  1.51345388e-02, -1.51345388e-02,
         -1.63564802e-01,  1.47203436e-01, -9.76900798e-03,
         -9.76900798e-03,  1.77002028e-01, -2.22297466e-04,
          2.22297466e-04,  2.22297466e-04, -2.22297466e-04,
         -3.31948825e-02, -3.31948825e-02,  2.20039773e-01,
          9.30341183e-03, -9.30341183e-03, -9.30341183e-03,
          9.30341183e-03,  5.79482164e-03,  5.79482164e-03,
          5.79482164e-03,  5.79482164e-03,  1.50011763e-01,
          1.52199186e-01,  1.39897517e-01, -1.22603108e-02,
         -1.22603108e-02,  1.51345388e-02,  1.51345388e-02,
         -1.51345388e-02, -1.51345388e-02,  1.50011763e-01,
          1.52110181e-01,  2.68920746e-02, -2.68920746e-02,
         -2.68920746e-02,  2.68920746e-02,  1.42120487e-01,
          1.47203436e-01, -9.76900798e-03, -9.76900798e-03,
          1.52199186e-01,  1.42120487e-01,  1.51532178e-01,
          2.78432259e-01,  2.64452527e-01,  2.69255875e-01,
          8.06903827e-03,  8.06903827e-03
   \end{itemize}
   \item[\#5] 
   \begin{itemize}
       \item \textbf{Geometry:}\\
        O 0.0 0.0 0.0\\
        H 0.0 0.0 0.8\\
        H 0.0 0.7391036260090295 -0.3061467458920718
        \item \textbf{Pauli strings:}
        'IIIIII', 'IIIIIZ', 'IIIZXX', 'IIIIYY', 'IIIIZZ', 'IIIIZI', 'IIIZZI', 'IIIZZZ', 'IIIZII', 'IIIZIZ', 'IIIZYY', 'IIIIXX', 'IIZIII', 'IIZIIZ', 'IIZZXX', 'IIZIYY', 'IZXIZX', 'IIXIZX', 'IZXIIX', 'IIXIIX', 'ZXXIII', 'ZXXIIZ', 'IYYIII', 'IYYIIZ', 'ZXXZXX', 'IYYZXX', 'ZXXIYY', 'IYYIYY', 'XXXXXX', 'YXYXXX', 'XXXYXY', 'YXYYXY', 'IZZIII', 'IZZIIZ', 'IZZZXX', 'IZZIYY', 'ZXZIZX', 'IXIIZX', 'ZXZIIX', 'IXIIIX', 'ZZIIII', 'ZZIIIZ', 'ZZIZXX', 'ZZIIYY', 'XZIXXX', 'XIIXXX', 'XZIYXY', 'XIIYXY', 'ZIIIII', 'ZIIIIZ', 'ZIIZXX', 'ZIIIYY', 'IIZIZZ', 'IZXZXZ', 'IIXZXZ', 'IZXIXI', 'IIXIXI', 'ZXXIZZ', 'IYYIZZ', 'IZZIZZ', 'ZXZZXZ', 'IXIZXZ', 'ZXZIXI', 'IXIIXI', 'XXZXXZ', 'YYIXXZ', 'XXZYYI', 'YYIYYI', 'ZZIIZZ', 'ZIIIZZ', 'IIZZZI', 'ZXXZZI', 'IYYZZI', 'XXXXZI', 'YXYXZI', 'XXXXII', 'YXYXII', 'IZZZZI', 'ZZIZZI', 'XZIXZI', 'XIIXZI', 'XZIXII', 'XIIXII', 'ZIIZZI', 'IIZZII', 'ZXXZII', 'IYYZII', 'IZZZII', 'ZZIZII', 'ZIIZII', 'IZIIII', 'ZZZIII', 'ZIZIII', 'ZYYIII', 'IXXIII'
        \item \textbf{Coefficients:}
        -4.14902563e+00,  2.42769065e-01,  2.35579836e-02,
          2.35579836e-02,  1.87499302e-01,  2.89929265e-01,
         -2.37870527e-01,  2.82450714e-01, -2.87815108e-01,
          2.82040298e-01, -6.43341754e-04, -6.43341754e-04,
          2.42769065e-01,  2.16725662e-01,  2.12397177e-02,
          2.12397177e-02,  1.50915092e-02, -1.50915092e-02,
         -1.50915092e-02,  1.50915092e-02,  2.35579836e-02,
          2.12397177e-02,  2.35579836e-02,  2.12397177e-02,
          1.00721340e-02,  1.00721340e-02,  1.00721340e-02,
          1.00721340e-02,  1.57026000e-02,  1.57026000e-02,
          1.57026000e-02,  1.57026000e-02,  1.87499302e-01,
          1.90899042e-01,  1.88105646e-02,  1.88105646e-02,
         -2.38008939e-03,  2.38008939e-03,  2.38008939e-03,
         -2.38008939e-03, -2.37870527e-01,  1.37260299e-01,
          9.77596827e-03,  9.77596827e-03, -1.31876400e-02,
          1.31876400e-02, -1.31876400e-02,  1.31876400e-02,
         -2.87815108e-01,  1.60230319e-01,  7.35967221e-03,
          7.35967221e-03,  1.90899042e-01, -2.38008939e-03,
          2.38008939e-03,  2.38008939e-03, -2.38008939e-03,
          1.88105646e-02,  1.88105646e-02,  2.20039773e-01,
          1.10535704e-02, -1.10535704e-02, -1.10535704e-02,
          1.10535704e-02,  5.88145960e-03,  5.88145960e-03,
          5.88145960e-03,  5.88145960e-03,  1.48566149e-01,
          1.61144008e-01,  1.37260299e-01,  9.77596827e-03,
          9.77596827e-03, -1.31876400e-02, -1.31876400e-02,
          1.31876400e-02,  1.31876400e-02,  1.48566149e-01,
          1.48713098e-01,  2.98601850e-02, -2.98601850e-02,
         -2.98601850e-02,  2.98601850e-02,  1.43981917e-01,
          1.60230319e-01,  7.35967221e-03,  7.35967221e-03,
          1.61144008e-01,  1.43981917e-01,  1.61377613e-01,
          2.89929265e-01,  2.82450714e-01,  2.82040298e-01,
         -6.43341754e-04, -6.43341754e-04
        
   \end{itemize}
   \item[\#6]
   \begin{itemize}
       \item \textbf{Geometry:}\\
        O 0.0 0.0 0.0\\
        H 0.0 0.0 0.8500000000000001\\
        H 0.0 0.7852976026345938 -0.3252809175103263
        \item \textbf{Pauli strings:}\\
        'IIIIII', 'IIIIIZ', 'IIIZXX', 'IIIIYY', 'IIIIZZ', 'IIIIZI', 'IIIZZI', 'IIIZZZ', 'IIIZII', 'IIIZIZ', 'IIIZYY', 'IIIIXX', 'IIZIII', 'IIZIIZ', 'IIZZXX', 'IIZIYY', 'IZXIZX', 'IIXIZX', 'IZXIIX', 'IIXIIX', 'ZXXIII', 'ZXXIIZ', 'IYYIII', 'IYYIIZ', 'ZXXZXX', 'IYYZXX', 'ZXXIYY', 'IYYIYY', 'XXXXXX', 'YXYXXX', 'XXXYXY', 'YXYYXY', 'IZZIII', 'IZZIIZ', 'IZZZXX', 'IZZIYY', 'ZXZIZX', 'IXIIZX', 'ZXZIIX', 'IXIIIX', 'ZZIIII', 'ZZIIIZ', 'ZZIZXX', 'ZZIIYY', 'XZIXXX', 'XIIXXX', 'XZIYXY', 'XIIYXY', 'ZIIIII', 'ZIIIIZ', 'ZIIZXX', 'ZIIIYY', 'IIZIZZ', 'IZXZXZ', 'IIXZXZ', 'IZXIXI', 'IIXIXI', 'ZXXIZZ', 'IYYIZZ', 'IZZIZZ', 'ZXZZXZ', 'IXIZXZ', 'ZXZIXI', 'IXIIXI', 'XXZXXZ', 'YYIXXZ', 'XXZYYI', 'YYIYYI', 'ZZIIZZ', 'ZIIIZZ', 'IIZZZI', 'ZXXZZI', 'IYYZZI', 'XXXXZI', 'YXYXZI', 'XXXXII', 'YXYXII', 'IZZZZI', 'ZZIZZI', 'XZIXZI', 'XIIXZI', 'XZIXII', 'XIIXII', 'ZIIZZI', 'IIZZII', 'ZXXZII', 'IYYZII', 'IZZZII', 'ZZIZII', 'ZIIZII', 'IZIIII', 'ZZZIII', 'ZIZIII', 'ZYYIII', 'IXXIII']
        \item \textbf{Coeffiecients:}\\
        -4.20708968e+00,  2.35873518e-01,  2.86645427e-02,
          2.86645427e-02,  1.82777308e-01,  2.87320321e-01,
         -2.10782194e-01,  2.79209856e-01, -2.54933693e-01,
          2.79317683e-01, -1.53129487e-03, -1.53129487e-03,
          2.35873518e-01,  2.11382784e-01,  2.40007156e-02,
          2.40007156e-02,  1.47253618e-02, -1.47253618e-02,
         -1.47253618e-02,  1.47253618e-02,  2.86645427e-02,
          2.40007156e-02,  2.86645427e-02,  2.40007156e-02,
          1.16745976e-02,  1.16745976e-02,  1.16745976e-02,
          1.16745976e-02,  1.59112952e-02,  1.59112952e-02,
          1.59112952e-02,  1.59112952e-02,  1.82777308e-01,
          1.88886896e-01,  2.16290437e-02,  2.16290437e-02,
         -1.80010327e-03,  1.80010327e-03,  1.80010327e-03,
         -1.80010327e-03, -2.10782194e-01,  1.36923314e-01,
          1.01305249e-02,  1.01305249e-02, -1.35318655e-02,
          1.35318655e-02, -1.35318655e-02,  1.35318655e-02,
         -2.54933693e-01,  1.58169430e-01,  8.36598660e-03,
          8.36598660e-03,  1.88886896e-01, -1.80010327e-03,
          1.80010327e-03,  1.80010327e-03, -1.80010327e-03,
          2.16290437e-02,  2.16290437e-02,  2.20039773e-01,
          1.06350123e-02, -1.06350123e-02, -1.06350123e-02,
          1.06350123e-02,  6.00132778e-03,  6.00132778e-03,
          6.00132778e-03,  6.00132778e-03,  1.47694561e-01,
          1.59962468e-01,  1.36923314e-01,  1.01305249e-02,
          1.01305249e-02, -1.35318655e-02, -1.35318655e-02,
          1.35318655e-02,  1.35318655e-02,  1.47694561e-01,
          1.48808126e-01,  3.00309985e-02, -3.00309985e-02,
         -3.00309985e-02,  3.00309985e-02,  1.43189786e-01,
          1.58169430e-01,  8.36598660e-03,  8.36598660e-03,
          1.59962468e-01,  1.43189786e-01,  1.59640489e-01,
          2.87320321e-01,  2.79209856e-01,  2.79317683e-01,
         -1.53129487e-03, -1.53129487e-03
   \end{itemize} 

   \item[\#7]
   \begin{itemize}
       \item \textbf{Geometry:}\\
        O 0.0 0.0 0.0\\
        H 0.0 0.0 0.9\\
        H 0.0 0.831491579260158 -0.34441508912858076
        \item \textbf{Pauli strings:}\\
        'IIIIII', 'IIIIIZ', 'IIIZXX', 'IIIIYY', 'IIIIZZ', 'IIIIZI', 'IIIZZI', 'IIIZZZ', 'IIIZII', 'IIIZIZ', 'IIIZYY', 'IIIIXX', 'IIZIII', 'IIZIIZ', 'IIZZXX', 'IIZIYY', 'IZXIZX', 'IIXIZX', 'IZXIIX', 'IIXIIX', 'ZXXIII', 'ZXXIIZ', 'IYYIII', 'IYYIIZ', 'ZXXZXX', 'IYYZXX', 'ZXXIYY', 'IYYIYY', 'XXXXXX', 'YXYXXX', 'XXXYXY', 'YXYYXY', 'IZZIII', 'IZZIIZ', 'IZZZXX', 'IZZIYY', 'ZXZIZX', 'IXIIZX', 'ZXZIIX', 'IXIIIX', 'ZZIIII', 'ZZIIIZ', 'ZZIZXX', 'ZZIIYY', 'XZIXXX', 'XIIXXX', 'XZIYXY', 'XIIYXY', 'ZIIIII', 'ZIIIIZ', 'ZIIZXX', 'ZIIIYY', 'IIZIZZ', 'IZXZXZ', 'IIXZXZ', 'IZXIXI', 'IIXIXI', 'ZXXIZZ', 'IYYIZZ', 'IZZIZZ', 'ZXZZXZ', 'IXIZXZ', 'ZXZIXI', 'IXIIXI', 'XXZXXZ', 'YYIXXZ', 'XXZYYI', 'YYIYYI', 'ZZIIZZ', 'ZIIIZZ', 'IIZZZI', 'ZXXZZI', 'IYYZZI', 'XXXXZI', 'YXYXZI', 'XXXXII', 'YXYXII', 'IZZZZI', 'ZZIZZI', 'XZIXZI', 'XIIXZI', 'XZIXII', 'XIIXII', 'ZIIZZI', 'IIZZII', 'ZXXZII', 'IYYZII', 'IZZZII', 'ZZIZII', 'ZIIZII', 'IZIIII', 'ZZZIII', 'ZIZIII', 'ZYYIII', 'IXXIII']
        \item \textbf{Coefficients:}\\
        -4.25733010e+00,  2.29370089e-01,  3.36107000e-02,
          3.36107000e-02,  1.79171785e-01,  2.84401694e-01,
         -1.84566966e-01,  2.75640434e-01, -2.24292132e-01,
          2.76349386e-01, -2.48896484e-03, -2.48896484e-03,
          2.29370089e-01,  2.05561975e-01,  2.64632839e-02,
          2.64632839e-02,  1.43016906e-02, -1.43016906e-02,
         -1.43016906e-02,  1.43016906e-02,  3.36107000e-02,
          2.64632839e-02,  3.36107000e-02,  2.64632839e-02,
          1.35149763e-02,  1.35149763e-02,  1.35149763e-02,
          1.35149763e-02,  1.61349449e-02,  1.61349449e-02,
          1.61349449e-02,  1.61349449e-02,  1.79171785e-01,
          1.86588052e-01,  2.44894599e-02,  2.44894599e-02,
         -1.20140996e-03,  1.20140996e-03,  1.20140996e-03,
         -1.20140996e-03, -1.84566966e-01,  1.36496513e-01,
          1.04721044e-02,  1.04721044e-02, -1.38430008e-02,
          1.38430008e-02, -1.38430008e-02,  1.38430008e-02,
         -2.24292132e-01,  1.56012590e-01,  9.35890424e-03,
          9.35890424e-03,  1.86588052e-01, -1.20140996e-03,
          1.20140996e-03,  1.20140996e-03, -1.20140996e-03,
          2.44894599e-02,  2.44894599e-02,  2.20039773e-01,
          1.02066882e-02, -1.02066882e-02, -1.02066882e-02,
          1.02066882e-02,  6.11200010e-03,  6.11200010e-03,
          6.11200010e-03,  6.11200010e-03,  1.46678429e-01,
          1.58770897e-01,  1.36496513e-01,  1.04721044e-02,
          1.04721044e-02, -1.38430008e-02, -1.38430008e-02,
          1.38430008e-02,  1.38430008e-02,  1.46678429e-01,
          1.48359829e-01,  3.00218123e-02, -3.00218123e-02,
         -3.00218123e-02,  3.00218123e-02,  1.42137145e-01,
          1.56012590e-01,  9.35890424e-03,  9.35890424e-03,
          1.58770897e-01,  1.42137145e-01,  1.57852695e-01,
          2.84401694e-01,  2.75640434e-01,  2.76349386e-01,
         -2.48896484e-03, -2.48896484e-03
   \end{itemize}

   \item[\#8]
   \begin{itemize}
       \item \textbf{Geometry:}\\
        O 0.0 0.0 0.0\\
        H 0.0 0.0 0.95\\
        H 0.0 0.8776855558857224 -0.36354926074683525
        \item \textbf{Pauli strings:}
        'IIIIII', 'IIIIIZ', 'IIIZXX', 'IIIIYY', 'IIIIZZ', 'IIIIZI', 'IIIZZI', 'IIIZZZ', 'IIIZII', 'IIIZIZ', 'IIIZYY', 'IIIIXX', 'IIZIII', 'IIZIIZ', 'IIZZXX', 'IIZIYY', 'IZXIZX', 'IIXIZX', 'IZXIIX', 'IIXIIX', 'ZXXIII', 'ZXXIIZ', 'IYYIII', 'IYYIIZ', 'ZXXZXX', 'IYYZXX', 'ZXXIYY', 'IYYIYY', 'XXXXXX', 'YXYXXX', 'XXXYXY', 'YXYYXY', 'IZZIII', 'IZZIIZ', 'IZZZXX', 'IZZIYY', 'ZXZIZX', 'IXIIZX', 'ZXZIIX', 'IXIIIX', 'ZZIIII', 'ZZIIIZ', 'ZZIZXX', 'ZZIIYY', 'XZIXXX', 'XIIXXX', 'XZIYXY', 'XIIYXY', 'ZIIIII', 'ZIIIIZ', 'ZIIZXX', 'ZIIIYY', 'IIZIZZ', 'IZXZXZ', 'IIXZXZ', 'IZXIXI', 'IIXIXI', 'ZXXIZZ', 'IYYIZZ', 'IZZIZZ', 'ZXZZXZ', 'IXIZXZ', 'ZXZIXI', 'IXIIXI', 'XXZXXZ', 'YYIXXZ', 'XXZYYI', 'YYIYYI', 'ZZIIZZ', 'ZIIIZZ', 'IIZZZI', 'ZXXZZI', 'IYYZZI', 'XXXXZI', 'YXYXZI', 'XXXXII', 'YXYXII', 'IZZZZI', 'ZZIZZI', 'XZIXZI', 'XIIXZI', 'XZIXII', 'XIIXII', 'ZIIZZI', 'IIZZII', 'ZXXZII', 'IYYZII', 'IZZZII', 'ZZIZII', 'ZIIZII', 'IZIIII', 'ZZZIII', 'ZIZIII', 'ZYYIII', 'IXXIII'
        \item \textbf{Coefficients:}
        -4.29906968e+00,  2.23042684e-01,  3.82639948e-02,
          3.82639948e-02,  1.76392755e-01,  2.81227973e-01,
         -1.59681654e-01,  2.71812599e-01, -1.95822315e-01,
          2.73212074e-01, -3.50141196e-03, -3.50141196e-03,
          2.23042684e-01,  1.99389329e-01,  2.85768646e-02,
          2.85768646e-02,  1.38245536e-02, -1.38245536e-02,
         -1.38245536e-02,  1.38245536e-02,  3.82639948e-02,
          2.85768646e-02,  3.82639948e-02,  2.85768646e-02,
          1.55633386e-02,  1.55633386e-02,  1.55633386e-02,
          1.55633386e-02,  1.63625133e-02,  1.63625133e-02,
          1.63625133e-02,  1.63625133e-02,  1.76392755e-01,
          1.84027940e-01,  2.73665079e-02,  2.73665079e-02,
         -5.93228945e-04,  5.93228945e-04,  5.93228945e-04,
         -5.93228945e-04, -1.59681654e-01,  1.36017301e-01,
          1.08425220e-02,  1.08425220e-02, -1.41200572e-02,
          1.41200572e-02, -1.41200572e-02,  1.41200572e-02,
         -1.95822315e-01,  1.53767568e-01,  1.03382677e-02,
          1.03382677e-02,  1.84027940e-01, -5.93228945e-04,
          5.93228945e-04,  5.93228945e-04, -5.93228945e-04,
          2.73665079e-02,  2.73665079e-02,  2.20039773e-01,
          9.78098456e-03, -9.78098456e-03, -9.78098456e-03,
          9.78098456e-03,  6.21138320e-03,  6.21138320e-03,
          6.21138320e-03,  6.21138320e-03,  1.45588003e-01,
          1.57570019e-01,  1.36017301e-01,  1.08425220e-02,
          1.08425220e-02, -1.41200572e-02, -1.41200572e-02,
          1.41200572e-02,  1.41200572e-02,  1.45588003e-01,
          1.47491434e-01,  2.98796649e-02, -2.98796649e-02,
         -2.98796649e-02,  2.98796649e-02,  1.40904252e-01,
          1.53767568e-01,  1.03382677e-02,  1.03382677e-02,
          1.57570019e-01,  1.40904252e-01,  1.56030750e-01,
          2.81227973e-01,  2.71812599e-01,  2.73212074e-01,
         -3.50141196e-03, -3.50141196e-03
   \end{itemize}
   \item[\#9] 
   \begin{itemize}
       \item \textbf{Geometry:}\\
        O 0.0 0.0 0.0\\
        H 0.0 0.0 1.0\\
        H 0.0 0.9238795325112867 -0.3826834323650897
        \item
        \textbf{Pauli strings:}\\
        'IIIIII', 'IIIIIZ', 'IIIZXX', 'IIIIYY', 'IIIIZZ', 'IIIIZI', 'IIIZZI', 'IIIZZZ', 'IIIZII', 'IIIZIZ', 'IIIZYY', 'IIIIXX', 'IIZIII', 'IIZIIZ', 'IIZZXX', 'IIZIYY', 'IZXIZX', 'IIXIZX', 'IZXIIX', 'IIXIIX', 'ZXXIII', 'ZXXIIZ', 'IYYIII', 'IYYIIZ', 'ZXXZXX', 'IYYZXX', 'ZXXIYY', 'IYYIYY', 'XXXXXX', 'YXYXXX', 'XXXYXY', 'YXYYXY', 'IZZIII', 'IZZIIZ', 'IZZZXX', 'IZZIYY', 'ZXZIZX', 'IXIIZX', 'ZXZIIX', 'IXIIIX', 'ZZIIII', 'ZZIIIZ', 'ZZIZXX', 'ZZIIYY', 'XZIXXX', 'XIIXXX', 'XZIYXY', 'XIIYXY', 'ZIIIII', 'ZIIIIZ', 'ZIIZXX', 'ZIIIYY', 'IIZIZZ', 'IZXZXZ', 'IIXZXZ', 'IZXIXI', 'IIXIXI', 'ZXXIZZ', 'IYYIZZ', 'IZZIZZ', 'ZXZZXZ', 'IXIZXZ', 'ZXZIXI', 'IXIIXI', 'XXZXXZ', 'YYIXXZ', 'XXZYYI', 'YYIYYI', 'ZZIIZZ', 'ZIIIZZ', 'IIZZZI', 'ZXXZZI', 'IYYZZI', 'XXXXZI', 'YXYXZI', 'XXXXII', 'YXYXII', 'IZZZZI', 'ZZIZZI', 'XZIXZI', 'XIIXZI', 'XZIXII', 'XIIXII', 'ZIIZZI', 'IIZZII', 'ZXXZII', 'IYYZII', 'IZZZII', 'ZZIZII', 'ZIIZII', 'IZIIII', 'ZZZIII', 'ZIZIII', 'ZYYIII', 'IXXIII'
        \item 
        \textbf{Coefficients:}\\
        -4.33226054e+00,  2.16784172e-01,  4.25242155e-02,
          4.25242155e-02,  1.74236143e-01,  2.77848545e-01,
         -1.36387133e-01,  2.67785165e-01, -1.69446847e-01,
          2.69965074e-01, -4.55375052e-03, -4.55375052e-03,
          2.16784172e-01,  1.93013854e-01,  3.03086793e-02,
          3.03086793e-02,  1.33024532e-02, -1.33024532e-02,
         -1.33024532e-02,  1.33024532e-02,  4.25242155e-02,
          3.03086793e-02,  4.25242155e-02,  3.03086793e-02,
          1.77739355e-02,  1.77739355e-02,  1.77739355e-02,
          1.77739355e-02,  1.65862607e-02,  1.65862607e-02,
          1.65862607e-02,  1.65862607e-02,  1.74236143e-01,
          1.81245843e-01,  3.02284376e-02,  3.02284376e-02,
          1.38047438e-05, -1.38047438e-05, -1.38047438e-05,
          1.38047438e-05, -1.36387133e-01,  1.35497296e-01,
          1.12687225e-02,  1.12687225e-02, -1.43629530e-02,
          1.43629530e-02, -1.43629530e-02,  1.43629530e-02,
         -1.69446847e-01,  1.51449364e-01,  1.12979294e-02,
          1.12979294e-02,  1.81245843e-01,  1.38047438e-05,
         -1.38047438e-05, -1.38047438e-05,  1.38047438e-05,
          3.02284376e-02,  3.02284376e-02,  2.20039773e-01,
          9.36788811e-03, -9.36788811e-03, -9.36788811e-03,
          9.36788811e-03,  6.29814479e-03,  6.29814479e-03,
          6.29814479e-03,  6.29814479e-03,  1.44469858e-01,
          1.56359949e-01,  1.35497296e-01,  1.12687225e-02,
          1.12687225e-02, -1.43629530e-02, -1.43629530e-02,
          1.43629530e-02,  1.43629530e-02,  1.44469858e-01,
          1.46304375e-01,  2.96450956e-02, -2.96450956e-02,
         -2.96450956e-02,  2.96450956e-02,  1.39550251e-01,
          1.51449364e-01,  1.12979294e-02,  1.12979294e-02,
          1.56359949e-01,  1.39550251e-01,  1.54187181e-01,
          2.77848545e-01,  2.67785165e-01,  2.69965074e-01,
         -4.55375052e-03, -4.55375052e-03
   \end{itemize}
   
   \item[\#10] 
   \begin{itemize}
       \item \textbf{Geometry:}\\
        O 0.0 0.0 0.0\\
        H 0.0 0.0 0.8\\
        H 0.0 0.5656854249492381 -0.565685424949238
        \item \textbf{Pauli strings:}\\
        'IIIIII', 'IIIIIZ', 'IIIZXX', 'IIIIYY', 'IIIIZZ', 'IIIIZI', 'IIIZZI', 'IIIZZZ', 'IIIZII', 'IIIZIZ', 'IIIZYY', 'IIIIXX', 'IIZIII', 'IIZIIZ', 'IIZZXX', 'IIZIYY', 'IZXIZX', 'IIXIZX', 'IZXIIX', 'IIXIIX', 'ZXXIII', 'ZXXIIZ', 'IYYIII', 'IYYIIZ', 'ZXXZXX', 'IYYZXX', 'ZXXIYY', 'IYYIYY', 'XXXXXX', 'YXYXXX', 'XXXYXY', 'YXYYXY', 'IZZIII', 'IZZIIZ', 'IZZZXX', 'IZZIYY', 'ZXZIZX', 'IXIIZX', 'ZXZIIX', 'IXIIIX', 'ZZIIII', 'ZZIIIZ', 'ZZIZXX', 'ZZIIYY', 'XZIXXX', 'XIIXXX', 'XZIYXY', 'XIIYXY', 'ZIIIII', 'ZIIIIZ', 'ZIIZXX', 'ZIIIYY', 'IIZIZZ', 'IZXZXZ', 'IIXZXZ', 'IZXIXI', 'IIXIXI', 'ZXXIZZ', 'IYYIZZ', 'IZZIZZ', 'ZXZZXZ', 'IXIZXZ', 'ZXZIXI', 'IXIIXI', 'XXZXXZ', 'YYIXXZ', 'XXZYYI', 'YYIYYI', 'ZZIIZZ', 'ZIIIZZ', 'IIZZZI', 'ZXXZZI', 'IYYZZI', 'XXXXZI', 'YXYXZI', 'XXXXII', 'YXYXII', 'IZZZZI', 'ZZIZZI', 'XZIXZI', 'XIIXZI', 'XZIXII', 'XIIXII', 'ZIIZZI', 'IIZZII', 'ZXXZII', 'IYYZII', 'IZZZII', 'ZZIZII', 'ZIIZII', 'IZIIII', 'ZZZIII', 'ZIZIII', 'ZYYIII', 'IXXIII'
        \item \textbf{Coefficients:}\\
        -4.11538751e+00,  2.17895403e-01,  1.13531792e-02,
          1.13531792e-02,  1.87775691e-01,  2.88289455e-01,
         -2.05602070e-01,  2.84402772e-01, -3.14580780e-01,
          2.83784059e-01,  7.55621549e-04,  7.55621549e-04,
          2.17895403e-01,  2.19470930e-01,  1.33766932e-02,
          1.33766932e-02,  1.36980237e-02, -1.36980237e-02,
         -1.36980237e-02,  1.36980237e-02,  1.13531792e-02,
          1.33766932e-02,  1.13531792e-02,  1.33766932e-02,
          9.63633303e-03,  9.63633303e-03,  9.63633303e-03,
          9.63633303e-03,  1.20822103e-02,  1.20822103e-02,
          1.20822103e-02,  1.20822103e-02,  1.87775691e-01,
          1.93236463e-01,  1.41914873e-02,  1.41914873e-02,
         -2.09397446e-03,  2.09397446e-03,  2.09397446e-03,
         -2.09397446e-03, -2.05602070e-01,  1.37534445e-01,
          9.72307776e-03,  9.72307776e-03, -1.13047814e-02,
          1.13047814e-02, -1.13047814e-02,  1.13047814e-02,
         -3.14580780e-01,  1.62251187e-01,  5.73630193e-03,
          5.73630193e-03,  1.93236463e-01, -2.09397446e-03,
          2.09397446e-03,  2.09397446e-03, -2.09397446e-03,
          1.41914873e-02,  1.41914873e-02,  2.20039773e-01,
          1.16735953e-02, -1.16735953e-02, -1.16735953e-02,
          1.16735953e-02,  6.07016921e-03,  6.07016921e-03,
          6.07016921e-03,  6.07016921e-03,  1.45288677e-01,
          1.62574829e-01,  1.37534445e-01,  9.72307776e-03,
          9.72307776e-03, -1.13047814e-02, -1.13047814e-02,
          1.13047814e-02,  1.13047814e-02,  1.45288677e-01,
          1.43113624e-01,  3.35267529e-02, -3.35267529e-02,
         -3.35267529e-02,  3.35267529e-02,  1.42277769e-01,
          1.62251187e-01,  5.73630193e-03,  5.73630193e-03,
          1.62574829e-01,  1.42277769e-01,  1.62507380e-01,
          2.88289455e-01,  2.84402772e-01,  2.83784059e-01,
          7.55621549e-04,  7.55621549e-04
   \end{itemize}
   
   \item[\#11]
   \begin{itemize}
       \item \textbf{Geometry:}\\
        O 0.0 0.0 0.0\\
        H 0.0 0.0 0.8500000000000001\\
        H 0.0 0.6010407640085655 -0.6010407640085654
        \item \textbf{Pauli strings:}\\
        'IIIIII', 'IIIIIZ', 'IIIZXX', 'IIIIYY', 'IIIIZZ', 'IIIIZI', 'IIIZZI', 'IIIZZZ', 'IIIZII', 'IIIZIZ', 'IIIZYY', 'IIIIXX', 'IIZIII', 'IIZIIZ', 'IIZZXX', 'IIZIYY', 'IZXIZX', 'IIXIZX', 'IZXIIX', 'IIXIIX', 'ZXXIII', 'ZXXIIZ', 'IYYIII', 'IYYIIZ', 'ZXXZXX', 'IYYZXX', 'ZXXIYY', 'IYYIYY', 'XXXXXX', 'YXYXXX', 'XXXYXY', 'YXYYXY', 'IZZIII', 'IZZIIZ', 'IZZZXX', 'IZZIYY', 'ZXZIZX', 'IXIIZX', 'ZXZIIX', 'IXIIIX', 'ZZIIII', 'ZZIIIZ', 'ZZIZXX', 'ZZIIYY', 'XZIXXX', 'XIIXXX', 'XZIYXY', 'XIIYXY', 'ZIIIII', 'ZIIIIZ', 'ZIIZXX', 'ZIIIYY', 'IIZIZZ', 'IZXZXZ', 'IIXZXZ', 'IZXIXI', 'IIXIXI', 'ZXXIZZ', 'IYYIZZ', 'IZZIZZ', 'ZXZZXZ', 'IXIZXZ', 'ZXZIXI', 'IXIIXI', 'XXZXXZ', 'YYIXXZ', 'XXZYYI', 'YYIYYI', 'ZZIIZZ', 'ZIIIZZ', 'IIZZZI', 'ZXXZZI', 'IYYZZI', 'XXXXZI', 'YXYXZI', 'XXXXII', 'YXYXII', 'IZZZZI', 'ZZIZZI', 'XZIXZI', 'XIIXZI', 'XZIXII', 'XIIXII', 'ZIIZZI', 'IIZZII', 'ZXXZII', 'IYYZII', 'IZZZII', 'ZZIZII', 'ZIIZII', 'IZIIII', 'ZZZIII', 'ZIZIII', 'ZYYIII', 'IXXIII'
        \item \textbf{Coefficients:}\\
        -4.18421016e+00,  2.12925633e-01,  1.65128088e-02,
          1.65128088e-02,  1.83688588e-01,  2.86059594e-01,
         -1.78590823e-01,  2.81186536e-01, -2.79600548e-01,
          2.80826471e-01,  4.03065401e-05,  4.03065401e-05,
          2.12925633e-01,  2.16209852e-01,  1.63876072e-02,
          1.63876072e-02,  1.35540038e-02, -1.35540038e-02,
         -1.35540038e-02,  1.35540038e-02,  1.65128088e-02,
          1.63876072e-02,  1.65128088e-02,  1.63876072e-02,
          1.03514259e-02,  1.03514259e-02,  1.03514259e-02,
          1.03514259e-02,  1.24079578e-02,  1.24079578e-02,
          1.24079578e-02,  1.24079578e-02,  1.83688588e-01,
          1.91934511e-01,  1.67218277e-02,  1.67218277e-02,
         -1.68233400e-03,  1.68233400e-03,  1.68233400e-03,
         -1.68233400e-03, -1.78590823e-01,  1.36055808e-01,
          9.77911827e-03,  9.77911827e-03, -1.16672670e-02,
          1.16672670e-02, -1.16672670e-02,  1.16672670e-02,
         -2.79600548e-01,  1.60721737e-01,  6.77720124e-03,
          6.77720124e-03,  1.91934511e-01, -1.68233400e-03,
          1.68233400e-03,  1.68233400e-03, -1.68233400e-03,
          1.67218277e-02,  1.67218277e-02,  2.20039773e-01,
          1.11818191e-02, -1.11818191e-02, -1.11818191e-02,
          1.11818191e-02,  6.21866198e-03,  6.21866198e-03,
          6.21866198e-03,  6.21866198e-03,  1.43694511e-01,
          1.61700816e-01,  1.36055808e-01,  9.77911827e-03,
          9.77911827e-03, -1.16672670e-02, -1.16672670e-02,
          1.16672670e-02,  1.16672670e-02,  1.43694511e-01,
          1.42935168e-01,  3.36112423e-02, -3.36112423e-02,
         -3.36112423e-02,  3.36112423e-02,  1.41290329e-01,
          1.60721737e-01,  6.77720124e-03,  6.77720124e-03,
          1.61700816e-01,  1.41290329e-01,  1.61239045e-01,
          2.86059594e-01,  2.81186536e-01,  2.80826471e-01,
          4.03065401e-05,  4.03065401e-05
   \end{itemize}

   \item[\#12]
   \begin{itemize}
       \item \textbf{Geometry:}\\
        O 0.0 0.0 0.0\\
        H 0.0 0.0 0.9\\
        H 0.0 0.6363961030678928 -0.6363961030678927 
        \item \textbf{Pauli strings:}\\
        'IIIIII', 'IIIIIZ', 'IIIZXX', 'IIIIYY', 'IIIIZZ', 'IIIIZI', 'IIIZZI', 'IIIZZZ', 'IIIZII', 'IIIZIZ', 'IIIZYY', 'IIIIXX', 'IIZIII', 'IIZIIZ', 'IIZZXX', 'IIZIYY', 'IZXIZX', 'IIXIZX', 'IZXIIX', 'IIXIIX', 'ZXXIII', 'ZXXIIZ', 'IYYIII', 'IYYIIZ', 'ZXXZXX', 'IYYZXX', 'ZXXIYY', 'IYYIYY', 'XXXXXX', 'YXYXXX', 'XXXYXY', 'YXYYXY', 'IZZIII', 'IZZIIZ', 'IZZZXX', 'IZZIYY', 'ZXZIZX', 'IXIIZX', 'ZXZIIX', 'IXIIIX', 'ZZIIII', 'ZZIIIZ', 'ZZIZXX', 'ZZIIYY', 'XZIXXX', 'XIIXXX', 'XZIYXY', 'XIIYXY', 'ZIIIII', 'ZIIIIZ', 'ZIIZXX', 'ZIIIYY', 'IIZIZZ', 'IZXZXZ', 'IIXZXZ', 'IZXIXI', 'IIXIXI', 'ZXXIZZ', 'IYYIZZ', 'IZZIZZ', 'ZXZZXZ', 'IXIZXZ', 'ZXZIXI', 'IXIIXI', 'XXZXXZ', 'YYIXXZ', 'XXZYYI', 'YYIYYI', 'ZZIIZZ', 'ZIIIZZ', 'IIZZZI', 'ZXXZZI', 'IYYZZI', 'XXXXZI', 'YXYXZI', 'XXXXII', 'YXYXII', 'IZZZZI', 'ZZIZZI', 'XZIXZI', 'XIIXZI', 'XZIXII', 'XIIXII', 'ZIIZZI', 'IIZZII', 'ZXXZII', 'IYYZII', 'IZZZII', 'ZZIZII', 'ZIIZII', 'IZIIII', 'ZZZIII', 'ZIZIII', 'ZYYIII', 'IXXIII'
        \item \textbf{Coefficients:}\\
        -4.24527030e+00,  2.08534711e-01,  2.19732113e-02,
          2.19732113e-02,  1.80768886e-01,  2.83572464e-01,
         -1.52957835e-01,  2.77669369e-01, -2.46741044e-01,
          2.77637920e-01, -7.60640039e-04, -7.60640039e-04,
          2.08534711e-01,  2.12320277e-01,  1.94153038e-02,
          1.94153038e-02,  1.33687190e-02, -1.33687190e-02,
         -1.33687190e-02,  1.33687190e-02,  2.19732113e-02,
          1.94153038e-02,  2.19732113e-02,  1.94153038e-02,
          1.13493149e-02,  1.13493149e-02,  1.13493149e-02,
          1.13493149e-02,  1.27799522e-02,  1.27799522e-02,
          1.27799522e-02,  1.27799522e-02,  1.80768886e-01,
          1.90347528e-01,  1.94272411e-02,  1.94272411e-02,
         -1.23403835e-03,  1.23403835e-03,  1.23403835e-03,
         -1.23403835e-03, -1.52957835e-01,  1.34580758e-01,
          9.76190567e-03,  9.76190567e-03, -1.20325715e-02,
          1.20325715e-02, -1.20325715e-02,  1.20325715e-02,
         -2.46741044e-01,  1.59059066e-01,  7.86806794e-03,
          7.86806794e-03,  1.90347528e-01, -1.23403835e-03,
          1.23403835e-03,  1.23403835e-03, -1.23403835e-03,
          1.94272411e-02,  1.94272411e-02,  2.20039773e-01,
          1.06607502e-02, -1.06607502e-02, -1.06607502e-02,
          1.06607502e-02,  6.35764925e-03,  6.35764925e-03,
          6.35764925e-03,  6.35764925e-03,  1.42019557e-01,
          1.60795576e-01,  1.34580758e-01,  9.76190567e-03,
          9.76190567e-03, -1.20325715e-02, -1.20325715e-02,
          1.20325715e-02,  1.20325715e-02,  1.42019557e-01,
          1.42352955e-01,  3.34682426e-02, -3.34682426e-02,
         -3.34682426e-02,  3.34682426e-02,  1.40061898e-01,
          1.59059066e-01,  7.86806794e-03,  7.86806794e-03,
          1.60795576e-01,  1.40061898e-01,  1.59859584e-01,
          2.83572464e-01,  2.77669369e-01,  2.77637920e-01,
         -7.60640039e-04, -7.60640039e-04
   \end{itemize}
   
   \item[\#13]
   \begin{itemize}
       \item \textbf{Geometry:}\\
        O 0.0 0.0 0.0\\
        H 0.0 0.0 0.95\\
        H 0.0 0.6717514421272202 -0.67175144212722
        \item \textbf{Pauli strings:}\\
        'IIIIII', 'IIIIIZ', 'IIIZXX', 'IIIIYY', 'IIIIZZ', 'IIIIZI', 'IIIZZI', 'IIIZZZ', 'IIIZII', 'IIIZIZ', 'IIIZYY', 'IIIIXX', 'IIZIII', 'IIZIIZ', 'IIZZXX', 'IIZIYY', 'IZXIZX', 'IIXIZX', 'IZXIIX', 'IIXIIX', 'ZXXIII', 'ZXXIIZ', 'IYYIII', 'IYYIIZ', 'ZXXZXX', 'IYYZXX', 'ZXXIYY', 'IYYIYY', 'XXXXXX', 'YXYXXX', 'XXXYXY', 'YXYYXY', 'IZZIII', 'IZZIIZ', 'IZZZXX', 'IZZIYY', 'ZXZIZX', 'IXIIZX', 'ZXZIIX', 'IXIIIX', 'ZZIIII', 'ZZIIIZ', 'ZZIZXX', 'ZZIIYY', 'XZIXXX', 'XIIXXX', 'XZIYXY', 'XIIYXY', 'ZIIIII', 'ZIIIIZ', 'ZIIZXX', 'ZIIIYY', 'IIZIZZ', 'IZXZXZ', 'IIXZXZ', 'IZXIXI', 'IIXIXI', 'ZXXIZZ', 'IYYIZZ', 'IZZIZZ', 'ZXZZXZ', 'IXIZXZ', 'ZXZIXI', 'IXIIXI', 'XXZXXZ', 'YYIXXZ', 'XXZYYI', 'YYIYYI', 'ZZIIZZ', 'ZIIIZZ', 'IIZZZI', 'ZXXZZI', 'IYYZZI', 'XXXXZI', 'YXYXZI', 'XXXXII', 'YXYXII', 'IZZZZI', 'ZZIZZI', 'XZIXZI', 'XIIXZI', 'XZIXII', 'XIIXII', 'ZIIZZI', 'IIZZII', 'ZXXZII', 'IYYZII', 'IZZZII', 'ZZIZII', 'ZIIZII', 'IZIIII', 'ZZZIII', 'ZIZIII', 'ZYYIII', 'IXXIII'
        \item \textbf{Coefficients:}\\
        -4.29751152e+00,  2.04360520e-01,  2.75821889e-02,
          2.75821889e-02,  1.78694818e-01,  2.80854611e-01,
         -1.29024015e-01,  2.73913983e-01, -2.15984312e-01,
          2.74294495e-01, -1.63841304e-03, -1.63841304e-03,
          2.04360520e-01,  2.07788658e-01,  2.23814173e-02,
          2.23814173e-02,  1.31349391e-02, -1.31349391e-02,
         -1.31349391e-02,  1.31349391e-02,  2.75821889e-02,
          2.23814173e-02,  2.75821889e-02,  2.23814173e-02,
          1.26649690e-02,  1.26649690e-02,  1.26649690e-02,
          1.26649690e-02,  1.31893596e-02,  1.31893596e-02,
          1.31893596e-02,  1.31893596e-02,  1.78694818e-01,
          1.88454057e-01,  2.22937970e-02,  2.22937970e-02,
         -7.50206924e-04,  7.50206924e-04,  7.50206924e-04,
         -7.50206924e-04, -1.29024015e-01,  1.33198921e-01,
          9.71875623e-03,  9.71875623e-03, -1.23929083e-02,
          1.23929083e-02, -1.23929083e-02,  1.23929083e-02,
         -2.15984312e-01,  1.57256889e-01,  9.01268259e-03,
          9.01268259e-03,  1.88454057e-01, -7.50206924e-04,
          7.50206924e-04,  7.50206924e-04, -7.50206924e-04,
          2.22937970e-02,  2.22937970e-02,  2.20039773e-01,
          1.01252951e-02, -1.01252951e-02, -1.01252951e-02,
          1.01252951e-02,  6.48455356e-03,  6.48455356e-03,
          6.48455356e-03,  6.48455356e-03,  1.40352261e-01,
          1.59864584e-01,  1.33198921e-01,  9.71875623e-03,
          9.71875623e-03, -1.23929083e-02, -1.23929083e-02,
          1.23929083e-02,  1.23929083e-02,  1.40352261e-01,
          1.41452422e-01,  3.31449055e-02, -3.31449055e-02,
         -3.31449055e-02,  3.31449055e-02,  1.38680399e-01,
          1.57256889e-01,  9.01268259e-03,  9.01268259e-03,
          1.59864584e-01,  1.38680399e-01,  1.58396241e-01,
          2.80854611e-01,  2.73913983e-01,  2.74294495e-01,
         -1.63841304e-03, -1.63841304e-03
   \end{itemize}
   
   \item[\#14]
   \begin{itemize}
       \item \textbf{Geometry:}\\
        O 0.0 0.0 0.0\\
        H 0.0 0.0 1.0\\
        H 0.0 0.7071067811865476 -0.7071067811865475
        \item \textbf{Pauli strings:}\\
        'IIIIII', 'IIIIIZ', 'IIIZXX', 'IIIIYY', 'IIIIZZ', 'IIIZZI', 'IIIZII', 'IIZIII', 'ZXXIII', 'IYYIII', 'IZZIII', 'ZZIIII', 'ZIIIII', 'IIIIZI', 'IIIZZZ', 'IIIZIZ', 'IIIZYY', 'IIIIXX', 'IIZIIZ', 'IIZZXX', 'IIZIYY', 'IZXIZX', 'IIXIZX', 'IZXIIX', 'IIXIIX', 'ZXXIIZ', 'IYYIIZ', 'ZXXZXX', 'IYYZXX', 'ZXXIYY', 'IYYIYY', 'XXXXXX', 'YXYXXX', 'XXXYXY', 'YXYYXY', 'IZZIIZ', 'IZZZXX', 'IZZIYY', 'ZXZIZX', 'IXIIZX', 'ZXZIIX', 'IXIIIX', 'ZZIIIZ', 'ZZIZXX', 'ZZIIYY', 'XZIXXX', 'XIIXXX', 'XZIYXY', 'XIIYXY', 'ZIIIIZ', 'ZIIZXX', 'ZIIIYY', 'IIZIZZ', 'IZXZXZ', 'IIXZXZ', 'IZXIXI', 'IIXIXI', 'ZXXIZZ', 'IYYIZZ', 'IZZIZZ', 'ZXZZXZ', 'IXIZXZ', 'ZXZIXI', 'IXIIXI', 'XXZXXZ', 'YYIXXZ', 'XXZYYI', 'YYIYYI', 'ZZIIZZ', 'ZIIIZZ', 'IIZZZI', 'ZXXZZI', 'IYYZZI', 'XXXXZI', 'YXYXZI', 'XXXXII', 'YXYXII', 'IZZZZI', 'ZZIZZI', 'XZIXZI', 'XIIXZI', 'XZIXII', 'XIIXII', 'ZIIZZI', 'IIZZII', 'ZXXZII', 'IYYZII', 'IZZZII', 'ZZIZII', 'ZIIZII', 'IZIIII', 'ZZZIII', 'ZIZIII', 'ZYYIII', 'IXXIII'
        \item \textbf{Coefficients:}
        -4.34044048e+00,  2.00156300e-01,  3.31586149e-02,
          3.31586149e-02,  1.77224925e-01, -1.06919570e-01,
         -1.87323329e-01,  2.00156300e-01,  3.31586149e-02,
          3.31586149e-02,  1.77224925e-01, -1.06919570e-01,
         -1.87323329e-01,  2.77926721e-01,  2.69967545e-01,
          2.70853968e-01, -2.58301184e-03, -2.58301184e-03,
          2.02651846e-01,  2.51875326e-02,  2.51875326e-02,
          1.28480595e-02, -1.28480595e-02, -1.28480595e-02,
          1.28480595e-02,  2.51875326e-02,  2.51875326e-02,
          1.43093264e-02,  1.43093264e-02,  1.43093264e-02,
          1.43093264e-02,  1.36252332e-02,  1.36252332e-02,
          1.36252332e-02,  1.36252332e-02,  1.86248325e-01,
          2.52877722e-02,  2.52877722e-02, -2.35918259e-04,
          2.35918259e-04,  2.35918259e-04, -2.35918259e-04,
          1.31966794e-01,  9.69791055e-03,  9.69791055e-03,
         -1.27388868e-02,  1.27388868e-02, -1.27388868e-02,
          1.27388868e-02,  1.55311641e-01,  1.02017918e-02,
          1.02017918e-02,  1.86248325e-01, -2.35918259e-04,
          2.35918259e-04,  2.35918259e-04, -2.35918259e-04,
          2.52877722e-02,  2.52877722e-02,  2.20039773e-01,
          9.59106147e-03, -9.59106147e-03, -9.59106147e-03,
          9.59106147e-03,  6.59719606e-03,  6.59719606e-03,
          6.59719606e-03,  6.59719606e-03,  1.38758621e-01,
          1.58907274e-01,  1.31966794e-01,  9.69791055e-03,
          9.69791055e-03, -1.27388868e-02, -1.27388868e-02,
          1.27388868e-02,  1.27388868e-02,  1.38758621e-01,
          1.40298739e-01,  3.26864549e-02, -3.26864549e-02,
         -3.26864549e-02,  3.26864549e-02,  1.37212910e-01,
          1.55311641e-01,  1.02017918e-02,  1.02017918e-02,
          1.58907274e-01,  1.37212910e-01,  1.56865872e-01,
          2.77926721e-01,  2.69967545e-01,  2.70853968e-01,
         -2.58301184e-03, -2.58301184e-03
   \end{itemize}
   \item[\#15]
   \begin{itemize}
       \item \textbf{Geometry:}\\
        O 0.0 0.0 0.0\\
        H 0.0 0.0 0.8\\
        H 0.0 0.3061467458920719 -0.7391036260090295
        \item \textbf{Pauli strings:}\\
        'IIIIII', 'IIIIIZ', 'IIIZXX', 'IIIIYY', 'IIIIZZ', 'IIIZZI', 'IIIZII', 'IIZIII', 'ZXXIII', 'IYYIII', 'IZZIII', 'ZZIIII', 'ZIIIII', 'IIIIZI', 'IIIZZZ', 'IIIZIZ', 'IIIZYY', 'IIIIXX', 'IIZIIZ', 'IIZZXX', 'IIZIYY', 'IZXIZX', 'IIXIZX', 'IZXIIX', 'IIXIIX', 'ZXXIIZ', 'IYYIIZ', 'ZXXZXX', 'IYYZXX', 'ZXXIYY', 'IYYIYY', 'XXXXXX', 'YXYXXX', 'XXXYXY', 'YXYYXY', 'IZZIIZ', 'IZZZXX', 'IZZIYY', 'ZXZIZX', 'IXIIZX', 'ZXZIIX', 'IXIIIX', 'ZZIIIZ', 'ZZIZXX', 'ZZIIYY', 'XZIXXX', 'XIIXXX', 'XZIYXY', 'XIIYXY', 'ZIIIIZ', 'ZIIZXX', 'ZIIIYY', 'IIZIZZ', 'IZXZXZ', 'IIXZXZ', 'IZXIXI', 'IIXIXI', 'ZXXIZZ', 'IYYIZZ', 'IZZIZZ', 'ZXZZXZ', 'IXIZXZ', 'ZXZIXI', 'IXIIXI', 'XXZXXZ', 'YYIXXZ', 'XXZYYI', 'YYIYYI', 'ZZIIZZ', 'ZIIIZZ', 'IIZZZI', 'ZXXZZI', 'IYYZZI', 'XXXXZI', 'YXYXZI', 'XXXXII', 'YXYXII', 'IZZZZI', 'ZZIZZI', 'XZIXZI', 'XIIXZI', 'XZIXII', 'XIIXII', 'ZIIZZI', 'IIZZII', 'ZXXZII', 'IYYZII', 'IZZZII', 'ZZIZII', 'ZIIZII', 'IZIIII', 'ZZZIII', 'ZIZIII', 'ZYYIII', 'IXXIII'
        \item \textbf{Coefficients:} 
        -4.08632279e+00,  1.96205234e-01,  2.83382424e-03,
          2.83382424e-03,  1.87268744e-01, -1.73812359e-01,
         -3.41165436e-01,  1.96205234e-01,  2.83382424e-03,
          2.83382424e-03,  1.87268744e-01, -1.73812359e-01,
         -3.41165436e-01,  2.87156835e-01,  2.85716257e-01,
          2.85464227e-01,  8.50712163e-04,  8.50712163e-04,
          2.20340092e-01,  5.67651883e-03,  5.67651883e-03,
          1.24331864e-02, -1.24331864e-02, -1.24331864e-02,
          1.24331864e-02,  5.67651883e-03,  5.67651883e-03,
          1.13456381e-02,  1.13456381e-02,  1.13456381e-02,
          1.13456381e-02,  8.07756972e-03,  8.07756972e-03,
          8.07756972e-03,  8.07756972e-03,  1.95405897e-01,
          7.73459300e-03,  7.73459300e-03, -1.33642229e-03,
          1.33642229e-03,  1.33642229e-03, -1.33642229e-03,
          1.39930154e-01,  6.71811240e-03,  6.71811240e-03,
         -6.91326441e-03,  6.91326441e-03, -6.91326441e-03,
          6.91326441e-03,  1.63256947e-01,  3.00846304e-03,
          3.00846304e-03,  1.95405897e-01, -1.33642229e-03,
          1.33642229e-03,  1.33642229e-03, -1.33642229e-03,
          7.73459300e-03,  7.73459300e-03,  2.20039773e-01,
          1.23691125e-02, -1.23691125e-02, -1.23691125e-02,
          1.23691125e-02,  6.16159981e-03,  6.16159981e-03,
          6.16159981e-03,  6.16159981e-03,  1.42653962e-01,
          1.63293341e-01,  1.39930154e-01,  6.71811240e-03,
          6.71811240e-03, -6.91326441e-03, -6.91326441e-03,
          6.91326441e-03,  6.91326441e-03,  1.42653962e-01,
          1.36904593e-01,  3.72343089e-02, -3.72343089e-02,
         -3.72343089e-02,  3.72343089e-02,  1.41418432e-01,
          1.63256947e-01,  3.00846304e-03,  3.00846304e-03,
          1.63293341e-01,  1.41418432e-01,  1.63453864e-01,
          2.87156835e-01,  2.85716257e-01,  2.85464227e-01,
          8.50712163e-04,  8.50712163e-04
   \end{itemize}
   \item[\#16]
   \begin{itemize}
       \item \textbf{Geometry:}\\
        O 0.0 0.0 0.0\\
        H 0.0 0.0 0.8500000000000001\\
        H 0.0 0.32528091751032645 -0.7852976026345938
        \item \textbf{Pauli strings:}\\
        'IIIIII', 'IIIIIZ', 'IIIZXX', 'IIIIYY', 'IIIIZZ', 'IIIZZI', 'IIIZII', 'IIZIII', 'ZXXIII', 'IYYIII', 'IZZIII', 'ZZIIII', 'ZIIIII', 'IIIIZI', 'IIIZZZ', 'IIIZIZ', 'IIIZYY', 'IIIIXX', 'IIZIIZ', 'IIZZXX', 'IIZIYY', 'IZXIZX', 'IIXIZX', 'IZXIIX', 'IIXIIX', 'ZXXIIZ', 'IYYIIZ', 'ZXXZXX', 'IYYZXX', 'ZXXIYY', 'IYYIYY', 'XXXXXX', 'YXYXXX', 'XXXYXY', 'YXYYXY', 'IZZIIZ', 'IZZZXX', 'IZZIYY', 'ZXZIZX', 'IXIIZX', 'ZXZIIX', 'IXIIIX', 'ZZIIIZ', 'ZZIZXX', 'ZZIIYY', 'XZIXXX', 'XIIXXX', 'XZIYXY', 'XIIYXY', 'ZIIIIZ', 'ZIIZXX', 'ZIIIYY', 'IIZIZZ', 'IZXZXZ', 'IIXZXZ', 'IZXIXI', 'IIXIXI', 'ZXXIZZ', 'IYYIZZ', 'IZZIZZ', 'ZXZZXZ', 'IXIZXZ', 'ZXZIXI', 'IXIIXI', 'XXZXXZ', 'YYIXXZ', 'XXZYYI', 'YYIYYI', 'ZZIIZZ', 'ZIIIZZ', 'IIZZZI', 'ZXXZZI', 'IYYZZI', 'XXXXZI', 'YXYXZI', 'XXXXII', 'YXYXII', 'IZZZZI', 'ZZIZZI', 'XZIXZI', 'XIIXZI', 'XZIXII', 'XIIXII', 'ZIIZZI', 'IIZZII', 'ZXXZII', 'IYYZII', 'IZZZII', 'ZZIZII', 'ZIIZII', 'IZIIII', 'ZZZIII', 'ZIZIII', 'ZYYIII', 'IXXIII'
        \item \textbf{Coefficients:}
        -4.16628458e+00,  1.92746067e-01,  6.03249785e-03,
          6.03249785e-03,  1.83891130e-01, -1.47036211e-01,
         -3.03538946e-01,  1.92746067e-01,  6.03249785e-03,
          6.03249785e-03,  1.83891130e-01, -1.47036211e-01,
         -3.03538946e-01,  2.85269047e-01,  2.82563442e-01,
          2.82379184e-01,  4.46507743e-04,  4.46507743e-04,
          2.19374172e-01,  7.65933707e-03,  7.65933707e-03,
          1.24163643e-02, -1.24163643e-02, -1.24163643e-02,
          1.24163643e-02,  7.65933707e-03,  7.65933707e-03,
          1.11748550e-02,  1.11748550e-02,  1.11748550e-02,
          1.11748550e-02,  8.35858145e-03,  8.35858145e-03,
          8.35858145e-03,  8.35858145e-03,  1.94968175e-01,
          9.35855304e-03,  9.35855304e-03, -1.15236236e-03,
          1.15236236e-03,  1.15236236e-03, -1.15236236e-03,
          1.37424141e-01,  6.81557487e-03,  6.81557487e-03,
         -7.23411348e-03,  7.23411348e-03, -7.23411348e-03,
          7.23411348e-03,  1.62380988e-01,  3.72330965e-03,
          3.72330965e-03,  1.94968175e-01, -1.15236236e-03,
          1.15236236e-03,  1.15236236e-03, -1.15236236e-03,
          9.35855304e-03,  9.35855304e-03,  2.20039773e-01,
          1.18871594e-02, -1.18871594e-02, -1.18871594e-02,
          1.18871594e-02,  6.33196177e-03,  6.33196177e-03,
          6.33196177e-03,  6.33196177e-03,  1.40243936e-01,
          1.62646118e-01,  1.37424141e-01,  6.81557487e-03,
          6.81557487e-03, -7.23411348e-03, -7.23411348e-03,
          7.23411348e-03,  7.23411348e-03,  1.40243936e-01,
          1.36235089e-01,  3.73969277e-02, -3.73969277e-02,
         -3.73969277e-02,  3.73969277e-02,  1.40114164e-01,
          1.62380988e-01,  3.72330965e-03,  3.72330965e-03,
          1.62646118e-01,  1.40114164e-01,  1.62533072e-01,
          2.85269047e-01,  2.82563442e-01,  2.82379184e-01,
          4.46507743e-04,  4.46507743e-04
   \end{itemize}
   
   \item[\#17]
   \begin{itemize}
       \item \textbf{Geometry:}\\
        O 0.0 0.0 0.0\\
        H 0.0 0.0 0.9\\
        H 0.0 0.34441508912858093 -0.831491579260158
        \item \textbf{Pauli strings:}\\
        'IIIIII', 'IIIIIZ', 'IIIZXX', 'IIIIYY', 'IIIIZZ', 'IIIZZI', 'IIIZII', 'IIZIII', 'ZXXIII', 'IYYIII', 'IZZIII', 'ZZIIII', 'ZIIIII', 'IIIIZI', 'IIIZZZ', 'IIIZIZ', 'IIIZYY', 'IIIIXX', 'IIZIIZ', 'IIZZXX', 'IIZIYY', 'IZXIZX', 'IIXIZX', 'IZXIIX', 'IIXIIX', 'ZXXIIZ', 'IYYIIZ', 'ZXXZXX', 'IYYZXX', 'ZXXIYY', 'IYYIYY', 'XXXXXX', 'YXYXXX', 'XXXYXY', 'YXYYXY', 'IZZIIZ', 'IZZZXX', 'IZZIYY', 'ZXZIZX', 'IXIIZX', 'ZXZIIX', 'IXIIIX', 'ZZIIIZ', 'ZZIZXX', 'ZZIIYY', 'XZIXXX', 'XIIXXX', 'XZIYXY', 'XIIYXY', 'ZIIIIZ', 'ZIIZXX', 'ZIIIYY', 'IIZIZZ', 'IZXZXZ', 'IIXZXZ', 'IZXIXI', 'IIXIXI', 'ZXXIZZ', 'IYYIZZ', 'IZZIZZ', 'ZXZZXZ', 'IXIZXZ', 'ZXZIXI', 'IXIIXI', 'XXZXXZ', 'YYIXXZ', 'XXZYYI', 'YYIYYI', 'ZZIIZZ', 'ZIIIZZ', 'IIZZZI', 'ZXXZZI', 'IYYZZI', 'XXXXZI', 'YXYXZI', 'XXXXII', 'YXYXII', 'IZZZZI', 'ZZIZZI', 'XZIXZI', 'XIIXZI', 'XZIXII', 'XIIXII', 'ZIIZZI', 'IIZZII', 'ZXXZII', 'IYYZII', 'IZZZII', 'ZZIZII', 'ZIIZII', 'IZIIII', 'ZZZIII', 'ZIZIII', 'ZYYIII', 'IXXIII'
        \item \textbf{Coefficients:}
        -4.23972575e+00,  1.90353433e-01,  9.77259265e-03,
          9.77259265e-03,  1.81797388e-01, -1.22171585e-01,
         -2.67909485e-01,  1.90353433e-01,  9.77259265e-03,
          9.77259265e-03,  1.81797388e-01, -1.22171585e-01,
         -2.67909485e-01,  2.83221840e-01,  2.79180236e-01,
          2.79090104e-01, -2.28222707e-05, -2.28222707e-05,
          2.18091262e-01,  9.89379447e-03,  9.89379447e-03,
          1.23902551e-02, -1.23902551e-02, -1.23902551e-02,
          1.23902551e-02,  9.89379447e-03,  9.89379447e-03,
          1.10959874e-02,  1.10959874e-02,  1.10959874e-02,
          1.10959874e-02,  8.68199438e-03,  8.68199438e-03,
          8.68199438e-03,  8.68199438e-03,  1.94388438e-01,
          1.12204704e-02,  1.12204704e-02, -9.42271482e-04,
          9.42271482e-04,  9.42271482e-04, -9.42271482e-04,
          1.34780686e-01,  6.82971950e-03,  6.82971950e-03,
         -7.60514464e-03,  7.60514464e-03, -7.60514464e-03,
          7.60514464e-03,  1.61443379e-01,  4.53477494e-03,
          4.53477494e-03,  1.94388438e-01, -9.42271482e-04,
          9.42271482e-04,  9.42271482e-04, -9.42271482e-04,
          1.12204704e-02,  1.12204704e-02,  2.20039773e-01,
          1.13538449e-02, -1.13538449e-02, -1.13538449e-02,
          1.13538449e-02,  6.49804035e-03,  6.49804035e-03,
          6.49804035e-03,  6.49804035e-03,  1.37682564e-01,
          1.61993578e-01,  1.34780686e-01,  6.82971950e-03,
          6.82971950e-03, -7.60514464e-03, -7.60514464e-03,
          7.60514464e-03,  7.60514464e-03,  1.37682564e-01,
          1.35320112e-01,  3.73129624e-02, -3.73129624e-02,
         -3.73129624e-02,  3.73129624e-02,  1.38536620e-01,
          1.61443379e-01,  4.53477494e-03,  4.53477494e-03,
          1.61993578e-01,  1.38536620e-01,  1.61500209e-01,
          2.83221840e-01,  2.79180236e-01,  2.79090104e-01,
         -2.28222707e-05, -2.28222707e-05
   \end{itemize}
   \item[\#18]
   \begin{itemize}
       \item \textbf{Geometry:}\\
        O 0.0 0.0 0.0\\
        H 0.0 0.0 0.95\\
        H 0.0 0.36354926074683536 -0.8776855558857224
        \item \textbf{Pauli strings:}\\
        'IIIIII', 'IIIIIZ', 'IIIZXX', 'IIIIYY', 'IIIIZZ', 'IIIZZI', 'IIIZII', 'IIZIII', 'ZXXIII', 'IYYIII', 'IZZIII', 'ZZIIII', 'ZIIIII', 'IIIIZI', 'IIIZZZ', 'IIIZIZ', 'IIIZYY', 'IIIIXX', 'IIZIIZ', 'IIZZXX', 'IIZIYY', 'IZXIZX', 'IIXIZX', 'IZXIIX', 'IIXIIX', 'ZXXIIZ', 'IYYIIZ', 'ZXXZXX', 'IYYZXX', 'ZXXIYY', 'IYYIYY', 'XXXXXX', 'YXYXXX', 'XXXYXY', 'YXYYXY', 'IZZIIZ', 'IZZZXX', 'IZZIYY', 'ZXZIZX', 'IXIIZX', 'ZXZIIX', 'IXIIIX', 'ZZIIIZ', 'ZZIZXX', 'ZZIIYY', 'XZIXXX', 'XIIXXX', 'XZIYXY', 'XIIYXY', 'ZIIIIZ', 'ZIIZXX', 'ZIIIYY', 'IIZIZZ', 'IZXZXZ', 'IIXZXZ', 'IZXIXI', 'IIXIXI', 'ZXXIZZ', 'IYYIZZ', 'IZZIZZ', 'ZXZZXZ', 'IXIZXZ', 'ZXZIXI', 'IXIIXI', 'XXZXXZ', 'YYIXXZ', 'XXZYYI', 'YYIYYI', 'ZZIIZZ', 'ZIIIZZ', 'IIZZZI', 'ZXXZZI', 'IYYZZI', 'XXXXZI', 'YXYXZI', 'XXXXII', 'YXYXII', 'IZZZZI', 'ZZIZZI', 'XZIXZI', 'XIIXZI', 'XZIXII', 'XIIXII', 'ZIIZZI', 'IIZZII', 'ZXXZII', 'IYYZII', 'IZZZII', 'ZZIZII', 'ZIIZII', 'IZIIII', 'ZZZIII', 'ZIZIII', 'ZYYIII', 'IXXIII'
        \item \textbf{Coefficients:}\\
        -4.30565124e+00,  1.88633320e-01,  1.41067776e-02,
          1.41067776e-02,  1.80673167e-01, -9.94636052e-02,
         -2.34264366e-01,  1.88633320e-01,  1.41067776e-02,
          1.41067776e-02,  1.80673167e-01, -9.94636052e-02,
         -2.34264366e-01,  2.81035959e-01,  2.75633850e-01,
          2.75674220e-01, -5.62118770e-04, -5.62118770e-04,
          2.16385936e-01,  1.24159791e-02,  1.24159791e-02,
          1.23496801e-02, -1.23496801e-02, -1.23496801e-02,
          1.23496801e-02,  1.24159791e-02,  1.24159791e-02,
          1.11699437e-02,  1.11699437e-02,  1.11699437e-02,
          1.11699437e-02,  9.05524453e-03,  9.05524453e-03,
          9.05524453e-03,  9.05524453e-03,  1.93620176e-01,
          1.33686063e-02,  1.33686063e-02, -6.98664524e-04,
          6.98664524e-04,  6.98664524e-04, -6.98664524e-04,
          1.32105247e-01,  6.76451603e-03,  6.76451603e-03,
         -8.02974175e-03,  8.02974175e-03, -8.02974175e-03,
          8.02974175e-03,  1.60441829e-01,  5.47541027e-03,
          5.47541027e-03,  1.93620176e-01, -6.98664524e-04,
          6.98664524e-04,  6.98664524e-04, -6.98664524e-04,
          1.33686063e-02,  1.33686063e-02,  2.20039773e-01,
          1.07758169e-02, -1.07758169e-02, -1.07758169e-02,
          1.07758169e-02,  6.65804605e-03,  6.65804605e-03,
          6.65804605e-03,  6.65804605e-03,  1.35063452e-01,
          1.61356593e-01,  1.32105247e-01,  6.76451603e-03,
          6.76451603e-03, -8.02974175e-03, -8.02974175e-03,
          8.02974175e-03,  8.02974175e-03,  1.35063452e-01,
          1.34247149e-01,  3.70098970e-02, -3.70098970e-02,
         -3.70098970e-02,  3.70098970e-02,  1.36775360e-01,
          1.60441829e-01,  5.47541027e-03,  5.47541027e-03,
          1.61356593e-01,  1.36775360e-01,  1.60407978e-01,
          2.81035959e-01,  2.75633850e-01,  2.75674220e-01,
         -5.62118770e-04, -5.62118770e-04
   \end{itemize}
   \item[\#19]
   \begin{itemize}
       \item \textbf{Geometry:}\\
        O 0.0 0.0 0.0\\
        H 0.0 0.0 1.0\\
        H 0.0 0.3826834323650899 -0.9238795325112867
        \item \textbf{Pauli strings:}\\
        'IIIIII', 'IIIIIZ', 'IIIZXX', 'IIIIYY', 'IIIIZZ', 'IIIZZI', 'IIIZII', 'IIZIII', 'ZXXIII', 'IYYIII', 'IZZIII', 'ZZIIII', 'ZIIIII', 'IIIIZI', 'IIIZZZ', 'IIIZIZ', 'IIIZYY', 'IIIIXX', 'IIZIIZ', 'IIZZXX', 'IIZIYY', 'IZXIZX', 'IIXIZX', 'IZXIIX', 'IIXIIX', 'ZXXIIZ', 'IYYIIZ', 'ZXXZXX', 'IYYZXX', 'ZXXIYY', 'IYYIYY', 'XXXXXX', 'YXYXXX', 'XXXYXY', 'YXYYXY', 'IZZIIZ', 'IZZZXX', 'IZZIYY', 'ZXZIZX', 'IXIIZX', 'ZXZIIX', 'IXIIIX', 'ZZIIIZ', 'ZZIZXX', 'ZZIIYY', 'XZIXXX', 'XIIXXX', 'XZIYXY', 'XIIYXY', 'ZIIIIZ', 'ZIIZXX', 'ZIIIYY', 'IIZIZZ', 'IZXZXZ', 'IIXZXZ', 'IZXIXI', 'IIXIXI', 'ZXXIZZ', 'IYYIZZ', 'IZZIZZ', 'ZXZZXZ', 'IXIZXZ', 'ZXZIXI', 'IXIIXI', 'XXZXXZ', 'YYIXXZ', 'XXZYYI', 'YYIYYI', 'ZZIIZZ', 'ZIIIZZ', 'IIZZZI', 'ZXXZZI', 'IYYZZI', 'XXXXZI', 'YXYXZI', 'XXXXII', 'YXYXII', 'IZZZZI', 'ZZIZZI', 'XZIXZI', 'XIIXZI', 'XZIXII', 'XIIXII', 'ZIIZZI', 'IIZZII', 'ZXXZII', 'IYYZII', 'IZZZII', 'ZZIZII', 'ZIIZII', 'IZIIII', 'ZZZIII', 'ZIZIII', 'ZYYIII', 'IXXIII'
        \item \textbf{Coefficients:}\\
        -4.36323874e+00,  1.87219172e-01,  1.90726262e-02,
          1.90726262e-02,  1.80257439e-01, -7.89419562e-02,
         -2.02636420e-01,  1.87219172e-01,  1.90726262e-02,
          1.90726262e-02,  1.80257439e-01, -7.89419562e-02,
         -2.02636420e-01,  2.78714514e-01,  2.71964120e-01,
          2.72184935e-01, -1.17946032e-03, -1.17946032e-03,
          2.14123049e-01,  1.52461308e-02,  1.52461308e-02,
          1.22873102e-02, -1.22873102e-02, -1.22873102e-02,
          1.22873102e-02,  1.52461308e-02,  1.52461308e-02,
          1.14761113e-02,  1.14761113e-02,  1.14761113e-02,
          1.14761113e-02,  9.48707549e-03,  9.48707549e-03,
          9.48707549e-03,  9.48707549e-03,  1.92602344e-01,
          1.58495422e-02,  1.58495422e-02, -4.13232170e-04,
          4.13232170e-04,  4.13232170e-04, -4.13232170e-04,
          1.29509240e-01,  6.62790475e-03,  6.62790475e-03,
         -8.50871199e-03,  8.50871199e-03, -8.50871199e-03,
          8.50871199e-03,  1.59352050e-01,  6.57460209e-03,
          6.57460209e-03,  1.92602344e-01, -4.13232170e-04,
          4.13232170e-04,  4.13232170e-04, -4.13232170e-04,
          1.58495422e-02,  1.58495422e-02,  2.20039773e-01,
          1.01618242e-02, -1.01618242e-02, -1.01618242e-02,
          1.01618242e-02,  6.80958211e-03,  6.80958211e-03,
          6.80958211e-03,  6.80958211e-03,  1.32481785e-01,
          1.60740574e-01,  1.29509240e-01,  6.62790475e-03,
          6.62790475e-03, -8.50871199e-03, -8.50871199e-03,
          8.50871199e-03,  8.50871199e-03,  1.32481785e-01,
          1.33076897e-01,  3.65092358e-02, -3.65092358e-02,
         -3.65092358e-02,  3.65092358e-02,  1.34908717e-01,
          1.59352050e-01,  6.57460209e-03,  6.57460209e-03,
          1.60740574e-01,  1.34908717e-01,  1.59285988e-01,
          2.78714514e-01,  2.71964120e-01,  2.72184935e-01,
         -1.17946032e-03, -1.17946032e-03
   \end{itemize}
   
   \item[\#20]
   \begin{itemize}
       \item \textbf{Geometry:}\\
        O 0.0 0.0 0.0\\
        H 0.0 0.0 0.8\\
        H 0.0 9.797174393178826e-17 -0.8
        \item \textbf{Pauli strings:}\\
        'IIIIII', 'IIIIIZ', 'IIIIZZ', 'IIIZZI', 'IIIZII', 'IIZIII', 'IZZIII', 'ZZIIII', 'ZIIIII', 'IIIIZI', 'IIIZZZ', 'IIIZIZ', 'IIZIIZ', 'IZXIZX', 'IIXIZX', 'IZXIIX', 'IIXIIX', 'ZXXZXX', 'IYYZXX', 'ZXXIYY', 'IYYIYY', 'XXXXXX', 'YXYXXX', 'XXXYXY', 'YXYYXY', 'IZZIIZ', 'ZZIIIZ', 'ZIIIIZ', 'IIZIZZ', 'IZZIZZ', 'ZXZZXZ', 'IXIZXZ', 'ZXZIXI', 'IXIIXI', 'XXZXXZ', 'YYIXXZ', 'XXZYYI', 'YYIYYI', 'ZZIIZZ', 'ZIIIZZ', 'IIZZZI', 'IZZZZI', 'ZZIZZI', 'XZIXZI', 'XIIXZI', 'XZIXII', 'XIIXII', 'ZIIZZI', 'IIZZII', 'IZZZII', 'ZZIZII', 'ZIIZII', 'IZIIII', 'ZZZIII', 'ZIZIII'
        \item \textbf{Coefficients:}\\
        -4.15892301,  0.18389841,  0.18389841, -0.13360864,
         -0.31338097,  0.18389841,  0.18389841, -0.13360864,
         -0.31338097,  0.28496167,  0.28310775,  0.28310775,
          0.22003977,  0.01186111, -0.01186111, -0.01186111,
          0.01186111,  0.01227444,  0.01227444,  0.01227444,
          0.01227444,  0.00636644,  0.00636644,  0.00636644,
          0.00636644,  0.19631755,  0.13880676,  0.16294186,
          0.19631755,  0.22003977,  0.01227444, -0.01227444,
         -0.01227444,  0.01227444,  0.00636644,  0.00636644,
          0.00636644,  0.00636644,  0.13880676,  0.16294186,
          0.13880676,  0.13880676,  0.1327679 ,  0.03922261,
         -0.03922261, -0.03922261,  0.03922261,  0.13972784,
          0.16294186,  0.16294186,  0.13972784,  0.16302154,
          0.28496167,  0.28310775,  0.28310775
   \end{itemize}
   
   \item[\#21]
   \begin{itemize}
       \item \textbf{Geometry:}\\
        O 0.0 0.0 0.0\\
        H 0.0 0.0 0.8500000000000001\\
        H 0.0 1.0409497792752504e-16 -0.8500000000000001
        \item \textbf{Pauli strings:}\\
        'IIIIII', 'IIIIIZ', 'IIIIZZ', 'IIIZZI', 'IIIZII', 'IIZIII', 'IZZIII', 'ZZIIII', 'ZIIIII', 'IIIIZI', 'IIIZZZ', 'IIIZIZ', 'IIZIIZ', 'IZXIZX', 'IIXIZX', 'IZXIIX', 'IIXIIX', 'ZXXZXX', 'IYYZXX', 'ZXXIYY', 'IYYIYY', 'XXXXXX', 'YXYXXX', 'XXXYXY', 'YXYYXY', 'IZZIIZ', 'ZZIIIZ', 'ZIIIIZ', 'IIZIZZ', 'IZZIZZ', 'ZXZZXZ', 'IXIZXZ', 'ZXZIXI', 'IXIIXI', 'XXZXXZ', 'YYIXXZ', 'XXZYYI', 'YYIYYI', 'ZZIIZZ', 'ZIIIZZ', 'IIZZZI', 'IZZZZI', 'ZZIZZI', 'XZIXZI', 'XIIXZI', 'XZIXII', 'XIIXII', 'ZIIZZI', 'IIZZII', 'IZZZII', 'ZZIZII', 'ZIIZII', 'IZIIII', 'ZZZIII', 'ZIZIII'
        \item \textbf{Coefficients:}\\
        -4.15892301,  0.18389841,  0.18389841, -0.13360864,
         -0.31338097,  0.18389841,  0.18389841, -0.13360864,
         -0.31338097,  0.28496167,  0.28310775,  0.28310775,
          0.22003977,  0.01186111, -0.01186111, -0.01186111,
          0.01186111,  0.01227444,  0.01227444,  0.01227444,
          0.01227444,  0.00636644,  0.00636644,  0.00636644,
          0.00636644,  0.19631755,  0.13880676,  0.16294186,
          0.19631755,  0.22003977,  0.01227444, -0.01227444,
         -0.01227444,  0.01227444,  0.00636644,  0.00636644,
          0.00636644,  0.00636644,  0.13880676,  0.16294186,
          0.13880676,  0.13880676,  0.1327679 ,  0.03922261,
         -0.03922261, -0.03922261,  0.03922261,  0.13972784,
          0.16294186,  0.16294186,  0.13972784,  0.16302154,
          0.28496167,  0.28310775,  0.28310775
   \end{itemize}
   
   \item[\#22] 
   \begin{itemize}
       \item \textbf{Geometry:}\\
        O 0.0 0.0 0.0\\
        H 0.0 0.0 0.9\\
        H 0.0 1.1021821192326179e-16 -0.9
        \item \textbf{Pauli strings:}\\
        'IIIIII', 'IIIIIZ', 'IIIIZZ', 'IIIZZI', 'IIIZII', 'IIZIII', 'IZZIII', 'ZZIIII', 'ZIIIII', 'IIIIZI', 'IIIZZZ', 'IIIZIZ', 'IIZIIZ', 'IZXIZX', 'IIXIZX', 'IZXIIX', 'IIXIIX', 'ZXXZXX', 'IYYZXX', 'ZXXIYY', 'IYYIYY', 'XXXXXX', 'YXYXXX', 'XXXYXY', 'YXYYXY', 'IZZIIZ', 'ZZIIIZ', 'ZIIIIZ', 'IIZIZZ', 'IZZIZZ', 'ZXZZXZ', 'IXIZXZ', 'ZXZIXI', 'IXIIXI', 'XXZXXZ', 'YYIXXZ', 'XXZYYI', 'YYIYYI', 'ZZIIZZ', 'ZIIIZZ', 'IIZZZI', 'IZZZZI', 'ZZIZZI', 'XZIXZI', 'XIIXZI', 'XZIXII', 'XIIXII', 'ZIIZZI', 'IIZZII', 'IZZZII', 'ZZIZII', 'ZIIZII', 'IZIIII', 'ZZZIII', 'ZIZIII'
        \item \textbf{Coefficients:}\\
        -4.23874008,  0.18223838,  0.18223838, -0.10886628,
         -0.27642339,  0.18223838,  0.18223838, -0.10886628,
         -0.27642339,  0.2831215 ,  0.27982418,  0.27982418,
          0.22003977,  0.01186111, -0.01186111, -0.01186111,
          0.01186111,  0.01178326,  0.01178326,  0.01178326,
          0.01178326,  0.00654525,  0.00654525,  0.00654525,
          0.00654525,  0.19631755,  0.13574887,  0.16240381,
          0.19631755,  0.22003977,  0.01178326, -0.01178326,
         -0.01178326,  0.01178326,  0.00654525,  0.00654525,
          0.00654525,  0.00654525,  0.13574887,  0.16240381,
          0.13574887,  0.13574887,  0.13147267,  0.0392786 ,
         -0.0392786 , -0.0392786 ,  0.0392786 ,  0.13794366,
          0.16240381,  0.16240381,  0.13794366,  0.16214081,
          0.2831215 ,  0.27982418,  0.27982418
   \end{itemize}
   
   \item[\#23]
   \begin{itemize}
       \item \textbf{Geometry:}\\
        O 0.0 0.0 0.0\\
        H 0.0 0.0 0.95\\
        H 0.0 1.1634144591899854e-16 -0.95
        \item \textbf{Pauli strings:}\\
        'IIIIII', 'IIIIIZ', 'IIIIZZ', 'IIIZZI', 'IIIZII', 'IIZIII', 'IZZIII', 'ZZIIII', 'ZIIIII', 'IIIIZI', 'IIIZZZ', 'IIIZIZ', 'IIZIIZ', 'IZXIZX', 'IIXIZX', 'IZXIIX', 'IIXIIX', 'ZXXZXX', 'IYYZXX', 'ZXXIYY', 'IYYIYY', 'XXXXXX', 'YXYXXX', 'XXXYXY', 'YXYYXY', 'IZZIIZ', 'ZZIIIZ', 'ZIIIIZ', 'IIZIZZ', 'IZZIZZ', 'ZXZZXZ', 'IXIZXZ', 'ZXZIXI', 'IXIIXI', 'XXZXXZ', 'YYIXXZ', 'XXZYYI', 'YYIYYI', 'ZZIIZZ', 'ZIIIZZ', 'IIZZZI', 'IZZZZI', 'ZZIZZI', 'XZIXZI', 'XIIXZI', 'XZIXII', 'XIIXII', 'ZIIZZI', 'IIZZII', 'IZZZII', 'ZZIZII', 'ZIIZII', 'IZIIII', 'ZZZIII', 'ZIZIII'
        \item \textbf{Coefficients:}\\
        -4.31289927,  0.18169445,  0.18169445, -0.08651672,
         -0.24131155,  0.18169445,  0.18169445, -0.08651672,
         -0.24131155,  0.28122599,  0.27645092,  0.27645092,
          0.22003977,  0.01186111, -0.01186111, -0.01186111,
          0.01186111,  0.0112474 ,  0.0112474 ,  0.0112474 ,
          0.0112474 ,  0.00672415,  0.00672415,  0.00672415,
          0.00672415,  0.19631755,  0.13249362,  0.16192885,
          0.19631755,  0.22003977,  0.0112474 , -0.0112474 ,
         -0.0112474 ,  0.0112474 ,  0.00672415,  0.00672415,
          0.00672415,  0.00672415,  0.13249362,  0.16192885,
          0.13249362,  0.13249362,  0.13003905,  0.03914436,
         -0.03914436, -0.03914436,  0.03914436,  0.13591391,
          0.16192885,  0.16192885,  0.13591391,  0.16125096,
          0.28122599,  0.27645092,  0.27645092
   \end{itemize}
   \item[\#24] 
   \begin{itemize}
       \item \textbf{Geometry:}\\
        O 0.0 0.0 0.0\\
        H 0.0 0.0 1.0\\
        H 0.0 1.2246467991473532e-16 -1.0
        \item \textbf{Pauli strings:}\\
        'IIIIII', 'IIIIIZ', 'IIIIZZ', 'IIIZZI', 'IIIZII', 'IIZIII', 'IZZIII', 'ZZIIII', 'ZIIIII', 'IIIIZI', 'IIIZZZ', 'IIIZIZ', 'IIZIIZ', 'IZXIZX', 'IIXIZX', 'IZXIIX', 'IIXIIX', 'ZXXZXX', 'IYYZXX', 'ZXXIYY', 'IYYIYY', 'XXXXXX', 'YXYXXX', 'XXXYXY', 'YXYYXY', 'IZZIIZ', 'ZZIIIZ', 'ZIIIIZ', 'IIZIZZ', 'IZZIZZ', 'ZXZZXZ', 'IXIZXZ', 'ZXZIXI', 'IXIIXI', 'XXZXXZ', 'YYIXXZ', 'XXZYYI', 'YYIYYI', 'ZZIIZZ', 'ZIIIZZ', 'IIZZZI', 'IZZZZI', 'ZZIZZI', 'XZIXZI', 'XIIXZI', 'XZIXII', 'XIIXII', 'ZIIZZI', 'IIZZII', 'IZZZII', 'ZZIZII', 'ZIIZII', 'IZIIII', 'ZZZIII', 'ZIZIII'
        \item \textbf{Coefficients:}\\
        -4.38140985,  0.18207071,  0.18207071, -0.06664236,
         -0.20799744,  0.18207071,  0.18207071, -0.06664236,
         -0.20799744,  0.27930622,  0.27305561,  0.27305561,
          0.22003977,  0.01186111, -0.01186111, -0.01186111,
          0.01186111,  0.01067147,  0.01067147,  0.01067147,
          0.01067147,  0.00690341,  0.00690341,  0.00690341,
          0.00690341,  0.19631755,  0.12908167,  0.16154881,
          0.19631755,  0.22003977,  0.01067147, -0.01067147,
         -0.01067147,  0.01067147,  0.00690341,  0.00690341,
          0.00690341,  0.00690341,  0.12908167,  0.16154881,
          0.12908167,  0.12908167,  0.12853852,  0.03883979,
         -0.03883979, -0.03883979,  0.03883979,  0.13368957,
          0.16154881,  0.16154881,  0.13368957,  0.16041746,
          0.27930622,  0.27305561,  0.27305561
       \end{itemize}
\end{description}

\subsection*{$\mathrm{H_2O}$ asymmetric}

\begin{description}
   \item[\#0] 
       \begin{itemize}
            \item \textbf{Geometry:}\\
            O 0.0 0.0 0.0\\
            H 0.0 0.0 0.8\\
            H 0.0 0.7933318483619749 -0.10307559543641971
            \item \textbf{Pauli strings:}\\
            'IIIIII', 'IIIIIZ', 'IIIZXX', 'IIIIYY', 'IIIIZZ', 'IIIZZI', 'IIIZII', 'IIZIII', 'ZXXIII', 'IYYIII', 'IZZIII', 'ZZIIII', 'ZIIIII', 'IIIIZI', 'IIIZZZ', 'IIIZIZ', 'IIIZYY', 'IIIIXX', 'IIZIIZ', 'IIZZXX', 'IIZIYY', 'IZXIZX', 'IIXIZX', 'IZXIIX', 'IIXIIX', 'ZXXIIZ', 'IYYIIZ', 'ZXXZXX', 'IYYZXX', 'ZXXIYY', 'IYYIYY', 'XXXXXX', 'YXYXXX', 'XXXYXY', 'YXYYXY', 'IZZIIZ', 'IZZZXX', 'IZZIYY', 'ZXZIZX', 'IXIIZX', 'ZXZIIX', 'IXIIIX', 'ZZIIIZ', 'ZZIZXX', 'ZZIIYY', 'XZIXXX', 'XIIXXX', 'XZIYXY', 'XIIYXY', 'ZIIIIZ', 'ZIIZXX', 'ZIIIYY', 'IIZIZZ', 'IZXZXZ', 'IIXZXZ', 'IZXIXI', 'IIXIXI', 'ZXXIZZ', 'IYYIZZ', 'IZZIZZ', 'ZXZZXZ', 'IXIZXZ', 'ZXZIXI', 'IXIIXI', 'XXZXXZ', 'YYIXXZ', 'XXZYYI', 'YYIYYI', 'ZZIIZZ', 'ZIIIZZ', 'IIZZZI', 'ZXXZZI', 'IYYZZI', 'XXXXZI', 'YXYXZI', 'XXXXII', 'YXYXII', 'IZZZZI', 'ZZIZZI', 'XZIXZI', 'XIIXZI', 'XZIXII', 'XIIXII', 'ZIIZZI', 'IIZZII', 'ZXXZII', 'IYYZII', 'IZZZII', 'ZZIZII', 'ZIIZII', 'IZIIII', 'ZZZIII', 'ZIZIII', 'ZYYIII', 'IXXIII'
            \item \textbf{Coefficients:}\\
            -4.17069100e+00,  2.58470991e-01, -3.29604990e-02,
             -3.29604990e-02,  1.86463799e-01, -2.50181961e-01,
             -2.78364178e-01,  2.58470991e-01, -3.29604990e-02,
             -3.29604990e-02,  1.86463799e-01, -2.50181961e-01,
             -2.78364178e-01,  2.91466262e-01,  2.80681805e-01,
              2.80956764e-01,  2.46397724e-03,  2.46397724e-03,
              2.14262526e-01, -2.58062619e-02, -2.58062619e-02,
              1.60454450e-02, -1.60454450e-02, -1.60454450e-02,
              1.60454450e-02, -2.58062619e-02, -2.58062619e-02,
              1.16972978e-02,  1.16972978e-02,  1.16972978e-02,
              1.16972978e-02,  1.74440381e-02,  1.74440381e-02,
              1.74440381e-02,  1.74440381e-02,  1.89466864e-01,
             -2.10099414e-02, -2.10099414e-02,  2.43807446e-03,
             -2.43807446e-03, -2.43807446e-03,  2.43807446e-03,
              1.38471816e-01, -8.96551343e-03, -8.96551343e-03,
              1.36298705e-02, -1.36298705e-02,  1.36298705e-02,
             -1.36298705e-02,  1.58186447e-01, -7.35416806e-03,
             -7.35416806e-03,  1.89466864e-01,  2.43807446e-03,
             -2.43807446e-03, -2.43807446e-03,  2.43807446e-03,
             -2.10099414e-02, -2.10099414e-02,  2.20039773e-01,
              1.06933227e-02, -1.06933227e-02, -1.06933227e-02,
              1.06933227e-02,  5.66748072e-03,  5.66748072e-03,
              5.66748072e-03,  5.66748072e-03,  1.50907677e-01,
              1.59574767e-01,  1.38471816e-01, -8.96551343e-03,
             -8.96551343e-03,  1.36298705e-02,  1.36298705e-02,
             -1.36298705e-02, -1.36298705e-02,  1.50907677e-01,
              1.52052590e-01,  2.76401498e-02, -2.76401498e-02,
             -2.76401498e-02,  2.76401498e-02,  1.45684992e-01,
              1.58186447e-01, -7.35416806e-03, -7.35416806e-03,
              1.59574767e-01,  1.45684992e-01,  1.60764020e-01,
              2.91466262e-01,  2.80681805e-01,  2.80956764e-01,
              2.46397724e-03,  2.46397724e-03
       \end{itemize} 
   \item[\#1] 
       \begin{itemize}
            \item \textbf{Geometry:}\\
            O 0.0 0.0 0.0\\
            H 0.0 0.0 0.8\\
            H 0.0 0.8924983294072217 -0.11596004486597218
            \item \textbf{Pauli strings:}\\
            'IIIIII', 'IIIIIZ', 'IIIZXX', 'IIIIYY', 'IIIXXX', 'IIIYXY', 'IIIIZZ', 'IIIZZI', 'IIIXZI', 'IIIXII', 'IIIZII', 'IIZIII', 'ZXXIII', 'IYYIII', 'XXXIII', 'YXYIII', 'IZZIII', 'ZZIIII', 'XZIIII', 'XIIIII', 'ZIIIII', 'IIIIZI', 'IIIZZZ', 'IIIXZZ', 'IIIXIZ', 'IIIZIZ', 'IIIZYY', 'IIIIXX', 'IIIXYY', 'IIIYYX', 'IIZIIZ', 'IIZZXX', 'IIZIYY', 'IIZXXX', 'IIZYXY', 'IZXIZX', 'IIXIZX', 'IZXIIX', 'IIXIIX', 'ZXXIIZ', 'IYYIIZ', 'ZXXZXX', 'IYYZXX', 'ZXXIYY', 'IYYIYY', 'ZXXXXX', 'IYYXXX', 'ZXXYXY', 'IYYYXY', 'XXXIIZ', 'YXYIIZ', 'XXXZXX', 'YXYZXX', 'XXXIYY', 'YXYIYY', 'XXXXXX', 'YXYXXX', 'XXXYXY', 'YXYYXY', 'IZZIIZ', 'IZZZXX', 'IZZIYY', 'IZZXXX', 'IZZYXY', 'ZXZIZX', 'IXIIZX', 'ZXZIIX', 'IXIIIX', 'XXZIZX', 'YYIIZX', 'XXZIIX', 'YYIIIX', 'ZZIIIZ', 'ZZIZXX', 'ZZIIYY', 'ZZIXXX', 'ZZIYXY', 'XZIIIZ', 'XIIIIZ', 'XZIZXX', 'XIIZXX', 'XZIIYY', 'XIIIYY', 'XZIXXX', 'XIIXXX', 'XZIYXY', 'XIIYXY', 'ZIIIIZ', 'ZIIZXX', 'ZIIIYY', 'ZIIXXX', 'ZIIYXY', 'IIZIZZ', 'IZXZXZ', 'IIXZXZ', 'IZXIXI', 'IIXIXI', 'IZXXXZ', 'IIXXXZ', 'IZXYYI', 'IIXYYI', 'ZXXIZZ', 'IYYIZZ', 'XXXIZZ', 'YXYIZZ', 'IZZIZZ', 'ZXZZXZ', 'IXIZXZ', 'ZXZIXI', 'IXIIXI', 'ZXZXXZ', 'IXIXXZ', 'ZXZYYI', 'IXIYYI', 'XXZZXZ', 'YYIZXZ', 'XXZIXI', 'YYIIXI', 'XXZXXZ', 'YYIXXZ', 'XXZYYI', 'YYIYYI', 'ZZIIZZ', 'XZIIZZ', 'XIIIZZ', 'ZIIIZZ', 'IIZZZI', 'IIZXZI', 'IIZXII', 'ZXXZZI', 'IYYZZI', 'ZXXXZI', 'IYYXZI', 'ZXXXII', 'IYYXII', 'XXXZZI', 'YXYZZI', 'XXXXZI', 'YXYXZI', 'XXXXII', 'YXYXII', 'IZZZZI', 'IZZXZI', 'IZZXII', 'ZZIZZI', 'ZZIXZI', 'ZZIXII', 'XZIZZI', 'XIIZZI', 'XZIXZI', 'XIIXZI', 'XZIXII', 'XIIXII', 'ZIIZZI', 'ZIIXZI', 'ZIIXII', 'IIZZII', 'ZXXZII', 'IYYZII', 'XXXZII', 'YXYZII', 'IZZZII', 'ZZIZII', 'XZIZII', 'XIIZII', 'ZIIZII', 'IZIIII', 'ZZZIII', 'XZZIII', 'XIZIII', 'ZIZIII', 'ZYYIII', 'IXXIII', 'XYYIII', 'YYXIII'
            \item \textbf{Coefficients:}\\
            -4.21500520e+00,  2.50023601e-01, -3.63955100e-02,
             -3.63955100e-02, -9.47845413e-03, -9.47845413e-03,
              1.81717596e-01, -2.06479067e-01, -1.35790214e-02,
              1.35790214e-02, -2.64851448e-01,  2.50023601e-01,
             -3.63955100e-02, -3.63955100e-02, -9.47845413e-03,
             -9.47845413e-03,  1.81717596e-01, -2.06479067e-01,
             -1.35790214e-02,  1.35790214e-02, -2.64851448e-01,
              2.88315753e-01,  2.77156624e-01, -1.65017674e-04,
              1.65017674e-04,  2.78145615e-01,  3.50816078e-03,
              3.50816078e-03,  5.27772292e-04, -5.27772292e-04,
              2.07393885e-01, -2.70383804e-02, -2.70383804e-02,
             -7.51433912e-03, -7.51433912e-03,  1.54550107e-02,
             -1.54550107e-02, -1.54550107e-02,  1.54550107e-02,
             -2.70383804e-02, -2.70383804e-02,  1.54175644e-02,
              1.54175644e-02,  1.54175644e-02,  1.54175644e-02,
             -2.55292385e-03, -2.55292385e-03, -2.55292385e-03,
             -2.55292385e-03, -7.51433912e-03, -7.51433912e-03,
             -2.55292385e-03, -2.55292385e-03, -2.55292385e-03,
             -2.55292385e-03,  1.59923180e-02,  1.59923180e-02,
              1.59923180e-02,  1.59923180e-02,  1.86854972e-01,
             -2.34028215e-02, -2.34028215e-02, -5.98088978e-03,
             -5.98088978e-03,  1.42160978e-03, -1.42160978e-03,
             -1.42160978e-03,  1.42160978e-03,  1.21388373e-03,
              1.21388373e-03, -1.21388373e-03, -1.21388373e-03,
              1.41298210e-01, -4.75338539e-03, -4.75338539e-03,
             -1.15608505e-02, -1.15608505e-02, -6.23420261e-03,
              6.23420261e-03, -4.89384932e-03,  4.89384932e-03,
             -4.89384932e-03,  4.89384932e-03,  8.51581559e-03,
             -8.51581559e-03,  8.51581559e-03, -8.51581559e-03,
              1.53107606e-01, -1.28004550e-02, -1.28004550e-02,
              7.01629715e-03,  7.01629715e-03,  1.86854972e-01,
              1.42160978e-03, -1.42160978e-03, -1.42160978e-03,
              1.42160978e-03,  1.21388373e-03, -1.21388373e-03,
              1.21388373e-03, -1.21388373e-03, -2.34028215e-02,
             -2.34028215e-02, -5.98088978e-03, -5.98088978e-03,
              2.20039773e-01,  9.58115816e-03, -9.58115816e-03,
             -9.58115816e-03,  9.58115816e-03,  1.72258815e-03,
             -1.72258815e-03,  1.72258815e-03, -1.72258815e-03,
              1.72258815e-03,  1.72258815e-03, -1.72258815e-03,
             -1.72258815e-03,  6.53785307e-03,  6.53785307e-03,
              6.53785307e-03,  6.53785307e-03,  1.50611485e-01,
             -1.79367293e-03,  1.79367293e-03,  1.57813831e-01,
              1.41298210e-01, -6.23420261e-03,  6.23420261e-03,
             -4.75338539e-03, -4.75338539e-03, -4.89384932e-03,
             -4.89384932e-03,  4.89384932e-03,  4.89384932e-03,
             -1.15608505e-02, -1.15608505e-02,  8.51581559e-03,
              8.51581559e-03, -8.51581559e-03, -8.51581559e-03,
              1.50611485e-01, -1.79367293e-03,  1.79367293e-03,
              1.63312900e-01, -1.20827009e-02,  1.20827009e-02,
             -1.20827009e-02,  1.20827009e-02,  1.69744520e-02,
             -1.69744520e-02, -1.69744520e-02,  1.69744520e-02,
              1.33890244e-01,  1.04363082e-02, -1.04363082e-02,
              1.53107606e-01, -1.28004550e-02, -1.28004550e-02,
              7.01629715e-03,  7.01629715e-03,  1.57813831e-01,
              1.33890244e-01,  1.04363082e-02, -1.04363082e-02,
              1.69048590e-01,  2.88315753e-01,  2.77156624e-01,
             -1.65017674e-04,  1.65017674e-04,  2.78145615e-01,
              3.50816078e-03,  3.50816078e-03,  5.27772292e-04,
             -5.27772292e-04
       \end{itemize}
    \item[\#2] 
       \begin{itemize}
            \item \textbf{Geometry:}\\
            O 0.0 0.0 0.0\\
            H 0.0 0.0 0.8\\
            H 0.0 0.9916648104524686 -0.12884449429552464
            \item \textbf{Pauli strings:}\\
            'IIIIII', 'IIIIIZ', 'IIIZXX', 'IIIIYY', 'IIIXXX', 'IIIYXY', 'IIIIZZ', 'IIIZZI', 'IIIXZI', 'IIIXII', 'IIIZII', 'IIZIII', 'ZXXIII', 'IYYIII', 'XXXIII', 'YXYIII', 'IZZIII', 'ZZIIII', 'XZIIII', 'XIIIII', 'ZIIIII', 'IIIIZI', 'IIIZZZ', 'IIIXZZ', 'IIIXIZ', 'IIIZIZ', 'IIIZYY', 'IIIIXX', 'IIIXYY', 'IIIYYX', 'IIZIIZ', 'IIZZXX', 'IIZIYY', 'IIZXXX', 'IIZYXY', 'IZXIZX', 'IIXIZX', 'IZXIIX', 'IIXIIX', 'ZXXIIZ', 'IYYIIZ', 'ZXXZXX', 'IYYZXX', 'ZXXIYY', 'IYYIYY', 'ZXXXXX', 'IYYXXX', 'ZXXYXY', 'IYYYXY', 'XXXIIZ', 'YXYIIZ', 'XXXZXX', 'YXYZXX', 'XXXIYY', 'YXYIYY', 'XXXXXX', 'YXYXXX', 'XXXYXY', 'YXYYXY', 'IZZIIZ', 'IZZZXX', 'IZZIYY', 'IZZXXX', 'IZZYXY', 'ZXZIZX', 'IXIIZX', 'ZXZIIX', 'IXIIIX', 'XXZIZX', 'YYIIZX', 'XXZIIX', 'YYIIIX', 'ZZIIIZ', 'ZZIZXX', 'ZZIIYY', 'ZZIXXX', 'ZZIYXY', 'XZIIIZ', 'XIIIIZ', 'XZIZXX', 'XIIZXX', 'XZIIYY', 'XIIIYY', 'XZIXXX', 'XIIXXX', 'XZIYXY', 'XIIYXY', 'ZIIIIZ', 'ZIIZXX', 'ZIIIYY', 'ZIIXXX', 'ZIIYXY', 'IIZIZZ', 'IZXZXZ', 'IIXZXZ', 'IZXIXI', 'IIXIXI', 'IZXXXZ', 'IIXXXZ', 'IZXYYI', 'IIXYYI', 'ZXXIZZ', 'IYYIZZ', 'XXXIZZ', 'YXYIZZ', 'IZZIZZ', 'ZXZZXZ', 'IXIZXZ', 'ZXZIXI', 'IXIIXI', 'ZXZXXZ', 'IXIXXZ', 'ZXZYYI', 'IXIYYI', 'XXZZXZ', 'YYIZXZ', 'XXZIXI', 'YYIIXI', 'XXZXXZ', 'YYIXXZ', 'XXZYYI', 'YYIYYI', 'ZZIIZZ', 'XZIIZZ', 'XIIIZZ', 'ZIIIZZ', 'IIZZZI', 'IIZXZI', 'IIZXII', 'ZXXZZI', 'IYYZZI', 'ZXXXZI', 'IYYXZI', 'ZXXXII', 'IYYXII', 'XXXZZI', 'YXYZZI', 'XXXXZI', 'YXYXZI', 'XXXXII', 'YXYXII', 'IZZZZI', 'IZZXZI', 'IZZXII', 'ZZIZZI', 'ZZIXZI', 'ZZIXII', 'XZIZZI', 'XIIZZI', 'XZIXZI', 'XIIXZI', 'XZIXII', 'XIIXII', 'ZIIZZI', 'ZIIXZI', 'ZIIXII', 'IIZZII', 'ZXXZII', 'IYYZII', 'XXXZII', 'YXYZII', 'IZZZII', 'ZZIZII', 'XZIZII', 'XIIZII', 'ZIIZII', 'IZIIII', 'ZZZIII', 'XZZIII', 'XIZIII', 'ZIZIII', 'ZYYIII', 'IXXIII', 'XYYIII', 'YYXIII'
            \item \textbf{Coefficients:}\\
            -4.22840830e+00,  2.39652977e-01, -4.02927172e-02,
             -4.02927172e-02, -1.19441127e-02, -1.19441127e-02,
              1.77421588e-01, -1.55806329e-01, -1.62708828e-02,
              1.62708828e-02, -2.65693013e-01,  2.39652977e-01,
             -4.02927172e-02, -4.02927172e-02, -1.19441127e-02,
             -1.19441127e-02,  1.77421588e-01, -1.55806329e-01,
             -1.62708828e-02,  1.62708828e-02, -2.65693013e-01,
              2.84108271e-01,  2.72751167e-01, -3.32376722e-04,
              3.32376722e-04,  2.74552276e-01,  4.87971987e-03,
              4.87971987e-03,  6.05627952e-04, -6.05627952e-04,
              1.98594103e-01, -2.82817395e-02, -2.82817395e-02,
             -9.29508383e-03, -9.29508383e-03,  1.45967128e-02,
             -1.45967128e-02, -1.45967128e-02,  1.45967128e-02,
             -2.82817395e-02, -2.82817395e-02,  2.05293889e-02,
              2.05293889e-02,  2.05293889e-02,  2.05293889e-02,
             -2.93344630e-03, -2.93344630e-03, -2.93344630e-03,
             -2.93344630e-03, -9.29508383e-03, -9.29508383e-03,
             -2.93344630e-03, -2.93344630e-03, -2.93344630e-03,
             -2.93344630e-03,  1.38630231e-02,  1.38630231e-02,
              1.38630231e-02,  1.38630231e-02,  1.83146645e-01,
             -2.70507266e-02, -2.70507266e-02, -7.56292184e-03,
             -7.56292184e-03,  2.16220887e-04, -2.16220887e-04,
             -2.16220887e-04,  2.16220887e-04,  1.65130758e-03,
              1.65130758e-03, -1.65130758e-03, -1.65130758e-03,
              1.43791781e-01, -1.22662681e-03, -1.22662681e-03,
             -1.31064614e-02, -1.31064614e-02, -7.20572735e-03,
              7.20572735e-03, -4.49785993e-03,  4.49785993e-03,
             -4.49785993e-03,  4.49785993e-03,  3.69437239e-03,
             -3.69437239e-03,  3.69437239e-03, -3.69437239e-03,
              1.47717196e-01, -1.86928552e-02, -1.86928552e-02,
              7.58690488e-03,  7.58690488e-03,  1.83146645e-01,
              2.16220887e-04, -2.16220887e-04, -2.16220887e-04,
              2.16220887e-04,  1.65130758e-03, -1.65130758e-03,
              1.65130758e-03, -1.65130758e-03, -2.70507266e-02,
             -2.70507266e-02, -7.56292184e-03, -7.56292184e-03,
              2.20039773e-01,  8.57075998e-03, -8.57075998e-03,
             -8.57075998e-03,  8.57075998e-03,  2.00014750e-03,
             -2.00014750e-03,  2.00014750e-03, -2.00014750e-03,
              2.00014750e-03,  2.00014750e-03, -2.00014750e-03,
             -2.00014750e-03,  7.31342466e-03,  7.31342466e-03,
              7.31342466e-03,  7.31342466e-03,  1.49268864e-01,
             -1.93975683e-03,  1.93975683e-03,  1.56802199e-01,
              1.43791781e-01, -7.20572735e-03,  7.20572735e-03,
             -1.22662681e-03, -1.22662681e-03, -4.49785993e-03,
             -4.49785993e-03,  4.49785993e-03,  4.49785993e-03,
             -1.31064614e-02, -1.31064614e-02,  3.69437239e-03,
              3.69437239e-03, -3.69437239e-03, -3.69437239e-03,
              1.49268864e-01, -1.93975683e-03,  1.93975683e-03,
              1.68649805e-01, -9.67841791e-03,  9.67841791e-03,
             -9.67841791e-03,  9.67841791e-03,  9.86819933e-03,
             -9.86819933e-03, -9.86819933e-03,  9.86819933e-03,
              1.25426538e-01,  8.59163160e-03, -8.59163160e-03,
              1.47717196e-01, -1.86928552e-02, -1.86928552e-02,
              7.58690488e-03,  7.58690488e-03,  1.56802199e-01,
              1.25426538e-01,  8.59163160e-03, -8.59163160e-03,
              1.74844295e-01,  2.84108271e-01,  2.72751167e-01,
             -3.32376722e-04,  3.32376722e-04,  2.74552276e-01,
              4.87971987e-03,  4.87971987e-03,  6.05627952e-04,
             -6.05627952e-04
       \end{itemize} 
   \item[\#3] 
       \begin{itemize}
            \item \textbf{Geometry:}\\
            O 0.0 0.0 0.0\\
            H 0.0 0.0 0.9\\
            H 0.0 0.7933318483619749 -0.10307559543641971            
            \item \textbf{Pauli strings:}\\
            'IIIIII', 'IIIIIZ', 'IIIZXX', 'IIIIYY', 'IIIXXX', 'IIIYXY', 'IIIIZZ', 'IIIZZI', 'IIIXZI', 'IIIXII', 'IIIZII', 'IIZIII', 'ZXXIII', 'IYYIII', 'XXXIII', 'YXYIII', 'IZZIII', 'ZZIIII', 'XZIIII', 'XIIIII', 'ZIIIII', 'IIIIZI', 'IIIZZZ', 'IIIXZZ', 'IIIXIZ', 'IIIZIZ', 'IIIZYY', 'IIIIXX', 'IIIXYY', 'IIIYYX', 'IIZIIZ', 'IIZZXX', 'IIZIYY', 'IIZXXX', 'IIZYXY', 'IZXIZX', 'IIXIZX', 'IZXIIX', 'IIXIIX', 'ZXXIIZ', 'IYYIIZ', 'ZXXZXX', 'IYYZXX', 'ZXXIYY', 'IYYIYY', 'ZXXXXX', 'IYYXXX', 'ZXXYXY', 'IYYYXY', 'XXXIIZ', 'YXYIIZ', 'XXXZXX', 'YXYZXX', 'XXXIYY', 'YXYIYY', 'XXXXXX', 'YXYXXX', 'XXXYXY', 'YXYYXY', 'IZZIIZ', 'IZZZXX', 'IZZIYY', 'IZZXXX', 'IZZYXY', 'ZXZIZX', 'IXIIZX', 'ZXZIIX', 'IXIIIX', 'XXZIZX', 'YYIIZX', 'XXZIIX', 'YYIIIX', 'ZZIIIZ', 'ZZIZXX', 'ZZIIYY', 'ZZIXXX', 'ZZIYXY', 'XZIIIZ', 'XIIIIZ', 'XZIZXX', 'XIIZXX', 'XZIIYY', 'XIIIYY', 'XZIXXX', 'XIIXXX', 'XZIYXY', 'XIIYXY', 'ZIIIIZ', 'ZIIZXX', 'ZIIIYY', 'ZIIXXX', 'ZIIYXY', 'IIZIZZ', 'IZXZXZ', 'IIXZXZ', 'IZXIXI', 'IIXIXI', 'IZXXXZ', 'IIXXXZ', 'IZXYYI', 'IIXYYI', 'ZXXIZZ', 'IYYIZZ', 'XXXIZZ', 'YXYIZZ', 'IZZIZZ', 'ZXZZXZ', 'IXIZXZ', 'ZXZIXI', 'IXIIXI', 'ZXZXXZ', 'IXIXXZ', 'ZXZYYI', 'IXIYYI', 'XXZZXZ', 'YYIZXZ', 'XXZIXI', 'YYIIXI', 'XXZXXZ', 'YYIXXZ', 'XXZYYI', 'YYIYYI', 'ZZIIZZ', 'XZIIZZ', 'XIIIZZ', 'ZIIIZZ', 'IIZZZI', 'IIZXZI', 'IIZXII', 'ZXXZZI', 'IYYZZI', 'ZXXXZI', 'IYYXZI', 'ZXXXII', 'IYYXII', 'XXXZZI', 'YXYZZI', 'XXXXZI', 'YXYXZI', 'XXXXII', 'YXYXII', 'IZZZZI', 'IZZXZI', 'IZZXII', 'ZZIZZI', 'ZZIXZI', 'ZZIXII', 'XZIZZI', 'XIIZZI', 'XZIXZI', 'XIIXZI', 'XZIXII', 'XIIXII', 'ZIIZZI', 'ZIIXZI', 'ZIIXII', 'IIZZII', 'ZXXZII', 'IYYZII', 'XXXZII', 'YXYZII', 'IZZZII', 'ZZIZII', 'XZIZII', 'XIIZII', 'ZIIZII', 'IZIIII', 'ZZZIII', 'XZZIII', 'XIZIII', 'ZIZIII', 'ZYYIII', 'IXXIII', 'XYYIII', 'YYXIII'
            \item \textbf{Coefficients:}\\
            -4.21500520e+00,  2.50023601e-01,  3.63955100e-02,
              3.63955100e-02,  9.47845413e-03,  9.47845413e-03,
              1.81717596e-01, -2.06479067e-01, -1.35790214e-02,
              1.35790214e-02, -2.64851448e-01,  2.50023601e-01,
              3.63955100e-02,  3.63955100e-02,  9.47845413e-03,
              9.47845413e-03,  1.81717596e-01, -2.06479067e-01,
             -1.35790214e-02,  1.35790214e-02, -2.64851448e-01,
              2.88315753e-01,  2.77156624e-01, -1.65017674e-04,
              1.65017674e-04,  2.78145615e-01, -3.50816078e-03,
             -3.50816078e-03, -5.27772292e-04,  5.27772292e-04,
              2.07393885e-01,  2.70383804e-02,  2.70383804e-02,
              7.51433912e-03,  7.51433912e-03,  1.54550107e-02,
             -1.54550107e-02, -1.54550107e-02,  1.54550107e-02,
              2.70383804e-02,  2.70383804e-02,  1.54175644e-02,
              1.54175644e-02,  1.54175644e-02,  1.54175644e-02,
             -2.55292385e-03, -2.55292385e-03, -2.55292385e-03,
             -2.55292385e-03,  7.51433912e-03,  7.51433912e-03,
             -2.55292385e-03, -2.55292385e-03, -2.55292385e-03,
             -2.55292385e-03,  1.59923180e-02,  1.59923180e-02,
              1.59923180e-02,  1.59923180e-02,  1.86854972e-01,
              2.34028215e-02,  2.34028215e-02,  5.98088978e-03,
              5.98088978e-03, -1.42160978e-03,  1.42160978e-03,
              1.42160978e-03, -1.42160978e-03, -1.21388373e-03,
             -1.21388373e-03,  1.21388373e-03,  1.21388373e-03,
              1.41298210e-01,  4.75338539e-03,  4.75338539e-03,
              1.15608505e-02,  1.15608505e-02, -6.23420261e-03,
              6.23420261e-03,  4.89384932e-03, -4.89384932e-03,
              4.89384932e-03, -4.89384932e-03, -8.51581559e-03,
              8.51581559e-03, -8.51581559e-03,  8.51581559e-03,
              1.53107606e-01,  1.28004550e-02,  1.28004550e-02,
             -7.01629715e-03, -7.01629715e-03,  1.86854972e-01,
             -1.42160978e-03,  1.42160978e-03,  1.42160978e-03,
             -1.42160978e-03, -1.21388373e-03,  1.21388373e-03,
             -1.21388373e-03,  1.21388373e-03,  2.34028215e-02,
              2.34028215e-02,  5.98088978e-03,  5.98088978e-03,
              2.20039773e-01,  9.58115816e-03, -9.58115816e-03,
             -9.58115816e-03,  9.58115816e-03,  1.72258815e-03,
             -1.72258815e-03,  1.72258815e-03, -1.72258815e-03,
              1.72258815e-03,  1.72258815e-03, -1.72258815e-03,
             -1.72258815e-03,  6.53785307e-03,  6.53785307e-03,
              6.53785307e-03,  6.53785307e-03,  1.50611485e-01,
             -1.79367293e-03,  1.79367293e-03,  1.57813831e-01,
              1.41298210e-01, -6.23420261e-03,  6.23420261e-03,
              4.75338539e-03,  4.75338539e-03,  4.89384932e-03,
              4.89384932e-03, -4.89384932e-03, -4.89384932e-03,
              1.15608505e-02,  1.15608505e-02, -8.51581559e-03,
             -8.51581559e-03,  8.51581559e-03,  8.51581559e-03,
              1.50611485e-01, -1.79367293e-03,  1.79367293e-03,
              1.63312900e-01, -1.20827009e-02,  1.20827009e-02,
             -1.20827009e-02,  1.20827009e-02,  1.69744520e-02,
             -1.69744520e-02, -1.69744520e-02,  1.69744520e-02,
              1.33890244e-01,  1.04363082e-02, -1.04363082e-02,
              1.53107606e-01,  1.28004550e-02,  1.28004550e-02,
             -7.01629715e-03, -7.01629715e-03,  1.57813831e-01,
              1.33890244e-01,  1.04363082e-02, -1.04363082e-02,
              1.69048590e-01,  2.88315753e-01,  2.77156624e-01,
             -1.65017674e-04,  1.65017674e-04,  2.78145615e-01,
             -3.50816078e-03, -3.50816078e-03, -5.27772292e-04,
              5.27772292e-04
       \end{itemize}
   \item[\#4] 
       \begin{itemize}
            \item \textbf{Geometry:}\\
            O 0.0 0.0 0.0\\
            H 0.0 0.0 0.9\\
            H 0.0 0.8924983294072217 -0.11596004486597218
            \item \textbf{Pauli strings:}\\
            'IIIIII', 'IIIIIZ', 'IIIZXX', 'IIIIYY', 'IIIIZZ', 'IIIZZI', 'IIIZII', 'IIZIII', 'ZXXIII', 'IYYIII', 'IZZIII', 'ZZIIII', 'ZIIIII', 'IIIIZI', 'IIIZZZ', 'IIIZIZ', 'IIIZYY', 'IIIIXX', 'IIZIIZ', 'IIZZXX', 'IIZIYY', 'IZXIZX', 'IIXIZX', 'IZXIIX', 'IIXIIX', 'ZXXIIZ', 'IYYIIZ', 'ZXXZXX', 'IYYZXX', 'ZXXIYY', 'IYYIYY', 'XXXXXX', 'YXYXXX', 'XXXYXY', 'YXYYXY', 'IZZIIZ', 'IZZZXX', 'IZZIYY', 'ZXZIZX', 'IXIIZX', 'ZXZIIX', 'IXIIIX', 'ZZIIIZ', 'ZZIZXX', 'ZZIIYY', 'XZIXXX', 'XIIXXX', 'XZIYXY', 'XIIYXY', 'ZIIIIZ', 'ZIIZXX', 'ZIIIYY', 'IIZIZZ', 'IZXZXZ', 'IIXZXZ', 'IZXIXI', 'IIXIXI', 'ZXXIZZ', 'IYYIZZ', 'IZZIZZ', 'ZXZZXZ', 'IXIZXZ', 'ZXZIXI', 'IXIIXI', 'XXZXXZ', 'YYIXXZ', 'XXZYYI', 'YYIYYI', 'ZZIIZZ', 'ZIIIZZ', 'IIZZZI', 'ZXXZZI', 'IYYZZI', 'XXXXZI', 'YXYXZI', 'XXXXII', 'YXYXII', 'IZZZZI', 'ZZIZZI', 'XZIXZI', 'XIIXZI', 'XZIXII', 'XIIXII', 'ZIIZZI', 'IIZZII', 'ZXXZII', 'IYYZII', 'IZZZII', 'ZZIZII', 'ZIIZII', 'IZIIII', 'ZZZIII', 'ZIZIII', 'ZYYIII', 'IXXIII'
            \item \textbf{Coefficients:}\\
            -4.26532294e+00,  2.42533983e-01, -4.15925279e-02,
             -4.15925279e-02,  1.77344296e-01, -1.97586889e-01,
             -2.16224170e-01,  2.42533983e-01, -4.15925279e-02,
             -4.15925279e-02,  1.77344296e-01, -1.97586889e-01,
             -2.16224170e-01,  2.85358783e-01,  2.73753754e-01,
              2.75585636e-01,  4.47846624e-03,  4.47846624e-03,
              2.01201256e-01, -2.98353089e-02, -2.98353089e-02,
              1.49309938e-02, -1.49309938e-02, -1.49309938e-02,
              1.49309938e-02, -2.98353089e-02, -2.98353089e-02,
              1.58240088e-02,  1.58240088e-02,  1.58240088e-02,
              1.58240088e-02,  1.77387433e-02,  1.77387433e-02,
              1.77387433e-02,  1.77387433e-02,  1.84405523e-01,
             -2.67611432e-02, -2.67611432e-02,  1.13247736e-03,
             -1.13247736e-03, -1.13247736e-03,  1.13247736e-03,
              1.38744738e-01, -1.04219863e-02, -1.04219863e-02,
              1.43546335e-02, -1.43546335e-02,  1.43546335e-02,
             -1.43546335e-02,  1.53539151e-01, -9.06052076e-03,
             -9.06052076e-03,  1.84405523e-01,  1.13247736e-03,
             -1.13247736e-03, -1.13247736e-03,  1.13247736e-03,
             -2.67611432e-02, -2.67611432e-02,  2.20039773e-01,
              9.99783947e-03, -9.99783947e-03, -9.99783947e-03,
              9.99783947e-03,  5.84948740e-03,  5.84948740e-03,
              5.84948740e-03,  5.84948740e-03,  1.49783068e-01,
              1.56682512e-01,  1.38744738e-01, -1.04219863e-02,
             -1.04219863e-02,  1.43546335e-02,  1.43546335e-02,
             -1.43546335e-02, -1.43546335e-02,  1.49783068e-01,
              1.51973314e-01,  2.79694998e-02, -2.79694998e-02,
             -2.79694998e-02,  2.79694998e-02,  1.43853754e-01,
              1.53539151e-01, -9.06052076e-03, -9.06052076e-03,
              1.56682512e-01,  1.43853754e-01,  1.56560315e-01,
              2.85358783e-01,  2.73753754e-01,  2.75585636e-01,
              4.47846624e-03,  4.47846624e-03
       \end{itemize} 
   \item[\#5] 
       \begin{itemize}
            \item \textbf{Geometry:}\\
            O 0.0 0.0 0.0\\
            H 0.0 0.0 0.9\\
            H 0.0 0.9916648104524686 -0.12884449429552464
            \item \textbf{Pauli strings:}\\
            'IIIIII', 'IIIIIZ', 'IIIZXX', 'IIIIYY', 'IIIXXX', 'IIIYXY', 'IIIIZZ', 'IIIZZI', 'IIIXZI', 'IIIXII', 'IIIZII', 'IIZIII', 'ZXXIII', 'IYYIII', 'XXXIII', 'YXYIII', 'IZZIII', 'ZZIIII', 'XZIIII', 'XIIIII', 'ZIIIII', 'IIIIZI', 'IIIZZZ', 'IIIXZZ', 'IIIXIZ', 'IIIZIZ', 'IIIZYY', 'IIIIXX', 'IIIXYY', 'IIIYYX', 'IIZIIZ', 'IIZZXX', 'IIZIYY', 'IIZXXX', 'IIZYXY', 'IZXIZX', 'IIXIZX', 'IZXIIX', 'IIXIIX', 'ZXXIIZ', 'IYYIIZ', 'ZXXZXX', 'IYYZXX', 'ZXXIYY', 'IYYIYY', 'ZXXXXX', 'IYYXXX', 'ZXXYXY', 'IYYYXY', 'XXXIIZ', 'YXYIIZ', 'XXXZXX', 'YXYZXX', 'XXXIYY', 'YXYIYY', 'XXXXXX', 'YXYXXX', 'XXXYXY', 'YXYYXY', 'IZZIIZ', 'IZZZXX', 'IZZIYY', 'IZZXXX', 'IZZYXY', 'ZXZIZX', 'IXIIZX', 'ZXZIIX', 'IXIIIX', 'XXZIZX', 'YYIIZX', 'XXZIIX', 'YYIIIX', 'ZZIIIZ', 'ZZIZXX', 'ZZIIYY', 'ZZIXXX', 'ZZIYXY', 'XZIIIZ', 'XIIIIZ', 'XZIZXX', 'XIIZXX', 'XZIIYY', 'XIIIYY', 'XZIXXX', 'XIIXXX', 'XZIYXY', 'XIIYXY', 'ZIIIIZ', 'ZIIZXX', 'ZIIIYY', 'ZIIXXX', 'ZIIYXY', 'IIZIZZ', 'IZXZXZ', 'IIXZXZ', 'IZXIXI', 'IIXIXI', 'IZXXXZ', 'IIXXXZ', 'IZXYYI', 'IIXYYI', 'ZXXIZZ', 'IYYIZZ', 'XXXIZZ', 'YXYIZZ', 'IZZIZZ', 'ZXZZXZ', 'IXIZXZ', 'ZXZIXI', 'IXIIXI', 'ZXZXXZ', 'IXIXXZ', 'ZXZYYI', 'IXIYYI', 'XXZZXZ', 'YYIZXZ', 'XXZIXI', 'YYIIXI', 'XXZXXZ', 'YYIXXZ', 'XXZYYI', 'YYIYYI', 'ZZIIZZ', 'XZIIZZ', 'XIIIZZ', 'ZIIIZZ', 'IIZZZI', 'IIZXZI', 'IIZXII', 'ZXXZZI', 'IYYZZI', 'ZXXXZI', 'IYYXZI', 'ZXXXII', 'IYYXII', 'XXXZZI', 'YXYZZI', 'XXXXZI', 'YXYXZI', 'XXXXII', 'YXYXII', 'IZZZZI', 'IZZXZI', 'IZZXII', 'ZZIZZI', 'ZZIXZI', 'ZZIXII', 'XZIZZI', 'XIIZZI', 'XZIXZI', 'XIIXZI', 'XZIXII', 'XIIXII', 'ZIIZZI', 'ZIIXZI', 'ZIIXII', 'IIZZII', 'ZXXZII', 'IYYZII', 'XXXZII', 'YXYZII', 'IZZZII', 'ZZIZII', 'XZIZII', 'XIIZII', 'ZIIZII', 'IZIIII', 'ZZZIII', 'XZZIII', 'XIZIII', 'ZIZIII', 'ZYYIII', 'IXXIII', 'XYYIII', 'YYXIII'
            \item \textbf{Coefficients:}\\
            -4.28808082e+00,  2.33670012e-01, -4.38485129e-02,
             -4.38485129e-02, -1.15040611e-02, -1.15040611e-02,
              1.73742405e-01, -1.57174114e-01, -1.27208744e-02,
              1.27208744e-02, -2.06345005e-01,  2.33670012e-01,
             -4.38485129e-02, -4.38485129e-02, -1.15040611e-02,
             -1.15040611e-02,  1.73742405e-01, -1.57174114e-01,
             -1.27208744e-02,  1.27208744e-02, -2.06345005e-01,
              2.81472843e-01,  2.69696113e-01, -5.85036266e-04,
              5.85036266e-04,  2.72171249e-01,  5.60718440e-03,
              5.60718440e-03,  9.55846941e-04, -9.55846941e-04,
              1.93492226e-01, -2.99588058e-02, -2.99588058e-02,
             -8.35661001e-03, -8.35661001e-03,  1.41797993e-02,
             -1.41797993e-02, -1.41797993e-02,  1.41797993e-02,
             -2.99588058e-02, -2.99588058e-02,  2.01981379e-02,
              2.01981379e-02,  2.01981379e-02,  2.01981379e-02,
             -1.97909733e-03, -1.97909733e-03, -1.97909733e-03,
             -1.97909733e-03, -8.35661001e-03, -8.35661001e-03,
             -1.97909733e-03, -1.97909733e-03, -1.97909733e-03,
             -1.97909733e-03,  1.60692832e-02,  1.60692832e-02,
              1.60692832e-02,  1.60692832e-02,  1.81042425e-01,
             -2.91769426e-02, -2.91769426e-02, -7.12710883e-03,
             -7.12710883e-03,  1.34729080e-04, -1.34729080e-04,
             -1.34729080e-04,  1.34729080e-04,  8.51492850e-04,
              8.51492850e-04, -8.51492850e-04, -8.51492850e-04,
              1.41316009e-01, -5.62837830e-03, -5.62837830e-03,
             -1.22895990e-02, -1.22895990e-02, -5.35309247e-03,
              5.35309247e-03, -5.26684426e-03,  5.26684426e-03,
             -5.26684426e-03,  5.26684426e-03,  8.43828649e-03,
             -8.43828649e-03,  8.43828649e-03, -8.43828649e-03,
              1.48216654e-01, -1.52662008e-02, -1.52662008e-02,
              7.35401279e-03,  7.35401279e-03,  1.81042425e-01,
              1.34729080e-04, -1.34729080e-04, -1.34729080e-04,
              1.34729080e-04,  8.51492850e-04, -8.51492850e-04,
              8.51492850e-04, -8.51492850e-04, -2.91769426e-02,
             -2.91769426e-02, -7.12710883e-03, -7.12710883e-03,
              2.20039773e-01,  8.98527032e-03, -8.98527032e-03,
             -8.98527032e-03,  8.98527032e-03,  1.43796418e-03,
             -1.43796418e-03,  1.43796418e-03, -1.43796418e-03,
              1.43796418e-03,  1.43796418e-03, -1.43796418e-03,
             -1.43796418e-03,  6.60328359e-03,  6.60328359e-03,
              6.60328359e-03,  6.60328359e-03,  1.49009148e-01,
             -1.35099470e-03,  1.35099470e-03,  1.55181525e-01,
              1.41316009e-01, -5.35309247e-03,  5.35309247e-03,
             -5.62837830e-03, -5.62837830e-03, -5.26684426e-03,
             -5.26684426e-03,  5.26684426e-03,  5.26684426e-03,
             -1.22895990e-02, -1.22895990e-02,  8.43828649e-03,
              8.43828649e-03, -8.43828649e-03, -8.43828649e-03,
              1.49009148e-01, -1.35099470e-03,  1.35099470e-03,
              1.62129585e-01, -1.14730837e-02,  1.14730837e-02,
             -1.14730837e-02,  1.14730837e-02,  1.63358583e-02,
             -1.63358583e-02, -1.63358583e-02,  1.63358583e-02,
              1.30946076e-01,  1.13268293e-02, -1.13268293e-02,
              1.48216654e-01, -1.52662008e-02, -1.52662008e-02,
              7.35401279e-03,  7.35401279e-03,  1.55181525e-01,
              1.30946076e-01,  1.13268293e-02, -1.13268293e-02,
              1.66528550e-01,  2.81472843e-01,  2.69696113e-01,
             -5.85036266e-04,  5.85036266e-04,  2.72171249e-01,
              5.60718440e-03,  5.60718440e-03,  9.55846941e-04,
             -9.55846941e-04
       \end{itemize}
    \item[\#6] 
       \begin{itemize}
            \item \textbf{Geometry:}\\
            O 0.0 0.0 0.0\\
            H 0.0 0.0 1.0\\
            H 0.0 0.7933318483619749 -0.10307559543641971
            \item \textbf{Pauli strings:}\\
            'IIIIII', 'IIIIIZ', 'IIIZXX', 'IIIIYY', 'IIIXXX', 'IIIYXY', 'IIIIZZ', 'IIIZZI', 'IIIXZI', 'IIIXII', 'IIIZII', 'IIZIII', 'ZXXIII', 'IYYIII', 'XXXIII', 'YXYIII', 'IZZIII', 'ZZIIII', 'XZIIII', 'XIIIII', 'ZIIIII', 'IIIIZI', 'IIIZZZ', 'IIIXZZ', 'IIIXIZ', 'IIIZIZ', 'IIIZYY', 'IIIIXX', 'IIIXYY', 'IIIYYX', 'IIZIIZ', 'IIZZXX', 'IIZIYY', 'IIZXXX', 'IIZYXY', 'IZXIZX', 'IIXIZX', 'IZXIIX', 'IIXIIX', 'ZXXIIZ', 'IYYIIZ', 'ZXXZXX', 'IYYZXX', 'ZXXIYY', 'IYYIYY', 'ZXXXXX', 'IYYXXX', 'ZXXYXY', 'IYYYXY', 'XXXIIZ', 'YXYIIZ', 'XXXZXX', 'YXYZXX', 'XXXIYY', 'YXYIYY', 'XXXXXX', 'YXYXXX', 'XXXYXY', 'YXYYXY', 'IZZIIZ', 'IZZZXX', 'IZZIYY', 'IZZXXX', 'IZZYXY', 'ZXZIZX', 'IXIIZX', 'ZXZIIX', 'IXIIIX', 'XXZIZX', 'YYIIZX', 'XXZIIX', 'YYIIIX', 'ZZIIIZ', 'ZZIZXX', 'ZZIIYY', 'ZZIXXX', 'ZZIYXY', 'XZIIIZ', 'XIIIIZ', 'XZIZXX', 'XIIZXX', 'XZIIYY', 'XIIIYY', 'XZIXXX', 'XIIXXX', 'XZIYXY', 'XIIYXY', 'ZIIIIZ', 'ZIIZXX', 'ZIIIYY', 'ZIIXXX', 'ZIIYXY', 'IIZIZZ', 'IZXZXZ', 'IIXZXZ', 'IZXIXI', 'IIXIXI', 'IZXXXZ', 'IIXXXZ', 'IZXYYI', 'IIXYYI', 'ZXXIZZ', 'IYYIZZ', 'XXXIZZ', 'YXYIZZ', 'IZZIZZ', 'ZXZZXZ', 'IXIZXZ', 'ZXZIXI', 'IXIIXI', 'ZXZXXZ', 'IXIXXZ', 'ZXZYYI', 'IXIYYI', 'XXZZXZ', 'YYIZXZ', 'XXZIXI', 'YYIIXI', 'XXZXXZ', 'YYIXXZ', 'XXZYYI', 'YYIYYI', 'ZZIIZZ', 'XZIIZZ', 'XIIIZZ', 'ZIIIZZ', 'IIZZZI', 'IIZXZI', 'IIZXII', 'ZXXZZI', 'IYYZZI', 'ZXXXZI', 'IYYXZI', 'ZXXXII', 'IYYXII', 'XXXZZI', 'YXYZZI', 'XXXXZI', 'YXYXZI', 'XXXXII', 'YXYXII', 'IZZZZI', 'IZZXZI', 'IZZXII', 'ZZIZZI', 'ZZIXZI', 'ZZIXII', 'XZIZZI', 'XIIZZI', 'XZIXZI', 'XIIXZI', 'XZIXII', 'XIIXII', 'ZIIZZI', 'ZIIXZI', 'ZIIXII', 'IIZZII', 'ZXXZII', 'IYYZII', 'XXXZII', 'YXYZII', 'IZZZII', 'ZZIZII', 'XZIZII', 'XIIZII', 'ZIIZII', 'IZIIII', 'ZZZIII', 'XZZIII', 'XIZIII', 'ZIZIII', 'ZYYIII', 'IXXIII', 'XYYIII', 'YYXIII'
            \item \textbf{Coefficients:}\\
            -4.22840830e+00,  2.39652977e-01,  4.02927172e-02,
              4.02927172e-02,  1.19441127e-02,  1.19441127e-02,
              1.77421588e-01, -1.55806329e-01, -1.62708828e-02,
              1.62708828e-02, -2.65693013e-01,  2.39652977e-01,
              4.02927172e-02,  4.02927172e-02,  1.19441127e-02,
              1.19441127e-02,  1.77421588e-01, -1.55806329e-01,
             -1.62708828e-02,  1.62708828e-02, -2.65693013e-01,
              2.84108271e-01,  2.72751167e-01, -3.32376722e-04,
              3.32376722e-04,  2.74552276e-01, -4.87971987e-03,
             -4.87971987e-03, -6.05627952e-04,  6.05627952e-04,
              1.98594103e-01,  2.82817395e-02,  2.82817395e-02,
              9.29508383e-03,  9.29508383e-03,  1.45967128e-02,
             -1.45967128e-02, -1.45967128e-02,  1.45967128e-02,
              2.82817395e-02,  2.82817395e-02,  2.05293889e-02,
              2.05293889e-02,  2.05293889e-02,  2.05293889e-02,
             -2.93344630e-03, -2.93344630e-03, -2.93344630e-03,
             -2.93344630e-03,  9.29508383e-03,  9.29508383e-03,
             -2.93344630e-03, -2.93344630e-03, -2.93344630e-03,
             -2.93344630e-03,  1.38630231e-02,  1.38630231e-02,
              1.38630231e-02,  1.38630231e-02,  1.83146645e-01,
              2.70507266e-02,  2.70507266e-02,  7.56292184e-03,
              7.56292184e-03, -2.16220887e-04,  2.16220887e-04,
              2.16220887e-04, -2.16220887e-04, -1.65130758e-03,
             -1.65130758e-03,  1.65130758e-03,  1.65130758e-03,
              1.43791781e-01,  1.22662681e-03,  1.22662681e-03,
              1.31064614e-02,  1.31064614e-02, -7.20572735e-03,
              7.20572735e-03,  4.49785993e-03, -4.49785993e-03,
              4.49785993e-03, -4.49785993e-03, -3.69437239e-03,
              3.69437239e-03, -3.69437239e-03,  3.69437239e-03,
              1.47717196e-01,  1.86928552e-02,  1.86928552e-02,
             -7.58690488e-03, -7.58690488e-03,  1.83146645e-01,
             -2.16220887e-04,  2.16220887e-04,  2.16220887e-04,
             -2.16220887e-04, -1.65130758e-03,  1.65130758e-03,
             -1.65130758e-03,  1.65130758e-03,  2.70507266e-02,
              2.70507266e-02,  7.56292184e-03,  7.56292184e-03,
              2.20039773e-01,  8.57075998e-03, -8.57075998e-03,
             -8.57075998e-03,  8.57075998e-03,  2.00014750e-03,
             -2.00014750e-03,  2.00014750e-03, -2.00014750e-03,
              2.00014750e-03,  2.00014750e-03, -2.00014750e-03,
             -2.00014750e-03,  7.31342465e-03,  7.31342465e-03,
              7.31342465e-03,  7.31342465e-03,  1.49268864e-01,
             -1.93975683e-03,  1.93975683e-03,  1.56802199e-01,
              1.43791781e-01, -7.20572735e-03,  7.20572735e-03,
              1.22662681e-03,  1.22662681e-03,  4.49785993e-03,
              4.49785993e-03, -4.49785993e-03, -4.49785993e-03,
              1.31064614e-02,  1.31064614e-02, -3.69437239e-03,
             -3.69437239e-03,  3.69437239e-03,  3.69437239e-03,
              1.49268864e-01, -1.93975683e-03,  1.93975683e-03,
              1.68649805e-01, -9.67841791e-03,  9.67841791e-03,
             -9.67841791e-03,  9.67841791e-03,  9.86819933e-03,
             -9.86819933e-03, -9.86819933e-03,  9.86819933e-03,
              1.25426538e-01,  8.59163160e-03, -8.59163160e-03,
              1.47717196e-01,  1.86928552e-02,  1.86928552e-02,
             -7.58690488e-03, -7.58690488e-03,  1.56802199e-01,
              1.25426538e-01,  8.59163160e-03, -8.59163160e-03,
              1.74844295e-01,  2.84108271e-01,  2.72751167e-01,
             -3.32376722e-04,  3.32376722e-04,  2.74552276e-01,
             -4.87971987e-03, -4.87971987e-03, -6.05627952e-04,
              6.05627952e-04
       \end{itemize} 
   \item[\#7] 
       \begin{itemize}
            \item \textbf{Geometry:}\\
            O 0.0 0.0 0.0\\
            H 0.0 0.0 1.0\\
            H 0.0 0.8924983294072217 -0.11596004486597218
            \item \textbf{Pauli strings:}\\
            'IIIIII', 'IIIIIZ', 'IIIZXX', 'IIIIYY', 'IIIXXX', 'IIIYXY', 'IIIIZZ', 'IIIZZI', 'IIIXZI', 'IIIXII', 'IIIZII', 'IIZIII', 'ZXXIII', 'IYYIII', 'XXXIII', 'YXYIII', 'IZZIII', 'ZZIIII', 'XZIIII', 'XIIIII', 'ZIIIII', 'IIIIZI', 'IIIZZZ', 'IIIXZZ', 'IIIXIZ', 'IIIZIZ', 'IIIZYY', 'IIIIXX', 'IIIXYY', 'IIIYYX', 'IIZIIZ', 'IIZZXX', 'IIZIYY', 'IIZXXX', 'IIZYXY', 'IZXIZX', 'IIXIZX', 'IZXIIX', 'IIXIIX', 'ZXXIIZ', 'IYYIIZ', 'ZXXZXX', 'IYYZXX', 'ZXXIYY', 'IYYIYY', 'ZXXXXX', 'IYYXXX', 'ZXXYXY', 'IYYYXY', 'XXXIIZ', 'YXYIIZ', 'XXXZXX', 'YXYZXX', 'XXXIYY', 'YXYIYY', 'XXXXXX', 'YXYXXX', 'XXXYXY', 'YXYYXY', 'IZZIIZ', 'IZZZXX', 'IZZIYY', 'IZZXXX', 'IZZYXY', 'ZXZIZX', 'IXIIZX', 'ZXZIIX', 'IXIIIX', 'XXZIZX', 'YYIIZX', 'XXZIIX', 'YYIIIX', 'ZZIIIZ', 'ZZIZXX', 'ZZIIYY', 'ZZIXXX', 'ZZIYXY', 'XZIIIZ', 'XIIIIZ', 'XZIZXX', 'XIIZXX', 'XZIIYY', 'XIIIYY', 'XZIXXX', 'XIIXXX', 'XZIYXY', 'XIIYXY', 'ZIIIIZ', 'ZIIZXX', 'ZIIIYY', 'ZIIXXX', 'ZIIYXY', 'IIZIZZ', 'IZXZXZ', 'IIXZXZ', 'IZXIXI', 'IIXIXI', 'IZXXXZ', 'IIXXXZ', 'IZXYYI', 'IIXYYI', 'ZXXIZZ', 'IYYIZZ', 'XXXIZZ', 'YXYIZZ', 'IZZIZZ', 'ZXZZXZ', 'IXIZXZ', 'ZXZIXI', 'IXIIXI', 'ZXZXXZ', 'IXIXXZ', 'ZXZYYI', 'IXIYYI', 'XXZZXZ', 'YYIZXZ', 'XXZIXI', 'YYIIXI', 'XXZXXZ', 'YYIXXZ', 'XXZYYI', 'YYIYYI', 'ZZIIZZ', 'XZIIZZ', 'XIIIZZ', 'ZIIIZZ', 'IIZZZI', 'IIZXZI', 'IIZXII', 'ZXXZZI', 'IYYZZI', 'ZXXXZI', 'IYYXZI', 'ZXXXII', 'IYYXII', 'XXXZZI', 'YXYZZI', 'XXXXZI', 'YXYXZI', 'XXXXII', 'YXYXII', 'IZZZZI', 'IZZXZI', 'IZZXII', 'ZZIZZI', 'ZZIXZI', 'ZZIXII', 'XZIZZI', 'XIIZZI', 'XZIXZI', 'XIIXZI', 'XZIXII', 'XIIXII', 'ZIIZZI', 'ZIIXZI', 'ZIIXII', 'IIZZII', 'ZXXZII', 'IYYZII', 'XXXZII', 'YXYZII', 'IZZZII', 'ZZIZII', 'XZIZII', 'XIIZII', 'ZIIZII', 'IZIIII', 'ZZZIII', 'XZZIII', 'XIZIII', 'ZIZIII', 'ZYYIII', 'IXXIII', 'XYYIII', 'YYXIII'
            \item \textbf{Coefficients:}\\
              4.38485129e-02,  1.15040611e-02,  1.15040611e-02,
              1.73742405e-01, -1.57174114e-01, -1.27208744e-02,
              1.27208744e-02, -2.06345005e-01,  2.33670012e-01,
              4.38485129e-02,  4.38485129e-02,  1.15040611e-02,
              1.15040611e-02,  1.73742405e-01, -1.57174114e-01,
             -1.27208744e-02,  1.27208744e-02, -2.06345005e-01,
              2.81472843e-01,  2.69696113e-01, -5.85036266e-04,
              5.85036266e-04,  2.72171249e-01, -5.60718440e-03,
             -5.60718440e-03, -9.55846941e-04,  9.55846941e-04,
              1.93492226e-01,  2.99588058e-02,  2.99588058e-02,
              8.35661001e-03,  8.35661001e-03,  1.41797993e-02,
             -1.41797993e-02, -1.41797993e-02,  1.41797993e-02,
              2.99588058e-02,  2.99588058e-02,  2.01981379e-02,
              2.01981379e-02,  2.01981379e-02,  2.01981379e-02,
             -1.97909733e-03, -1.97909733e-03, -1.97909733e-03,
             -1.97909733e-03,  8.35661001e-03,  8.35661001e-03,
             -1.97909733e-03, -1.97909733e-03, -1.97909733e-03,
             -1.97909733e-03,  1.60692832e-02,  1.60692832e-02,
              1.60692832e-02,  1.60692832e-02,  1.81042425e-01,
              2.91769426e-02,  2.91769426e-02,  7.12710883e-03,
              7.12710883e-03, -1.34729080e-04,  1.34729080e-04,
              1.34729080e-04, -1.34729080e-04, -8.51492850e-04,
             -8.51492850e-04,  8.51492850e-04,  8.51492850e-04,
              1.41316009e-01,  5.62837830e-03,  5.62837830e-03,
              1.22895990e-02,  1.22895990e-02, -5.35309247e-03,
              5.35309247e-03,  5.26684426e-03, -5.26684426e-03,
              5.26684426e-03, -5.26684426e-03, -8.43828649e-03,
              8.43828649e-03, -8.43828649e-03,  8.43828649e-03,
              1.48216654e-01,  1.52662008e-02,  1.52662008e-02,
             -7.35401279e-03, -7.35401279e-03,  1.81042425e-01,
             -1.34729080e-04,  1.34729080e-04,  1.34729080e-04,
             -1.34729080e-04, -8.51492850e-04,  8.51492850e-04,
             -8.51492850e-04,  8.51492850e-04,  2.91769426e-02,
              2.91769426e-02,  7.12710883e-03,  7.12710883e-03,
              2.20039773e-01,  8.98527032e-03, -8.98527032e-03,
             -8.98527032e-03,  8.98527032e-03,  1.43796418e-03,
             -1.43796418e-03,  1.43796418e-03, -1.43796418e-03,
              1.43796418e-03,  1.43796418e-03, -1.43796418e-03,
             -1.43796418e-03,  6.60328359e-03,  6.60328359e-03,
              6.60328359e-03,  6.60328359e-03,  1.49009148e-01,
             -1.35099470e-03,  1.35099470e-03,  1.55181525e-01,
              1.41316009e-01, -5.35309247e-03,  5.35309247e-03,
              5.62837830e-03,  5.62837830e-03,  5.26684426e-03,
              5.26684426e-03, -5.26684426e-03, -5.26684426e-03,
              1.22895990e-02,  1.22895990e-02, -8.43828649e-03,
             -8.43828649e-03,  8.43828649e-03,  8.43828649e-03,
              1.49009148e-01, -1.35099470e-03,  1.35099470e-03,
              1.62129585e-01, -1.14730837e-02,  1.14730837e-02,
             -1.14730837e-02,  1.14730837e-02,  1.63358583e-02,
             -1.63358583e-02, -1.63358583e-02,  1.63358583e-02,
              1.30946076e-01,  1.13268293e-02, -1.13268293e-02,
              1.48216654e-01,  1.52662008e-02,  1.52662008e-02,
             -7.35401279e-03, -7.35401279e-03,  1.55181525e-01,
              1.30946076e-01,  1.13268293e-02, -1.13268293e-02,
              1.66528550e-01,  2.81472843e-01,  2.69696113e-01,
             -5.85036266e-04,  5.85036266e-04,  2.72171249e-01,
             -5.60718440e-03, -5.60718440e-03, -9.55846941e-04,
              9.55846941e-04
       \end{itemize}
   \item[\#8] 
       \begin{itemize}
            \item \textbf{Geometry:}\\
            O 0.0 0.0 0.0\\
            H 0.0 0.0 1.0\\
            H 0.0 0.9916648104524686 -0.12884449429552464
            \item \textbf{Pauli strings:}\\
            'IIIIII', 'IIIIIZ', 'IIIZXX', 'IIIIYY', 'IIIIZZ', 'IIIZZI', 'IIIZII', 'IIZIII', 'ZXXIII', 'IYYIII', 'IZZIII', 'ZZIIII', 'ZIIIII', 'IIIIZI', 'IIIZZZ', 'IIIZIZ', 'IIIZYY', 'IIIIXX', 'IIZIIZ', 'IIZZXX', 'IIZIYY', 'IZXIZX', 'IIXIZX', 'IZXIIX', 'IIXIIX', 'ZXXIIZ', 'IYYIIZ', 'ZXXZXX', 'IYYZXX', 'ZXXIYY', 'IYYIYY', 'XXXXXX', 'YXYXXX', 'XXXYXY', 'YXYYXY', 'IZZIIZ', 'IZZZXX', 'IZZIYY', 'ZXZIZX', 'IXIIZX', 'ZXZIIX', 'IXIIIX', 'ZZIIIZ', 'ZZIZXX', 'ZZIIYY', 'XZIXXX', 'XIIXXX', 'XZIYXY', 'XIIYXY', 'ZIIIIZ', 'ZIIZXX', 'ZIIIYY', 'IIZIZZ', 'IZXZXZ', 'IIXZXZ', 'IZXIXI', 'IIXIXI', 'ZXXIZZ', 'IYYIZZ', 'IZZIZZ', 'ZXZZXZ', 'IXIZXZ', 'ZXZIXI', 'IXIIXI', 'XXZXXZ', 'YYIXXZ', 'XXZYYI', 'YYIYYI', 'ZZIIZZ', 'ZIIIZZ', 'IIZZZI', 'ZXXZZI', 'IYYZZI', 'XXXXZI', 'YXYXZI', 'XXXXII', 'YXYXII', 'IZZZZI', 'ZZIZZI', 'XZIXZI', 'XIIXZI', 'XZIXII', 'XIIXII', 'ZIIZZI', 'IIZZII', 'ZXXZII', 'IYYZII', 'IZZZII', 'ZZIZII', 'ZIIZII', 'IZIIII', 'ZZZIII', 'ZIZIII', 'ZYYIII', 'IXXIII'
            \item \textbf{Coefficients:}\\
            -4.32707452e+00,  2.27304465e-01,  4.85807576e-02,
              4.85807576e-02,  1.71488678e-01, -1.49310434e-01,
             -1.63041098e-01,  2.27304465e-01,  4.85807576e-02,
              4.85807576e-02,  1.71488678e-01, -1.49310434e-01,
             -1.63041098e-01,  2.78185239e-01,  2.65770979e-01,
              2.69492966e-01, -6.64977694e-03, -6.64977694e-03,
              1.87524843e-01,  3.21905436e-02,  3.21905436e-02,
              1.36218870e-02, -1.36218870e-02, -1.36218870e-02,
              1.36218870e-02,  3.21905436e-02,  3.21905436e-02,
              2.03977556e-02,  2.03977556e-02,  2.03977556e-02,
              2.03977556e-02,  1.80243273e-02,  1.80243273e-02,
              1.80243273e-02,  1.80243273e-02,  1.78406006e-01,
              3.23693643e-02,  3.23693643e-02,  1.56823766e-04,
             -1.56823766e-04, -1.56823766e-04,  1.56823766e-04,
              1.38318420e-01,  1.19839131e-02,  1.19839131e-02,
             -1.49177497e-02,  1.49177497e-02, -1.49177497e-02,
              1.49177497e-02,  1.48706086e-01,  1.06450139e-02,
              1.06450139e-02,  1.78406006e-01,  1.56823766e-04,
             -1.56823766e-04, -1.56823766e-04,  1.56823766e-04,
              3.23693643e-02,  3.23693643e-02,  2.20039773e-01,
              9.31499457e-03, -9.31499457e-03, -9.31499457e-03,
              9.31499457e-03,  5.99582058e-03,  5.99582058e-03,
              5.99582058e-03,  5.99582058e-03,  1.48126201e-01,
              1.53846135e-01,  1.38318420e-01,  1.19839131e-02,
              1.19839131e-02, -1.49177497e-02, -1.49177497e-02,
              1.49177497e-02,  1.49177497e-02,  1.48126201e-01,
              1.50084690e-01,  2.78303895e-02, -2.78303895e-02,
             -2.78303895e-02,  2.78303895e-02,  1.41231510e-01,
              1.48706086e-01,  1.06450139e-02,  1.06450139e-02,
              1.53846135e-01,  1.41231510e-01,  1.52374907e-01,
              2.78185239e-01,  2.65770979e-01,  2.69492966e-01,
             -6.64977694e-03, -6.64977694e-03
       \end{itemize} 
    \item[\#9] 
       \begin{itemize}
            \item \textbf{Geometry:}\\
            O 0.0 0.0 0.0\\
            H 0.0 0.0 0.8\\
            H 0.0 0.7790781047025562 -0.18176167575446953
            \item \textbf{Pauli strings:}\\
            'IIIIII', 'IIIIIZ', 'IIIZXX', 'IIIIYY', 'IIIIZZ', 'IIIZZI', 'IIIZII', 'IIZIII', 'ZXXIII', 'IYYIII', 'IZZIII', 'ZZIIII', 'ZIIIII', 'IIIIZI', 'IIIZZZ', 'IIIZIZ', 'IIIZYY', 'IIIIXX', 'IIZIIZ', 'IIZZXX', 'IIZIYY', 'IZXIZX', 'IIXIZX', 'IZXIIX', 'IIXIIX', 'ZXXIIZ', 'IYYIIZ', 'ZXXZXX', 'IYYZXX', 'ZXXIYY', 'IYYIYY', 'XXXXXX', 'YXYXXX', 'XXXYXY', 'YXYYXY', 'IZZIIZ', 'IZZZXX', 'IZZIYY', 'ZXZIZX', 'IXIIZX', 'ZXZIIX', 'IXIIIX', 'ZZIIIZ', 'ZZIZXX', 'ZZIIYY', 'XZIXXX', 'XIIXXX', 'XZIYXY', 'XIIYXY', 'ZIIIIZ', 'ZIIZXX', 'ZIIIYY', 'IIZIZZ', 'IZXZXZ', 'IIXZXZ', 'IZXIXI', 'IIXIXI', 'ZXXIZZ', 'IYYIZZ', 'IZZIZZ', 'ZXZZXZ', 'IXIZXZ', 'ZXZIXI', 'IXIIXI', 'XXZXXZ', 'YYIXXZ', 'XXZYYI', 'YYIYYI', 'ZZIIZZ', 'ZIIIZZ', 'IIZZZI', 'ZXXZZI', 'IYYZZI', 'XXXXZI', 'YXYXZI', 'XXXXII', 'YXYXII', 'IZZZZI', 'ZZIZZI', 'XZIXZI', 'XIIXZI', 'XZIXII', 'XIIXII', 'ZIIZZI', 'IIZZII', 'ZXXZII', 'IYYZII', 'IZZZII', 'ZZIZII', 'ZIIZII', 'IZIIII', 'ZZZIII', 'ZIZIII', 'ZYYIII', 'IXXIII'
            \item \textbf{Coefficients:}\\
            -4.16280752e+00,  2.52696356e-01,  2.92680371e-02,
              2.92680371e-02,  1.86951846e-01, -2.46802581e-01,
             -2.80705951e-01,  2.52696356e-01,  2.92680371e-02,
              2.92680371e-02,  1.86951846e-01, -2.46802581e-01,
             -2.80705951e-01,  2.90826437e-01,  2.81407509e-01,
              2.81364244e-01, -1.67514452e-03, -1.67514452e-03,
              2.15249721e-01,  2.41596506e-02,  2.41596506e-02,
              1.56791424e-02, -1.56791424e-02, -1.56791424e-02,
              1.56791424e-02,  2.41596506e-02,  2.41596506e-02,
              1.09607847e-02,  1.09607847e-02,  1.09607847e-02,
              1.09607847e-02,  1.68491361e-02,  1.68491361e-02,
              1.68491361e-02,  1.68491361e-02,  1.90000756e-01,
              2.02503168e-02,  2.02503168e-02, -2.42507261e-03,
              2.42507261e-03,  2.42507261e-03, -2.42507261e-03,
              1.37882527e-01,  9.33136622e-03,  9.33136622e-03,
             -1.35145293e-02,  1.35145293e-02, -1.35145293e-02,
              1.35145293e-02,  1.59038751e-01,  7.48571564e-03,
              7.48571564e-03,  1.90000756e-01, -2.42507261e-03,
              2.42507261e-03,  2.42507261e-03, -2.42507261e-03,
              2.02503168e-02,  2.02503168e-02,  2.20039773e-01,
              1.08279493e-02, -1.08279493e-02, -1.08279493e-02,
              1.08279493e-02,  5.75965442e-03,  5.75965442e-03,
              5.75965442e-03,  5.75965442e-03,  1.50002579e-01,
              1.60245421e-01,  1.37882527e-01,  9.33136622e-03,
              9.33136622e-03, -1.35145293e-02, -1.35145293e-02,
              1.35145293e-02,  1.35145293e-02,  1.50002579e-01,
              1.50776585e-01,  2.84731462e-02, -2.84731462e-02,
             -2.84731462e-02,  2.84731462e-02,  1.44977969e-01,
              1.59038751e-01,  7.48571564e-03,  7.48571564e-03,
              1.60245421e-01,  1.44977969e-01,  1.60969719e-01,
              2.90826437e-01,  2.81407509e-01,  2.81364244e-01,
             -1.67514452e-03, -1.67514452e-03
       \end{itemize} 
   \item[\#10] 
       \begin{itemize}
            \item \textbf{Geometry:}\\
            O 0.0 0.0 0.0\\
            H 0.0 0.0 0.8\\
            H 0.0 0.8764628677903757 -0.2044818852237782
            \item \textbf{Pauli strings:}\\
            'IIIIII', 'IIIIIZ', 'IIIZXX', 'IIIIYY', 'IIIXXX', 'IIIYXY', 'IIIIZZ', 'IIIZZI', 'IIIXZI', 'IIIXII', 'IIIZII', 'IIZIII', 'ZXXIII', 'IYYIII', 'XXXIII', 'YXYIII', 'IZZIII', 'ZZIIII', 'XZIIII', 'XIIIII', 'ZIIIII', 'IIIIZI', 'IIIZZZ', 'IIIXZZ', 'IIIXIZ', 'IIIZIZ', 'IIIZYY', 'IIIIXX', 'IIIXYY', 'IIIYYX', 'IIZIIZ', 'IIZZXX', 'IIZIYY', 'IIZXXX', 'IIZYXY', 'IZXIZX', 'IIXIZX', 'IZXIIX', 'IIXIIX', 'ZXXIIZ', 'IYYIIZ', 'ZXXZXX', 'IYYZXX', 'ZXXIYY', 'IYYIYY', 'ZXXXXX', 'IYYXXX', 'ZXXYXY', 'IYYYXY', 'XXXIIZ', 'YXYIIZ', 'XXXZXX', 'YXYZXX', 'XXXIYY', 'YXYIYY', 'XXXXXX', 'YXYXXX', 'XXXYXY', 'YXYYXY', 'IZZIIZ', 'IZZZXX', 'IZZIYY', 'IZZXXX', 'IZZYXY', 'ZXZIZX', 'IXIIZX', 'ZXZIIX', 'IXIIIX', 'XXZIZX', 'YYIIZX', 'XXZIIX', 'YYIIIX', 'ZZIIIZ', 'ZZIZXX', 'ZZIIYY', 'ZZIXXX', 'ZZIYXY', 'XZIIIZ', 'XIIIIZ', 'XZIZXX', 'XIIZXX', 'XZIIYY', 'XIIIYY', 'XZIXXX', 'XIIXXX', 'XZIYXY', 'XIIYXY', 'ZIIIIZ', 'ZIIZXX', 'ZIIIYY', 'ZIIXXX', 'ZIIYXY', 'IIZIZZ', 'IZXZXZ', 'IIXZXZ', 'IZXIXI', 'IIXIXI', 'IZXXXZ', 'IIXXXZ', 'IZXYYI', 'IIXYYI', 'ZXXIZZ', 'IYYIZZ', 'XXXIZZ', 'YXYIZZ', 'IZZIZZ', 'ZXZZXZ', 'IXIZXZ', 'ZXZIXI', 'IXIIXI', 'ZXZXXZ', 'IXIXXZ', 'ZXZYYI', 'IXIYYI', 'XXZZXZ', 'YYIZXZ', 'XXZIXI', 'YYIIXI', 'XXZXXZ', 'YYIXXZ', 'XXZYYI', 'YYIYYI', 'ZZIIZZ', 'XZIIZZ', 'XIIIZZ', 'ZIIIZZ', 'IIZZZI', 'IIZXZI', 'IIZXII', 'ZXXZZI', 'IYYZZI', 'ZXXXZI', 'IYYXZI', 'ZXXXII', 'IYYXII', 'XXXZZI', 'YXYZZI', 'XXXXZI', 'YXYXZI', 'XXXXII', 'YXYXII', 'IZZZZI', 'IZZXZI', 'IZZXII', 'ZZIZZI', 'ZZIXZI', 'ZZIXII', 'XZIZZI', 'XIIZZI', 'XZIXZI', 'XIIXZI', 'XZIXII', 'XIIXII', 'ZIIZZI', 'ZIIXZI', 'ZIIXII', 'IIZZII', 'ZXXZII', 'IYYZII', 'XXXZII', 'YXYZII', 'IZZZII', 'ZZIZII', 'XZIZII', 'XIIZII', 'ZIIZII', 'IZIIII', 'ZZZIII', 'XZZIII', 'XIZIII', 'ZIZIII', 'ZYYIII', 'IXXIII', 'XYYIII', 'YYXIII'
            \item \textbf{Coefficients:}\\
            -4.21057367e+00,  2.44869776e-01, -3.33651883e-02,
             -3.33651883e-02, -7.41037699e-03, -7.41037699e-03,
              1.82416184e-01, -2.04870177e-01, -1.36589245e-02,
              1.36589245e-02, -2.64959136e-01,  2.44869776e-01,
             -3.33651883e-02, -3.33651883e-02, -7.41037699e-03,
             -7.41037699e-03,  1.82416184e-01, -2.04870177e-01,
             -1.36589245e-02,  1.36589245e-02, -2.64959136e-01,
              2.87816187e-01,  2.77906746e-01, -5.69520436e-05,
              5.69520436e-05,  2.78525177e-01,  2.69690098e-03,
              2.69690098e-03,  2.63890432e-04, -2.63890432e-04,
              2.08829315e-01, -2.59130771e-02, -2.59130771e-02,
             -6.24290272e-03, -6.24290272e-03,  1.51616570e-02,
             -1.51616570e-02, -1.51616570e-02,  1.51616570e-02,
             -2.59130771e-02, -2.59130771e-02,  1.42527312e-02,
              1.42527312e-02,  1.42527312e-02,  1.42527312e-02,
             -2.44178434e-03, -2.44178434e-03, -2.44178434e-03,
             -2.44178434e-03, -6.24290272e-03, -6.24290272e-03,
             -2.44178434e-03, -2.44178434e-03, -2.44178434e-03,
             -2.44178434e-03,  1.56938226e-02,  1.56938226e-02,
              1.56938226e-02,  1.56938226e-02,  1.87569325e-01,
             -2.27729083e-02, -2.27729083e-02, -5.09961455e-03,
             -5.09961455e-03,  1.49726453e-03, -1.49726453e-03,
             -1.49726453e-03,  1.49726453e-03,  1.11196910e-03,
              1.11196910e-03, -1.11196910e-03, -1.11196910e-03,
              1.40148487e-01, -5.74364504e-03, -5.74364504e-03,
             -1.07521443e-02, -1.07521443e-02, -6.17546799e-03,
              6.17546799e-03, -4.80445109e-03,  4.80445109e-03,
             -4.80445109e-03,  4.80445109e-03,  9.29907678e-03,
             -9.29907678e-03,  9.29907678e-03, -9.29907678e-03,
              1.54406860e-01, -1.22741951e-02, -1.22741951e-02,
              6.55605548e-03,  6.55605548e-03,  1.87569325e-01,
              1.49726453e-03, -1.49726453e-03, -1.49726453e-03,
              1.49726453e-03,  1.11196910e-03, -1.11196910e-03,
              1.11196910e-03, -1.11196910e-03, -2.27729083e-02,
             -2.27729083e-02, -5.09961455e-03, -5.09961455e-03,
              2.20039773e-01,  9.81225660e-03, -9.81225660e-03,
             -9.81225660e-03,  9.81225660e-03,  1.60476046e-03,
             -1.60476046e-03,  1.60476046e-03, -1.60476046e-03,
              1.60476046e-03,  1.60476046e-03, -1.60476046e-03,
             -1.60476046e-03,  6.51355519e-03,  6.51355519e-03,
              6.51355519e-03,  6.51355519e-03,  1.49624396e-01,
             -2.07197114e-03,  2.07197114e-03,  1.58524545e-01,
              1.40148487e-01, -6.17546799e-03,  6.17546799e-03,
             -5.74364504e-03, -5.74364504e-03, -4.80445109e-03,
             -4.80445109e-03,  4.80445109e-03,  4.80445109e-03,
             -1.07521443e-02, -1.07521443e-02,  9.29907678e-03,
              9.29907678e-03, -9.29907678e-03, -9.29907678e-03,
              1.49624396e-01, -2.07197114e-03,  2.07197114e-03,
              1.60702492e-01, -1.24178688e-02,  1.24178688e-02,
             -1.24178688e-02,  1.24178688e-02,  1.91783391e-02,
             -1.91783391e-02, -1.91783391e-02,  1.91783391e-02,
              1.34586859e-01,  1.04189127e-02, -1.04189127e-02,
              1.54406860e-01, -1.22741951e-02, -1.22741951e-02,
              6.55605548e-03,  6.55605548e-03,  1.58524545e-01,
              1.34586859e-01,  1.04189127e-02, -1.04189127e-02,
              1.67871626e-01,  2.87816187e-01,  2.77906746e-01,
             -5.69520436e-05,  5.69520436e-05,  2.78525177e-01,
              2.69690098e-03,  2.69690098e-03,  2.63890432e-04,
             -2.63890432e-04
       \end{itemize}
   \item[\#11] 
       \begin{itemize}
            \item \textbf{Geometry:}\\
            O 0.0 0.0 0.0\\
            H 0.0 0.0 0.8\\
            H 0.0 0.9738476308781953 -0.2272020946930869
            \item \textbf{Pauli strings:}\\
            'IIIIII', 'IIIIIZ', 'IIIZXX', 'IIIIYY', 'IIIXXX', 'IIIYXY', 'IIIIZZ', 'IIIZZI', 'IIIXZI', 'IIIXII', 'IIIZII', 'IIZIII', 'ZXXIII', 'IYYIII', 'XXXIII', 'YXYIII', 'IZZIII', 'ZZIIII', 'XZIIII', 'XIIIII', 'ZIIIII', 'IIIIZI', 'IIIZZZ', 'IIIXZZ', 'IIIXIZ', 'IIIZIZ', 'IIIZYY', 'IIIIXX', 'IIIXYY', 'IIIYYX', 'IIZIIZ', 'IIZZXX', 'IIZIYY', 'IIZXXX', 'IIZYXY', 'IZXIZX', 'IIXIZX', 'IZXIIX', 'IIXIIX', 'ZXXIIZ', 'IYYIIZ', 'ZXXZXX', 'IYYZXX', 'ZXXIYY', 'IYYIYY', 'ZXXXXX', 'IYYXXX', 'ZXXYXY', 'IYYYXY', 'XXXIIZ', 'YXYIIZ', 'XXXZXX', 'YXYZXX', 'XXXIYY', 'YXYIYY', 'XXXXXX', 'YXYXXX', 'XXXYXY', 'YXYYXY', 'IZZIIZ', 'IZZZXX', 'IZZIYY', 'IZZXXX', 'IZZYXY', 'ZXZIZX', 'IXIIZX', 'ZXZIIX', 'IXIIIX', 'XXZIZX', 'YYIIZX', 'XXZIIX', 'YYIIIX', 'ZZIIIZ', 'ZZIZXX', 'ZZIIYY', 'ZZIXXX', 'ZZIYXY', 'XZIIIZ', 'XIIIIZ', 'XZIZXX', 'XIIZXX', 'XZIIYY', 'XIIIYY', 'XZIXXX', 'XIIXXX', 'XZIYXY', 'XIIYXY', 'ZIIIIZ', 'ZIIZXX', 'ZIIIYY', 'ZIIXXX', 'ZIIYXY', 'IIZIZZ', 'IZXZXZ', 'IIXZXZ', 'IZXIXI', 'IIXIXI', 'IZXXXZ', 'IIXXXZ', 'IZXYYI', 'IIXYYI', 'ZXXIZZ', 'IYYIZZ', 'XXXIZZ', 'YXYIZZ', 'IZZIZZ', 'ZXZZXZ', 'IXIZXZ', 'ZXZIXI', 'IXIIXI', 'ZXZXXZ', 'IXIXXZ', 'ZXZYYI', 'IXIYYI', 'XXZZXZ', 'YYIZXZ', 'XXZIXI', 'YYIIXI', 'XXZXXZ', 'YYIXXZ', 'XXZYYI', 'YYIYYI', 'ZZIIZZ', 'XZIIZZ', 'XIIIZZ', 'ZIIIZZ', 'IIZZZI', 'IIZXZI', 'IIZXII', 'ZXXZZI', 'IYYZZI', 'ZXXXZI', 'IYYXZI', 'ZXXXII', 'IYYXII', 'XXXZZI', 'YXYZZI', 'XXXXZI', 'YXYXZI', 'XXXXII', 'YXYXII', 'IZZZZI', 'IZZXZI', 'IZZXII', 'ZZIZZI', 'ZZIXZI', 'ZZIXII', 'XZIZZI', 'XIIZZI', 'XZIXZI', 'XIIXZI', 'XZIXII', 'XIIXII', 'ZIIZZI', 'ZIIXZI', 'ZIIXII', 'IIZZII', 'ZXXZII', 'IYYZII', 'XXXZII', 'YXYZII', 'IZZZII', 'ZZIZII', 'XZIZII', 'XIIZII', 'ZIIZII', 'IZIIII', 'ZZZIII', 'XZZIII', 'XIZIII', 'ZIZIII', 'ZYYIII', 'IXXIII', 'XYYIII', 'YYXIII'
            \item \textbf{Coefficients:}\\
            -4.23064680e+00,  2.35712842e-01, -3.76125834e-02,
             -3.76125834e-02, -1.01397121e-02, -1.01397121e-02,
              1.78637163e-01, -1.55268782e-01, -1.74910994e-02,
              1.74910994e-02, -2.64135521e-01,  2.35712842e-01,
             -3.76125834e-02, -3.76125834e-02, -1.01397121e-02,
             -1.01397121e-02,  1.78637163e-01, -1.55268782e-01,
             -1.74910994e-02,  1.74910994e-02, -2.64135521e-01,
              2.83875887e-01,  2.73621236e-01, -1.96710720e-04,
              1.96710720e-04,  2.74934867e-01,  3.96672647e-03,
              3.96672647e-03,  3.59301309e-04, -3.59301309e-04,
              2.00712467e-01, -2.75089457e-02, -2.75089457e-02,
             -8.28383437e-03, -8.28383437e-03,  1.44226373e-02,
             -1.44226373e-02, -1.44226373e-02,  1.44226373e-02,
             -2.75089457e-02, -2.75089457e-02,  1.88959351e-02,
              1.88959351e-02,  1.88959351e-02,  1.88959351e-02,
             -2.98259055e-03, -2.98259055e-03, -2.98259055e-03,
             -2.98259055e-03, -8.28383437e-03, -8.28383437e-03,
             -2.98259055e-03, -2.98259055e-03, -2.98259055e-03,
             -2.98259055e-03,  1.38303339e-02,  1.38303339e-02,
              1.38303339e-02,  1.38303339e-02,  1.84205088e-01,
             -2.62086255e-02, -2.62086255e-02, -6.87202717e-03,
             -6.87202717e-03,  3.78058389e-04, -3.78058389e-04,
             -3.78058389e-04,  3.78058389e-04,  1.55327487e-03,
              1.55327487e-03, -1.55327487e-03, -1.55327487e-03,
              1.42435706e-01, -2.07446737e-03, -2.07446737e-03,
             -1.28984054e-02, -1.28984054e-02, -7.61100328e-03,
              7.61100328e-03, -4.83240462e-03,  4.83240462e-03,
             -4.83240462e-03,  4.83240462e-03,  4.62268048e-03,
             -4.62268048e-03,  4.62268048e-03, -4.62268048e-03,
              1.49162161e-01, -1.79972769e-02, -1.79972769e-02,
              7.52291984e-03,  7.52291984e-03,  1.84205088e-01,
              3.78058389e-04, -3.78058389e-04, -3.78058389e-04,
              3.78058389e-04,  1.55327487e-03, -1.55327487e-03,
              1.55327487e-03, -1.55327487e-03, -2.62086255e-02,
             -2.62086255e-02, -6.87202717e-03, -6.87202717e-03,
              2.20039773e-01,  8.79524778e-03, -8.79524778e-03,
             -8.79524778e-03,  8.79524778e-03,  1.94041229e-03,
             -1.94041229e-03,  1.94041229e-03, -1.94041229e-03,
              1.94041229e-03,  1.94041229e-03, -1.94041229e-03,
             -1.94041229e-03,  7.27248350e-03,  7.27248350e-03,
              7.27248350e-03,  7.27248350e-03,  1.48398289e-01,
             -2.49128972e-03,  2.49128972e-03,  1.57353949e-01,
              1.42435706e-01, -7.61100328e-03,  7.61100328e-03,
             -2.07446737e-03, -2.07446737e-03, -4.83240462e-03,
             -4.83240462e-03,  4.83240462e-03,  4.83240462e-03,
             -1.28984054e-02, -1.28984054e-02,  4.62268048e-03,
              4.62268048e-03, -4.62268048e-03, -4.62268048e-03,
              1.48398289e-01, -2.49128972e-03,  2.49128972e-03,
              1.66880970e-01, -1.11105981e-02,  1.11105981e-02,
             -1.11105981e-02,  1.11105981e-02,  1.15013193e-02,
             -1.15013193e-02, -1.15013193e-02,  1.15013193e-02,
              1.25594756e-01,  9.43929717e-03, -9.43929717e-03,
              1.49162161e-01, -1.79972769e-02, -1.79972769e-02,
              7.52291984e-03,  7.52291984e-03,  1.57353949e-01,
              1.25594756e-01,  9.43929717e-03, -9.43929717e-03,
              1.73958313e-01,  2.83875887e-01,  2.73621236e-01,
             -1.96710720e-04,  1.96710720e-04,  2.74934867e-01,
              3.96672647e-03,  3.96672647e-03,  3.59301309e-04,
             -3.59301309e-04
       \end{itemize} 
   \item[\#12] 
       \begin{itemize}
            \item \textbf{Geometry:}\\
            O 0.0 0.0 0.0\\
            H 0.0 0.0 0.9\\
            H 0.0 0.7790781047025562 -0.18176167575446953
            \item \textbf{Pauli strings:}\\
            'IIIIII', 'IIIIIZ', 'IIIZXX', 'IIIIYY', 'IIIXXX', 'IIIYXY', 'IIIIZZ', 'IIIZZI', 'IIIXZI', 'IIIXII', 'IIIZII', 'IIZIII', 'ZXXIII', 'IYYIII', 'XXXIII', 'YXYIII', 'IZZIII', 'ZZIIII', 'XZIIII', 'XIIIII', 'ZIIIII', 'IIIIZI', 'IIIZZZ', 'IIIXZZ', 'IIIXIZ', 'IIIZIZ', 'IIIZYY', 'IIIIXX', 'IIIXYY', 'IIIYYX', 'IIZIIZ', 'IIZZXX', 'IIZIYY', 'IIZXXX', 'IIZYXY', 'IZXIZX', 'IIXIZX', 'IZXIIX', 'IIXIIX', 'ZXXIIZ', 'IYYIIZ', 'ZXXZXX', 'IYYZXX', 'ZXXIYY', 'IYYIYY', 'ZXXXXX', 'IYYXXX', 'ZXXYXY', 'IYYYXY', 'XXXIIZ', 'YXYIIZ', 'XXXZXX', 'YXYZXX', 'XXXIYY', 'YXYIYY', 'XXXXXX', 'YXYXXX', 'XXXYXY', 'YXYYXY', 'IZZIIZ', 'IZZZXX', 'IZZIYY', 'IZZXXX', 'IZZYXY', 'ZXZIZX', 'IXIIZX', 'ZXZIIX', 'IXIIIX', 'XXZIZX', 'YYIIZX', 'XXZIIX', 'YYIIIX', 'ZZIIIZ', 'ZZIZXX', 'ZZIIYY', 'ZZIXXX', 'ZZIYXY', 'XZIIIZ', 'XIIIIZ', 'XZIZXX', 'XIIZXX', 'XZIIYY', 'XIIIYY', 'XZIXXX', 'XIIXXX', 'XZIYXY', 'XIIYXY', 'ZIIIIZ', 'ZIIZXX', 'ZIIIYY', 'ZIIXXX', 'ZIIYXY', 'IIZIZZ', 'IZXZXZ', 'IIXZXZ', 'IZXIXI', 'IIXIXI', 'IZXXXZ', 'IIXXXZ', 'IZXYYI', 'IIXYYI', 'ZXXIZZ', 'IYYIZZ', 'XXXIZZ', 'YXYIZZ', 'IZZIZZ', 'ZXZZXZ', 'IXIZXZ', 'ZXZIXI', 'IXIIXI', 'ZXZXXZ', 'IXIXXZ', 'ZXZYYI', 'IXIYYI', 'XXZZXZ', 'YYIZXZ', 'XXZIXI', 'YYIIXI', 'XXZXXZ', 'YYIXXZ', 'XXZYYI', 'YYIYYI', 'ZZIIZZ', 'XZIIZZ', 'XIIIZZ', 'ZIIIZZ', 'IIZZZI', 'IIZXZI', 'IIZXII', 'ZXXZZI', 'IYYZZI', 'ZXXXZI', 'IYYXZI', 'ZXXXII', 'IYYXII', 'XXXZZI', 'YXYZZI', 'XXXXZI', 'YXYXZI', 'XXXXII', 'YXYXII', 'IZZZZI', 'IZZXZI', 'IZZXII', 'ZZIZZI', 'ZZIXZI', 'ZZIXII', 'XZIZZI', 'XIIZZI', 'XZIXZI', 'XIIXZI', 'XZIXII', 'XIIXII', 'ZIIZZI', 'ZIIXZI', 'ZIIXII', 'IIZZII', 'ZXXZII', 'IYYZII', 'XXXZII', 'YXYZII', 'IZZZII', 'ZZIZII', 'XZIZII', 'XIIZII', 'ZIIZII', 'IZIIII', 'ZZZIII', 'XZZIII', 'XIZIII', 'ZIZIII', 'ZYYIII', 'IXXIII', 'XYYIII', 'YYXIII'
            \item \textbf{Coefficients:}\\
            -4.21057367e+00,  2.44869776e-01, -3.33651883e-02,
             -3.33651883e-02, -7.41037699e-03, -7.41037699e-03,
              1.82416184e-01, -2.04870177e-01, -1.36589245e-02,
              1.36589245e-02, -2.64959136e-01,  2.44869776e-01,
             -3.33651883e-02, -3.33651883e-02, -7.41037699e-03,
             -7.41037699e-03,  1.82416184e-01, -2.04870177e-01,
             -1.36589245e-02,  1.36589245e-02, -2.64959136e-01,
              2.87816187e-01,  2.77906746e-01, -5.69520436e-05,
              5.69520436e-05,  2.78525177e-01,  2.69690098e-03,
              2.69690098e-03,  2.63890432e-04, -2.63890432e-04,
              2.08829315e-01, -2.59130771e-02, -2.59130771e-02,
             -6.24290272e-03, -6.24290272e-03,  1.51616570e-02,
             -1.51616570e-02, -1.51616570e-02,  1.51616570e-02,
             -2.59130771e-02, -2.59130771e-02,  1.42527312e-02,
              1.42527312e-02,  1.42527312e-02,  1.42527312e-02,
             -2.44178434e-03, -2.44178434e-03, -2.44178434e-03,
             -2.44178434e-03, -6.24290272e-03, -6.24290272e-03,
             -2.44178434e-03, -2.44178434e-03, -2.44178434e-03,
             -2.44178434e-03,  1.56938226e-02,  1.56938226e-02,
              1.56938226e-02,  1.56938226e-02,  1.87569325e-01,
             -2.27729083e-02, -2.27729083e-02, -5.09961455e-03,
             -5.09961455e-03,  1.49726453e-03, -1.49726453e-03,
             -1.49726453e-03,  1.49726453e-03,  1.11196910e-03,
              1.11196910e-03, -1.11196910e-03, -1.11196910e-03,
              1.40148487e-01, -5.74364504e-03, -5.74364504e-03,
             -1.07521443e-02, -1.07521443e-02, -6.17546799e-03,
              6.17546799e-03, -4.80445109e-03,  4.80445109e-03,
             -4.80445109e-03,  4.80445109e-03,  9.29907678e-03,
             -9.29907678e-03,  9.29907678e-03, -9.29907678e-03,
              1.54406860e-01, -1.22741951e-02, -1.22741951e-02,
              6.55605548e-03,  6.55605548e-03,  1.87569325e-01,
              1.49726453e-03, -1.49726453e-03, -1.49726453e-03,
              1.49726453e-03,  1.11196910e-03, -1.11196910e-03,
              1.11196910e-03, -1.11196910e-03, -2.27729083e-02,
             -2.27729083e-02, -5.09961455e-03, -5.09961455e-03,
              2.20039773e-01,  9.81225660e-03, -9.81225660e-03,
             -9.81225660e-03,  9.81225660e-03,  1.60476046e-03,
             -1.60476046e-03,  1.60476046e-03, -1.60476046e-03,
              1.60476046e-03,  1.60476046e-03, -1.60476046e-03,
             -1.60476046e-03,  6.51355519e-03,  6.51355519e-03,
              6.51355519e-03,  6.51355519e-03,  1.49624396e-01,
             -2.07197114e-03,  2.07197114e-03,  1.58524545e-01,
              1.40148487e-01, -6.17546799e-03,  6.17546799e-03,
             -5.74364504e-03, -5.74364504e-03, -4.80445109e-03,
             -4.80445109e-03,  4.80445109e-03,  4.80445109e-03,
             -1.07521443e-02, -1.07521443e-02,  9.29907678e-03,
              9.29907678e-03, -9.29907678e-03, -9.29907678e-03,
              1.49624396e-01, -2.07197114e-03,  2.07197114e-03,
              1.60702492e-01, -1.24178688e-02,  1.24178688e-02,
             -1.24178688e-02,  1.24178688e-02,  1.91783391e-02,
             -1.91783391e-02, -1.91783391e-02,  1.91783391e-02,
              1.34586859e-01,  1.04189127e-02, -1.04189127e-02,
              1.54406860e-01, -1.22741951e-02, -1.22741951e-02,
              6.55605548e-03,  6.55605548e-03,  1.58524545e-01,
              1.34586859e-01,  1.04189127e-02, -1.04189127e-02,
              1.67871626e-01,  2.87816187e-01,  2.77906746e-01,
             -5.69520436e-05,  5.69520436e-05,  2.78525177e-01,
              2.69690098e-03,  2.69690098e-03,  2.63890432e-04,
             -2.63890432e-04
       \end{itemize}
    \item[\#13] 
       \begin{itemize}
            \item \textbf{Geometry:}\\
            O 0.0 0.0 0.0\\
            H 0.0 0.0 0.9\\
            H 0.0 0.8764628677903757 -0.2044818852237782
            \item \textbf{Pauli strings:}\\
            'IIIIII', 'IIIIIZ', 'IIIZXX', 'IIIIYY', 'IIIIZZ', 'IIIZZI', 'IIIZII', 'IIZIII', 'ZXXIII', 'IYYIII', 'IZZIII', 'ZZIIII', 'ZIIIII', 'IIIIZI', 'IIIZZZ', 'IIIZIZ', 'IIIZYY', 'IIIIXX', 'IIZIIZ', 'IIZZXX', 'IIZIYY', 'IZXIZX', 'IIXIZX', 'IZXIIX', 'IIXIIX', 'ZXXIIZ', 'IYYIIZ', 'ZXXZXX', 'IYYZXX', 'ZXXIYY', 'IYYIYY', 'XXXXXX', 'YXYXXX', 'XXXYXY', 'YXYYXY', 'IZZIIZ', 'IZZZXX', 'IZZIYY', 'ZXZIZX', 'IXIIZX', 'ZXZIIX', 'IXIIIX', 'ZZIIIZ', 'ZZIZXX', 'ZZIIYY', 'XZIXXX', 'XIIXXX', 'XZIYXY', 'XIIYXY', 'ZIIIIZ', 'ZIIZXX', 'ZIIIYY', 'IIZIZZ', 'IZXZXZ', 'IIXZXZ', 'IZXIXI', 'IIXIXI', 'ZXXIZZ', 'IYYIZZ', 'IZZIZZ', 'ZXZZXZ', 'IXIZXZ', 'ZXZIXI', 'IXIIXI', 'XXZXXZ', 'YYIXXZ', 'XXZYYI', 'YYIYYI', 'ZZIIZZ', 'ZIIIZZ', 'IIZZZI', 'ZXXZZI', 'IYYZZI', 'XXXXZI', 'YXYXZI', 'XXXXII', 'YXYXII', 'IZZZZI', 'ZZIZZI', 'XZIXZI', 'XIIXZI', 'XZIXII', 'XIIXII', 'ZIIZZI', 'IIZZII', 'ZXXZII', 'IYYZII', 'IZZZII', 'ZZIZII', 'ZIIZII', 'IZIIII', 'ZZZIII', 'ZIZIII', 'ZYYIII', 'IXXIII'
            \item \textbf{Coefficients:}\\
            -4.26269975e+00,  2.37708799e-01,  3.84757998e-02,
              3.84757998e-02,  1.78139352e-01, -1.93823752e-01,
             -2.18184208e-01,  2.37708799e-01,  3.84757998e-02,
              3.84757998e-02,  1.78139352e-01, -1.93823752e-01,
             -2.18184208e-01,  2.84950857e-01,  2.74542158e-01,
              2.75874855e-01, -3.62075968e-03, -3.62075968e-03,
              2.02852995e-01,  2.86751841e-02,  2.86751841e-02,
              1.46874937e-02, -1.46874937e-02, -1.46874937e-02,
              1.46874937e-02,  2.86751841e-02,  2.86751841e-02,
              1.48678241e-02,  1.48678241e-02,  1.48678241e-02,
              1.48678241e-02,  1.71821446e-02,  1.71821446e-02,
              1.71821446e-02,  1.71821446e-02,  1.85217056e-01,
              2.59814913e-02,  2.59814913e-02, -1.16100144e-03,
              1.16100144e-03,  1.16100144e-03, -1.16100144e-03,
              1.37802660e-01,  1.04803606e-02,  1.04803606e-02,
             -1.42003541e-02,  1.42003541e-02, -1.42003541e-02,
              1.42003541e-02,  1.54534894e-01,  9.32137901e-03,
              9.32137901e-03,  1.85217056e-01, -1.16100144e-03,
              1.16100144e-03,  1.16100144e-03, -1.16100144e-03,
              2.59814913e-02,  2.59814913e-02,  2.20039773e-01,
              1.00724287e-02, -1.00724287e-02, -1.00724287e-02,
              1.00724287e-02,  5.96185659e-03,  5.96185659e-03,
              5.96185659e-03,  5.96185659e-03,  1.48594535e-01,
              1.57569179e-01,  1.37802660e-01,  1.04803606e-02,
              1.04803606e-02, -1.42003541e-02, -1.42003541e-02,
              1.42003541e-02,  1.42003541e-02,  1.48594535e-01,
              1.50591472e-01,  2.87438076e-02, -2.87438076e-02,
             -2.87438076e-02,  2.87438076e-02,  1.43165102e-01,
              1.54534894e-01,  9.32137901e-03,  9.32137901e-03,
              1.57569179e-01,  1.43165102e-01,  1.57032539e-01,
              2.84950857e-01,  2.74542158e-01,  2.75874855e-01,
             -3.62075968e-03, -3.62075968e-03
       \end{itemize} 
   \item[\#14] 
       \begin{itemize}
            \item \textbf{Geometry:}\\
            O 0.0 0.0 0.0\\
            H 0.0 0.0 0.9\\
            H 0.0 0.9738476308781953 -0.2272020946930869
            \item \textbf{Pauli strings:}\\
            'IIIIII', 'IIIIIZ', 'IIIZXX', 'IIIIYY', 'IIIXXX', 'IIIYXY', 'IIIIZZ', 'IIIZZI', 'IIIXZI', 'IIIXII', 'IIIZII', 'IIZIII', 'ZXXIII', 'IYYIII', 'XXXIII', 'YXYIII', 'IZZIII', 'ZZIIII', 'XZIIII', 'XIIIII', 'ZIIIII', 'IIIIZI', 'IIIZZZ', 'IIIXZZ', 'IIIXIZ', 'IIIZIZ', 'IIIZYY', 'IIIIXX', 'IIIXYY', 'IIIYYX', 'IIZIIZ', 'IIZZXX', 'IIZIYY', 'IIZXXX', 'IIZYXY', 'IZXIZX', 'IIXIZX', 'IZXIIX', 'IIXIIX', 'ZXXIIZ', 'IYYIIZ', 'ZXXZXX', 'IYYZXX', 'ZXXIYY', 'IYYIYY', 'ZXXXXX', 'IYYXXX', 'ZXXYXY', 'IYYYXY', 'XXXIIZ', 'YXYIIZ', 'XXXZXX', 'YXYZXX', 'XXXIYY', 'YXYIYY', 'XXXXXX', 'YXYXXX', 'XXXYXY', 'YXYYXY', 'IZZIIZ', 'IZZZXX', 'IZZIYY', 'IZZXXX', 'IZZYXY', 'ZXZIZX', 'IXIIZX', 'ZXZIIX', 'IXIIIX', 'XXZIZX', 'YYIIZX', 'XXZIIX', 'YYIIIX', 'ZZIIIZ', 'ZZIZXX', 'ZZIIYY', 'ZZIXXX', 'ZZIYXY', 'XZIIIZ', 'XIIIIZ', 'XZIZXX', 'XIIZXX', 'XZIIYY', 'XIIIYY', 'XZIXXX', 'XIIXXX', 'XZIYXY', 'XIIYXY', 'ZIIIIZ', 'ZIIZXX', 'ZIIIYY', 'ZIIXXX', 'ZIIYXY', 'IIZIZZ', 'IZXZXZ', 'IIXZXZ', 'IZXIXI', 'IIXIXI', 'IZXXXZ', 'IIXXXZ', 'IZXYYI', 'IIXYYI', 'ZXXIZZ', 'IYYIZZ', 'XXXIZZ', 'YXYIZZ', 'IZZIZZ', 'ZXZZXZ', 'IXIZXZ', 'ZXZIXI', 'IXIIXI', 'ZXZXXZ', 'IXIXXZ', 'ZXZYYI', 'IXIYYI', 'XXZZXZ', 'YYIZXZ', 'XXZIXI', 'YYIIXI', 'XXZXXZ', 'YYIXXZ', 'XXZYYI', 'YYIYYI', 'ZZIIZZ', 'XZIIZZ', 'XIIIZZ', 'ZIIIZZ', 'IIZZZI', 'IIZXZI', 'IIZXII', 'ZXXZZI', 'IYYZZI', 'ZXXXZI', 'IYYXZI', 'ZXXXII', 'IYYXII', 'XXXZZI', 'YXYZZI', 'XXXXZI', 'YXYXZI', 'XXXXII', 'YXYXII', 'IZZZZI', 'IZZXZI', 'IZZXII', 'ZZIZZI', 'ZZIXZI', 'ZZIXII', 'XZIZZI', 'XIIZZI', 'XZIXZI', 'XIIXZI', 'XZIXII', 'XIIXII', 'ZIIZZI', 'ZIIXZI', 'ZIIXII', 'IIZZII', 'ZXXZII', 'IYYZII', 'XXXZII', 'YXYZII', 'IZZZII', 'ZZIZII', 'XZIZII', 'XIIZII', 'ZIIZII', 'IZIIII', 'ZZZIII', 'XZZIII', 'XIZIII', 'ZIZIII', 'ZYYIII', 'IXXIII', 'XYYIII', 'YYXIII'
            \item \textbf{Coefficients:}\\
            -4.28992411e+00,  2.29665346e-01, -4.15944129e-02,
             -4.15944129e-02, -9.22038862e-03, -9.22038862e-03,
              1.74884246e-01, -1.55551731e-01, -1.30098547e-02,
              1.30098547e-02, -2.05740777e-01,  2.29665346e-01,
             -4.15944129e-02, -4.15944129e-02, -9.22038862e-03,
             -9.22038862e-03,  1.74884246e-01, -1.55551731e-01,
             -1.30098547e-02,  1.30098547e-02, -2.05740777e-01,
              2.81258739e-01,  2.70501923e-01, -3.90139842e-04,
              3.90139842e-04,  2.72491601e-01,  4.73232640e-03,
              4.73232640e-03,  6.28177510e-04, -6.28177510e-04,
              1.95541176e-01, -2.94771341e-02, -2.94771341e-02,
             -6.95723679e-03, -6.95723679e-03,  1.40230979e-02,
             -1.40230979e-02, -1.40230979e-02,  1.40230979e-02,
             -2.94771341e-02, -2.94771341e-02,  1.87519036e-02,
              1.87519036e-02,  1.87519036e-02,  1.87519036e-02,
             -1.98412698e-03, -1.98412698e-03, -1.98412698e-03,
             -1.98412698e-03, -6.95723679e-03, -6.95723679e-03,
             -1.98412698e-03, -1.98412698e-03, -1.98412698e-03,
             -1.98412698e-03,  1.59048816e-02,  1.59048816e-02,
              1.59048816e-02,  1.59048816e-02,  1.82081900e-01,
             -2.85765553e-02, -2.85765553e-02, -5.95484978e-03,
             -5.95484978e-03,  2.49759133e-04, -2.49759133e-04,
             -2.49759133e-04,  2.49759133e-04,  7.82871324e-04,
              7.82871324e-04, -7.82871324e-04, -7.82871324e-04,
              1.39803785e-01, -6.67209449e-03, -6.67209449e-03,
             -1.11507301e-02, -1.11507301e-02, -5.40878413e-03,
              5.40878413e-03, -5.04118655e-03,  5.04118655e-03,
             -5.04118655e-03,  5.04118655e-03,  9.57417574e-03,
             -9.57417574e-03,  9.57417574e-03, -9.57417574e-03,
              1.49736031e-01, -1.45198123e-02, -1.45198123e-02,
              6.83083830e-03,  6.83083830e-03,  1.82081900e-01,
              2.49759133e-04, -2.49759133e-04, -2.49759133e-04,
              2.49759133e-04,  7.82871324e-04, -7.82871324e-04,
              7.82871324e-04, -7.82871324e-04, -2.85765553e-02,
             -2.85765553e-02, -5.95484978e-03, -5.95484978e-03,
              2.20039773e-01,  9.17215967e-03, -9.17215967e-03,
             -9.17215967e-03,  9.17215967e-03,  1.29308324e-03,
             -1.29308324e-03,  1.29308324e-03, -1.29308324e-03,
              1.29308324e-03,  1.29308324e-03, -1.29308324e-03,
             -1.29308324e-03,  6.58094717e-03,  6.58094717e-03,
              6.58094717e-03,  6.58094717e-03,  1.47832611e-01,
             -1.74143407e-03,  1.74143407e-03,  1.56030988e-01,
              1.39803785e-01, -5.40878413e-03,  5.40878413e-03,
             -6.67209449e-03, -6.67209449e-03, -5.04118655e-03,
             -5.04118655e-03,  5.04118655e-03,  5.04118655e-03,
             -1.11507301e-02, -1.11507301e-02,  9.57417574e-03,
              9.57417574e-03, -9.57417574e-03, -9.57417574e-03,
              1.47832611e-01, -1.74143407e-03,  1.74143407e-03,
              1.59020151e-01, -1.18218703e-02,  1.18218703e-02,
             -1.18218703e-02,  1.18218703e-02,  1.90169592e-02,
             -1.90169592e-02, -1.90169592e-02,  1.90169592e-02,
              1.32216897e-01,  1.12223332e-02, -1.12223332e-02,
              1.49736031e-01, -1.45198123e-02, -1.45198123e-02,
              6.83083830e-03,  6.83083830e-03,  1.56030988e-01,
              1.32216897e-01,  1.12223332e-02, -1.12223332e-02,
              1.64899892e-01,  2.81258739e-01,  2.70501923e-01,
             -3.90139842e-04,  3.90139842e-04,  2.72491601e-01,
              4.73232640e-03,  4.73232640e-03,  6.28177510e-04,
             -6.28177510e-04
       \end{itemize}
   \item[\#15] 
       \begin{itemize}
            \item \textbf{Geometry:}\\
            O 0.0 0.0 0.0\\
            H 0.0 0.0 1.0\\
            H 0.0 0.7790781047025562 -0.18176167575446953
            \item \textbf{Pauli strings:}\\
            'IIIIII', 'IIIIIZ', 'IIIZXX', 'IIIIYY', 'IIIXXX', 'IIIYXY', 'IIIIZZ', 'IIIZZI', 'IIIXZI', 'IIIXII', 'IIIZII', 'IIZIII', 'ZXXIII', 'IYYIII', 'XXXIII', 'YXYIII', 'IZZIII', 'ZZIIII', 'XZIIII', 'XIIIII', 'ZIIIII', 'IIIIZI', 'IIIZZZ', 'IIIXZZ', 'IIIXIZ', 'IIIZIZ', 'IIIZYY', 'IIIIXX', 'IIIXYY', 'IIIYYX', 'IIZIIZ', 'IIZZXX', 'IIZIYY', 'IIZXXX', 'IIZYXY', 'IZXIZX', 'IIXIZX', 'IZXIIX', 'IIXIIX', 'ZXXIIZ', 'IYYIIZ', 'ZXXZXX', 'IYYZXX', 'ZXXIYY', 'IYYIYY', 'ZXXXXX', 'IYYXXX', 'ZXXYXY', 'IYYYXY', 'XXXIIZ', 'YXYIIZ', 'XXXZXX', 'YXYZXX', 'XXXIYY', 'YXYIYY', 'XXXXXX', 'YXYXXX', 'XXXYXY', 'YXYYXY', 'IZZIIZ', 'IZZZXX', 'IZZIYY', 'IZZXXX', 'IZZYXY', 'ZXZIZX', 'IXIIZX', 'ZXZIIX', 'IXIIIX', 'XXZIZX', 'YYIIZX', 'XXZIIX', 'YYIIIX', 'ZZIIIZ', 'ZZIZXX', 'ZZIIYY', 'ZZIXXX', 'ZZIYXY', 'XZIIIZ', 'XIIIIZ', 'XZIZXX', 'XIIZXX', 'XZIIYY', 'XIIIYY', 'XZIXXX', 'XIIXXX', 'XZIYXY', 'XIIYXY', 'ZIIIIZ', 'ZIIZXX', 'ZIIIYY', 'ZIIXXX', 'ZIIYXY', 'IIZIZZ', 'IZXZXZ', 'IIXZXZ', 'IZXIXI', 'IIXIXI', 'IZXXXZ', 'IIXXXZ', 'IZXYYI', 'IIXYYI', 'ZXXIZZ', 'IYYIZZ', 'XXXIZZ', 'YXYIZZ', 'IZZIZZ', 'ZXZZXZ', 'IXIZXZ', 'ZXZIXI', 'IXIIXI', 'ZXZXXZ', 'IXIXXZ', 'ZXZYYI', 'IXIYYI', 'XXZZXZ', 'YYIZXZ', 'XXZIXI', 'YYIIXI', 'XXZXXZ', 'YYIXXZ', 'XXZYYI', 'YYIYYI', 'ZZIIZZ', 'XZIIZZ', 'XIIIZZ', 'ZIIIZZ', 'IIZZZI', 'IIZXZI', 'IIZXII', 'ZXXZZI', 'IYYZZI', 'ZXXXZI', 'IYYXZI', 'ZXXXII', 'IYYXII', 'XXXZZI', 'YXYZZI', 'XXXXZI', 'YXYXZI', 'XXXXII', 'YXYXII', 'IZZZZI', 'IZZXZI', 'IZZXII', 'ZZIZZI', 'ZZIXZI', 'ZZIXII', 'XZIZZI', 'XIIZZI', 'XZIXZI', 'XIIXZI', 'XZIXII', 'XIIXII', 'ZIIZZI', 'ZIIXZI', 'ZIIXII', 'IIZZII', 'ZXXZII', 'IYYZII', 'XXXZII', 'YXYZII', 'IZZZII', 'ZZIZII', 'XZIZII', 'XIIZII', 'ZIIZII', 'IZIIII', 'ZZZIII', 'XZZIII', 'XIZIII', 'ZIZIII', 'ZYYIII', 'IXXIII', 'XYYIII', 'YYXIII'
            \item \textbf{Coefficients:}\\
            -4.23064680e+00,  2.35712842e-01, -3.76125834e-02,
             -3.76125834e-02, -1.01397121e-02, -1.01397121e-02,
              1.78637163e-01, -1.55268782e-01, -1.74910994e-02,
              1.74910994e-02, -2.64135521e-01,  2.35712842e-01,
             -3.76125834e-02, -3.76125834e-02, -1.01397121e-02,
             -1.01397121e-02,  1.78637163e-01, -1.55268782e-01,
             -1.74910994e-02,  1.74910994e-02, -2.64135521e-01,
              2.83875887e-01,  2.73621236e-01, -1.96710720e-04,
              1.96710720e-04,  2.74934867e-01,  3.96672647e-03,
              3.96672647e-03,  3.59301309e-04, -3.59301309e-04,
              2.00712467e-01, -2.75089457e-02, -2.75089457e-02,
             -8.28383437e-03, -8.28383437e-03,  1.44226373e-02,
             -1.44226373e-02, -1.44226373e-02,  1.44226373e-02,
             -2.75089457e-02, -2.75089457e-02,  1.88959351e-02,
              1.88959351e-02,  1.88959351e-02,  1.88959351e-02,
             -2.98259055e-03, -2.98259055e-03, -2.98259055e-03,
             -2.98259055e-03, -8.28383437e-03, -8.28383437e-03,
             -2.98259055e-03, -2.98259055e-03, -2.98259055e-03,
             -2.98259055e-03,  1.38303339e-02,  1.38303339e-02,
              1.38303339e-02,  1.38303339e-02,  1.84205088e-01,
             -2.62086255e-02, -2.62086255e-02, -6.87202717e-03,
             -6.87202717e-03,  3.78058389e-04, -3.78058389e-04,
             -3.78058389e-04,  3.78058389e-04,  1.55327487e-03,
              1.55327487e-03, -1.55327487e-03, -1.55327487e-03,
              1.42435706e-01, -2.07446737e-03, -2.07446737e-03,
             -1.28984054e-02, -1.28984054e-02, -7.61100328e-03,
              7.61100328e-03, -4.83240462e-03,  4.83240462e-03,
             -4.83240462e-03,  4.83240462e-03,  4.62268048e-03,
             -4.62268048e-03,  4.62268048e-03, -4.62268048e-03,
              1.49162161e-01, -1.79972769e-02, -1.79972769e-02,
              7.52291984e-03,  7.52291984e-03,  1.84205088e-01,
              3.78058389e-04, -3.78058389e-04, -3.78058389e-04,
              3.78058389e-04,  1.55327487e-03, -1.55327487e-03,
              1.55327487e-03, -1.55327487e-03, -2.62086255e-02,
             -2.62086255e-02, -6.87202717e-03, -6.87202717e-03,
              2.20039773e-01,  8.79524778e-03, -8.79524778e-03,
             -8.79524778e-03,  8.79524778e-03,  1.94041229e-03,
             -1.94041229e-03,  1.94041229e-03, -1.94041229e-03,
              1.94041229e-03,  1.94041229e-03, -1.94041229e-03,
             -1.94041229e-03,  7.27248350e-03,  7.27248350e-03,
              7.27248350e-03,  7.27248350e-03,  1.48398289e-01,
             -2.49128972e-03,  2.49128972e-03,  1.57353949e-01,
              1.42435706e-01, -7.61100328e-03,  7.61100328e-03,
             -2.07446737e-03, -2.07446737e-03, -4.83240462e-03,
             -4.83240462e-03,  4.83240462e-03,  4.83240462e-03,
             -1.28984054e-02, -1.28984054e-02,  4.62268048e-03,
              4.62268048e-03, -4.62268048e-03, -4.62268048e-03,
              1.48398289e-01, -2.49128972e-03,  2.49128972e-03,
              1.66880970e-01, -1.11105981e-02,  1.11105981e-02,
             -1.11105981e-02,  1.11105981e-02,  1.15013193e-02,
             -1.15013193e-02, -1.15013193e-02,  1.15013193e-02,
              1.25594756e-01,  9.43929717e-03, -9.43929717e-03,
              1.49162161e-01, -1.79972769e-02, -1.79972769e-02,
              7.52291984e-03,  7.52291984e-03,  1.57353949e-01,
              1.25594756e-01,  9.43929717e-03, -9.43929717e-03,
              1.73958313e-01,  2.83875887e-01,  2.73621236e-01,
             -1.96710720e-04,  1.96710720e-04,  2.74934867e-01,
              3.96672647e-03,  3.96672647e-03,  3.59301309e-04,
             -3.59301309e-04
       \end{itemize} 
   \item[\#16] 
       \begin{itemize}
            \item \textbf{Geometry:}\\
            O 0.0 0.0 0.0\\
            H 0.0 0.0 1.0\\
            H 0.0 0.8764628677903757 -0.2044818852237782
            \item \textbf{Pauli strings:}\\
            'IIIIII', 'IIIIIZ', 'IIIZXX', 'IIIIYY', 'IIIXXX', 'IIIYXY', 'IIIIZZ', 'IIIZZI', 'IIIXZI', 'IIIXII', 'IIIZII', 'IIZIII', 'ZXXIII', 'IYYIII', 'XXXIII', 'YXYIII', 'IZZIII', 'ZZIIII', 'XZIIII', 'XIIIII', 'ZIIIII', 'IIIIZI', 'IIIZZZ', 'IIIXZZ', 'IIIXIZ', 'IIIZIZ', 'IIIZYY', 'IIIIXX', 'IIIXYY', 'IIIYYX', 'IIZIIZ', 'IIZZXX', 'IIZIYY', 'IIZXXX', 'IIZYXY', 'IZXIZX', 'IIXIZX', 'IZXIIX', 'IIXIIX', 'ZXXIIZ', 'IYYIIZ', 'ZXXZXX', 'IYYZXX', 'ZXXIYY', 'IYYIYY', 'ZXXXXX', 'IYYXXX', 'ZXXYXY', 'IYYYXY', 'XXXIIZ', 'YXYIIZ', 'XXXZXX', 'YXYZXX', 'XXXIYY', 'YXYIYY', 'XXXXXX', 'YXYXXX', 'XXXYXY', 'YXYYXY', 'IZZIIZ', 'IZZZXX', 'IZZIYY', 'IZZXXX', 'IZZYXY', 'ZXZIZX', 'IXIIZX', 'ZXZIIX', 'IXIIIX', 'XXZIZX', 'YYIIZX', 'XXZIIX', 'YYIIIX', 'ZZIIIZ', 'ZZIZXX', 'ZZIIYY', 'ZZIXXX', 'ZZIYXY', 'XZIIIZ', 'XIIIIZ', 'XZIZXX', 'XIIZXX', 'XZIIYY', 'XIIIYY', 'XZIXXX', 'XIIXXX', 'XZIYXY', 'XIIYXY', 'ZIIIIZ', 'ZIIZXX', 'ZIIIYY', 'ZIIXXX', 'ZIIYXY', 'IIZIZZ', 'IZXZXZ', 'IIXZXZ', 'IZXIXI', 'IIXIXI', 'IZXXXZ', 'IIXXXZ', 'IZXYYI', 'IIXYYI', 'ZXXIZZ', 'IYYIZZ', 'XXXIZZ', 'YXYIZZ', 'IZZIZZ', 'ZXZZXZ', 'IXIZXZ', 'ZXZIXI', 'IXIIXI', 'ZXZXXZ', 'IXIXXZ', 'ZXZYYI', 'IXIYYI', 'XXZZXZ', 'YYIZXZ', 'XXZIXI', 'YYIIXI', 'XXZXXZ', 'YYIXXZ', 'XXZYYI', 'YYIYYI', 'ZZIIZZ', 'XZIIZZ', 'XIIIZZ', 'ZIIIZZ', 'IIZZZI', 'IIZXZI', 'IIZXII', 'ZXXZZI', 'IYYZZI', 'ZXXXZI', 'IYYXZI', 'ZXXXII', 'IYYXII', 'XXXZZI', 'YXYZZI', 'XXXXZI', 'YXYXZI', 'XXXXII', 'YXYXII', 'IZZZZI', 'IZZXZI', 'IZZXII', 'ZZIZZI', 'ZZIXZI', 'ZZIXII', 'XZIZZI', 'XIIZZI', 'XZIXZI', 'XIIXZI', 'XZIXII', 'XIIXII', 'ZIIZZI', 'ZIIXZI', 'ZIIXII', 'IIZZII', 'ZXXZII', 'IYYZII', 'XXXZII', 'YXYZII', 'IZZZII', 'ZZIZII', 'XZIZII', 'XIIZII', 'ZIIZII', 'IZIIII', 'ZZZIII', 'XZZIII', 'XIZIII', 'ZIZIII', 'ZYYIII', 'IXXIII', 'XYYIII', 'YYXIII'
            \item \textbf{Coefficients:}\\
            -4.28992411e+00,  2.29665346e-01, -4.15944129e-02,
             -4.15944129e-02, -9.22038862e-03, -9.22038862e-03,
              1.74884246e-01, -1.55551731e-01, -1.30098547e-02,
              1.30098547e-02, -2.05740777e-01,  2.29665346e-01,
             -4.15944129e-02, -4.15944129e-02, -9.22038862e-03,
             -9.22038862e-03,  1.74884246e-01, -1.55551731e-01,
             -1.30098547e-02,  1.30098547e-02, -2.05740777e-01,
              2.81258739e-01,  2.70501923e-01, -3.90139842e-04,
              3.90139842e-04,  2.72491601e-01,  4.73232640e-03,
              4.73232640e-03,  6.28177510e-04, -6.28177510e-04,
              1.95541176e-01, -2.94771341e-02, -2.94771341e-02,
             -6.95723679e-03, -6.95723679e-03,  1.40230979e-02,
             -1.40230979e-02, -1.40230979e-02,  1.40230979e-02,
             -2.94771341e-02, -2.94771341e-02,  1.87519036e-02,
              1.87519036e-02,  1.87519036e-02,  1.87519036e-02,
             -1.98412698e-03, -1.98412698e-03, -1.98412698e-03,
             -1.98412698e-03, -6.95723679e-03, -6.95723679e-03,
             -1.98412698e-03, -1.98412698e-03, -1.98412698e-03,
             -1.98412698e-03,  1.59048816e-02,  1.59048816e-02,
              1.59048816e-02,  1.59048816e-02,  1.82081900e-01,
             -2.85765553e-02, -2.85765553e-02, -5.95484978e-03,
             -5.95484978e-03,  2.49759133e-04, -2.49759133e-04,
             -2.49759133e-04,  2.49759133e-04,  7.82871324e-04,
              7.82871324e-04, -7.82871324e-04, -7.82871324e-04,
              1.39803785e-01, -6.67209449e-03, -6.67209449e-03,
             -1.11507301e-02, -1.11507301e-02, -5.40878413e-03,
              5.40878413e-03, -5.04118655e-03,  5.04118655e-03,
             -5.04118655e-03,  5.04118655e-03,  9.57417574e-03,
             -9.57417574e-03,  9.57417574e-03, -9.57417574e-03,
              1.49736031e-01, -1.45198123e-02, -1.45198123e-02,
              6.83083830e-03,  6.83083830e-03,  1.82081900e-01,
              2.49759133e-04, -2.49759133e-04, -2.49759133e-04,
              2.49759133e-04,  7.82871324e-04, -7.82871324e-04,
              7.82871324e-04, -7.82871324e-04, -2.85765553e-02,
             -2.85765553e-02, -5.95484978e-03, -5.95484978e-03,
              2.20039773e-01,  9.17215967e-03, -9.17215967e-03,
             -9.17215967e-03,  9.17215967e-03,  1.29308324e-03,
             -1.29308324e-03,  1.29308324e-03, -1.29308324e-03,
              1.29308324e-03,  1.29308324e-03, -1.29308324e-03,
             -1.29308324e-03,  6.58094717e-03,  6.58094717e-03,
              6.58094717e-03,  6.58094717e-03,  1.47832611e-01,
             -1.74143407e-03,  1.74143407e-03,  1.56030988e-01,
              1.39803785e-01, -5.40878413e-03,  5.40878413e-03,
             -6.67209449e-03, -6.67209449e-03, -5.04118655e-03,
             -5.04118655e-03,  5.04118655e-03,  5.04118655e-03,
             -1.11507301e-02, -1.11507301e-02,  9.57417574e-03,
              9.57417574e-03, -9.57417574e-03, -9.57417574e-03,
              1.47832611e-01, -1.74143407e-03,  1.74143407e-03,
              1.59020151e-01, -1.18218703e-02,  1.18218703e-02,
             -1.18218703e-02,  1.18218703e-02,  1.90169592e-02,
             -1.90169592e-02, -1.90169592e-02,  1.90169592e-02,
              1.32216897e-01,  1.12223332e-02, -1.12223332e-02,
              1.49736031e-01, -1.45198123e-02, -1.45198123e-02,
              6.83083830e-03,  6.83083830e-03,  1.56030988e-01,
              1.32216897e-01,  1.12223332e-02, -1.12223332e-02,
              1.64899892e-01,  2.81258739e-01,  2.70501923e-01,
             -3.90139842e-04,  3.90139842e-04,  2.72491601e-01,
              4.73232640e-03,  4.73232640e-03,  6.28177510e-04,
             -6.28177510e-04
       \end{itemize}
    \item[\#17] 
       \begin{itemize}
            \item \textbf{Geometry:}\\
            O 0.0 0.0 0.0\\
            H 0.0 0.0 1.0\\
            H 0.0 0.9738476308781953 -0.2272020946930869
            \item \textbf{Pauli strings:}\\
            'IIIIII', 'IIIIIZ', 'IIIZXX', 'IIIIYY', 'IIIIZZ', 'IIIZZI', 'IIIZII', 'IIZIII', 'ZXXIII', 'IYYIII', 'IZZIII', 'ZZIIII', 'ZIIIII', 'IIIIZI', 'IIIZZZ', 'IIIZIZ', 'IIIZYY', 'IIIIXX', 'IIZIIZ', 'IIZZXX', 'IIZIYY', 'IZXIZX', 'IIXIZX', 'IZXIIX', 'IIXIIX', 'ZXXIIZ', 'IYYIIZ', 'ZXXZXX', 'IYYZXX', 'ZXXIYY', 'IYYIYY', 'XXXXXX', 'YXYXXX', 'XXXYXY', 'YXYYXY', 'IZZIIZ', 'IZZZXX', 'IZZIYY', 'ZXZIZX', 'IXIIZX', 'ZXZIIX', 'IXIIIX', 'ZZIIIZ', 'ZZIZXX', 'ZZIIYY', 'XZIXXX', 'XIIXXX', 'XZIYXY', 'XIIYXY', 'ZIIIIZ', 'ZIIZXX', 'ZIIIYY', 'IIZIZZ', 'IZXZXZ', 'IIXZXZ', 'IZXIXI', 'IIXIXI', 'ZXXIZZ', 'IYYIZZ', 'IZZIZZ', 'ZXZZXZ', 'IXIZXZ', 'ZXZIXI', 'IXIIXI', 'XXZXXZ', 'YYIXXZ', 'XXZYYI', 'YYIYYI', 'ZZIIZZ', 'ZIIIZZ', 'IIZZZI', 'ZXXZZI', 'IYYZZI', 'XXXXZI', 'YXYXZI', 'XXXXII', 'YXYXII', 'IZZZZI', 'ZZIZZI', 'XZIXZI', 'XIIXZI', 'XZIXII', 'XIIXII', 'ZIIZZI', 'IIZZII', 'ZXXZII', 'IYYZII', 'IZZZII', 'ZZIZII', 'ZIIZII', 'IZIIII', 'ZZZIII', 'ZIZIII', 'ZYYIII', 'IXXIII'
            \item \textbf{Coefficients:}\\
            -4.32967475e+00,  2.23493405e-01,  4.61972308e-02,
              4.61972308e-02,  1.72656671e-01, -1.45434363e-01,
             -1.64508863e-01,  2.23493405e-01,  4.61972308e-02,
              4.61972308e-02,  1.72656671e-01, -1.45434363e-01,
             -1.64508863e-01,  2.78026706e-01,  2.66623639e-01,
              2.69671387e-01, -5.74748642e-03, -5.74748642e-03,
              1.89570522e-01,  3.16036068e-02,  3.16036068e-02,
              1.34959551e-02, -1.34959551e-02, -1.34959551e-02,
              1.34959551e-02,  3.16036068e-02,  3.16036068e-02,
              1.93571994e-02,  1.93571994e-02,  1.93571994e-02,
              1.93571994e-02,  1.75112966e-02,  1.75112966e-02,
              1.75112966e-02,  1.75112966e-02,  1.79470674e-01,
              3.16340843e-02,  3.16340843e-02,  1.04362117e-04,
             -1.04362117e-04, -1.04362117e-04,  1.04362117e-04,
              1.37193194e-01,  1.17370308e-02,  1.17370308e-02,
             -1.47313324e-02,  1.47313324e-02, -1.47313324e-02,
              1.47313324e-02,  1.49787346e-01,  1.10509033e-02,
              1.10509033e-02,  1.79470674e-01,  1.04362117e-04,
             -1.04362117e-04, -1.04362117e-04,  1.04362117e-04,
              3.16340843e-02,  3.16340843e-02,  2.20039773e-01,
              9.32981470e-03, -9.32981470e-03, -9.32981470e-03,
              9.32981470e-03,  6.12483503e-03,  6.12483503e-03,
              6.12483503e-03,  6.12483503e-03,  1.46725152e-01,
              1.54912480e-01,  1.37193194e-01,  1.17370308e-02,
              1.17370308e-02, -1.47313324e-02, -1.47313324e-02,
              1.47313324e-02,  1.47313324e-02,  1.46725152e-01,
              1.48623674e-01,  2.85224356e-02, -2.85224356e-02,
             -2.85224356e-02,  2.85224356e-02,  1.40574422e-01,
              1.49787346e-01,  1.10509033e-02,  1.10509033e-02,
              1.54912480e-01,  1.40574422e-01,  1.53057082e-01,
              2.78026706e-01,  2.66623639e-01,  2.69671387e-01,
             -5.74748642e-03, -5.74748642e-03
       \end{itemize} 
   \item[\#18] 
       \begin{itemize}
            \item \textbf{Geometry:}\\
            O 0.0 0.0 0.0\\
            H 0.0 0.0 0.8\\
            H 0.0 0.7570400701499316 -0.2586316534908027
            \item \textbf{Pauli strings:}\\
            'IIIIII', 'IIIIIZ', 'IIIZXX', 'IIIIYY', 'IIIIZZ', 'IIIZZI', 'IIIZII', 'IIZIII', 'ZXXIII', 'IYYIII', 'IZZIII', 'ZZIIII', 'ZIIIII', 'IIIIZI', 'IIIZZZ', 'IIIZIZ', 'IIIZYY', 'IIIIXX', 'IIZIIZ', 'IIZZXX', 'IIZIYY', 'IZXIZX', 'IIXIZX', 'IZXIIX', 'IIXIIX', 'ZXXIIZ', 'IYYIIZ', 'ZXXZXX', 'IYYZXX', 'ZXXIYY', 'IYYIYY', 'XXXXXX', 'YXYXXX', 'XXXYXY', 'YXYYXY', 'IZZIIZ', 'IZZZXX', 'IZZIYY', 'ZXZIZX', 'IXIIZX', 'ZXZIIX', 'IXIIIX', 'ZZIIIZ', 'ZZIZXX', 'ZZIIYY', 'XZIXXX', 'XIIXXX', 'XZIYXY', 'XIIYXY', 'ZIIIIZ', 'ZIIZXX', 'ZIIIYY', 'IIZIZZ', 'IZXZXZ', 'IIXZXZ', 'IZXIXI', 'IIXIXI', 'ZXXIZZ', 'IYYIZZ', 'IZZIZZ', 'ZXZZXZ', 'IXIZXZ', 'ZXZIXI', 'IXIIXI', 'XXZXXZ', 'YYIXXZ', 'XXZYYI', 'YYIYYI', 'ZZIIZZ', 'ZIIIZZ', 'IIZZZI', 'ZXXZZI', 'IYYZZI', 'XXXXZI', 'YXYXZI', 'XXXXII', 'YXYXII', 'IZZZZI', 'ZZIZZI', 'XZIXZI', 'XIIXZI', 'XZIXII', 'XIIXII', 'ZIIZZI', 'IIZZII', 'ZXXZII', 'IYYZII', 'IZZZII', 'ZZIZII', 'ZIIZII', 'IZIIII', 'ZZZIII', 'ZIZIII', 'ZYYIII', 'IXXIII'
            \item \textbf{Coefficients:}\\
            -4.15445279e+00,  2.46685374e-01,  2.57314065e-02,
              2.57314065e-02,  1.87321891e-01, -2.41795265e-01,
             -2.84634120e-01,  2.46685374e-01,  2.57314065e-02,
              2.57314065e-02,  1.87321891e-01, -2.41795265e-01,
             -2.84634120e-01,  2.90258226e-01,  2.82064937e-01,
              2.81775269e-01, -1.00873465e-03, -1.00873465e-03,
              2.16171801e-01,  2.24052463e-02,  2.24052463e-02,
              1.53185945e-02, -1.53185945e-02, -1.53185945e-02,
              1.53185945e-02,  2.24052463e-02,  2.24052463e-02,
              1.03698612e-02,  1.03698612e-02,  1.03698612e-02,
              1.03698612e-02,  1.61742218e-02,  1.61742218e-02,
              1.61742218e-02,  1.61742218e-02,  1.90545211e-01,
              1.94000876e-02,  1.94000876e-02, -2.40135936e-03,
              2.40135936e-03,  2.40135936e-03, -2.40135936e-03,
              1.37451575e-01,  9.62769507e-03,  9.62769507e-03,
             -1.33377403e-02,  1.33377403e-02, -1.33377403e-02,
              1.33377403e-02,  1.59795220e-01,  7.45497201e-03,
              7.45497201e-03,  1.90545211e-01, -2.40135936e-03,
              2.40135936e-03,  2.40135936e-03, -2.40135936e-03,
              1.94000876e-02,  1.94000876e-02,  2.20039773e-01,
              1.09646717e-02, -1.09646717e-02, -1.09646717e-02,
              1.09646717e-02,  5.83794489e-03,  5.83794489e-03,
              5.83794489e-03,  5.83794489e-03,  1.49118942e-01,
              1.60821167e-01,  1.37451575e-01,  9.62769507e-03,
              9.62769507e-03, -1.33377403e-02, -1.33377403e-02,
              1.33377403e-02,  1.33377403e-02,  1.49118942e-01,
              1.49520250e-01,  2.93150022e-02, -2.93150022e-02,
             -2.93150022e-02,  2.93150022e-02,  1.44346612e-01,
              1.59795220e-01,  7.45497201e-03,  7.45497201e-03,
              1.60821167e-01,  1.44346612e-01,  1.61210881e-01,
              2.90258226e-01,  2.82064937e-01,  2.81775269e-01,
             -1.00873465e-03, -1.00873465e-03
       \end{itemize}
   \item[\#19] 
       \begin{itemize}
            \item \textbf{Geometry:}\\
            O 0.0 0.0 0.0\\
            H 0.0 0.0 0.8\\
            H 0.0 0.8516700789186731 -0.290960610177153
            \item \textbf{Pauli strings:}\\
            'IIIIII', 'IIIIIZ', 'IIIZXX', 'IIIIYY', 'IIIXXX', 'IIIYXY', 'IIIIZZ', 'IIIZZI', 'IIIXZI', 'IIIXII', 'IIIZII', 'IIZIII', 'ZXXIII', 'IYYIII', 'XXXIII', 'YXYIII', 'IZZIII', 'ZZIIII', 'XZIIII', 'XIIIII', 'ZIIIII', 'IIIIZI', 'IIIZZZ', 'IIIXZZ', 'IIIXIZ', 'IIIZIZ', 'IIIZYY', 'IIIIXX', 'IIIXYY', 'IIIYYX', 'IIZIIZ', 'IIZZXX', 'IIZIYY', 'IIZXXX', 'IIZYXY', 'IZXIZX', 'IIXIZX', 'IZXIIX', 'IIXIIX', 'ZXXIIZ', 'IYYIIZ', 'ZXXZXX', 'IYYZXX', 'ZXXIYY', 'IYYIYY', 'ZXXXXX', 'IYYXXX', 'ZXXYXY', 'IYYYXY', 'XXXIIZ', 'YXYIIZ', 'XXXZXX', 'YXYZXX', 'XXXIYY', 'YXYIYY', 'XXXXXX', 'YXYXXX', 'XXXYXY', 'YXYYXY', 'IZZIIZ', 'IZZZXX', 'IZZIYY', 'IZZXXX', 'IZZYXY', 'ZXZIZX', 'IXIIZX', 'ZXZIIX', 'IXIIIX', 'XXZIZX', 'YYIIZX', 'XXZIIX', 'YYIIIX', 'ZZIIIZ', 'ZZIZXX', 'ZZIIYY', 'ZZIXXX', 'ZZIYXY', 'XZIIIZ', 'XIIIIZ', 'XZIZXX', 'XIIZXX', 'XZIIYY', 'XIIIYY', 'XZIXXX', 'XIIXXX', 'XZIYXY', 'XIIYXY', 'ZIIIIZ', 'ZIIZXX', 'ZIIIYY', 'ZIIXXX', 'ZIIYXY', 'IIZIZZ', 'IZXZXZ', 'IIXZXZ', 'IZXIXI', 'IIXIXI', 'IZXXXZ', 'IIXXXZ', 'IZXYYI', 'IIXYYI', 'ZXXIZZ', 'IYYIZZ', 'XXXIZZ', 'YXYIZZ', 'IZZIZZ', 'ZXZZXZ', 'IXIZXZ', 'ZXZIXI', 'IXIIXI', 'ZXZXXZ', 'IXIXXZ', 'ZXZYYI', 'IXIYYI', 'XXZZXZ', 'YYIZXZ', 'XXZIXI', 'YYIIXI', 'XXZXXZ', 'YYIXXZ', 'XXZYYI', 'YYIYYI', 'ZZIIZZ', 'XZIIZZ', 'XIIIZZ', 'ZIIIZZ', 'IIZZZI', 'IIZXZI', 'IIZXII', 'ZXXZZI', 'IYYZZI', 'ZXXXZI', 'IYYXZI', 'ZXXXII', 'IYYXII', 'XXXZZI', 'YXYZZI', 'XXXXZI', 'YXYXZI', 'XXXXII', 'YXYXII', 'IZZZZI', 'IZZXZI', 'IZZXII', 'ZZIZZI', 'ZZIXZI', 'ZZIXII', 'XZIZZI', 'XIIZZI', 'XZIXZI', 'XIIXZI', 'XZIXII', 'XIIXII', 'ZIIZZI', 'ZIIXZI', 'ZIIXII', 'IIZZII', 'ZXXZII', 'IYYZII', 'XXXZII', 'YXYZII', 'IZZZII', 'ZZIZII', 'XZIZII', 'XIIZII', 'ZIIZII', 'IZIIII', 'ZZZIII', 'XZZIII', 'XIZIII', 'ZIZIII', 'ZYYIII', 'IXXIII', 'XYYIII', 'YYXIII'
            \item \textbf{Coefficients:}\\
            -4.20545766e+00,  2.39457798e-01, -3.03870183e-02,
             -3.03870183e-02, -5.50469270e-03, -5.50469270e-03,
              1.82978910e-01, -2.01800682e-01, -1.31690228e-02,
              1.31690228e-02, -2.66476129e-01,  2.39457798e-01,
             -3.03870183e-02, -3.03870183e-02, -5.50469270e-03,
             -5.50469270e-03,  1.82978910e-01, -2.01800682e-01,
             -1.31690228e-02,  1.31690228e-02, -2.66476129e-01,
              2.87373749e-01,  2.78584086e-01,  1.54628390e-05,
             -1.54628390e-05,  2.78893217e-01,  2.00119947e-03,
              2.00119947e-03,  6.95224679e-05, -6.95224679e-05,
              2.10214340e-01, -2.46156495e-02, -2.46156495e-02,
             -4.97251725e-03, -4.97251725e-03,  1.48714915e-02,
             -1.48714915e-02, -1.48714915e-02,  1.48714915e-02,
             -2.46156495e-02, -2.46156495e-02,  1.32260186e-02,
              1.32260186e-02,  1.32260186e-02,  1.32260186e-02,
             -2.24996623e-03, -2.24996623e-03, -2.24996623e-03,
             -2.24996623e-03, -4.97251725e-03, -4.97251725e-03,
             -2.24996623e-03, -2.24996623e-03, -2.24996623e-03,
             -2.24996623e-03,  1.53142811e-02,  1.53142811e-02,
              1.53142811e-02,  1.53142811e-02,  1.88289818e-01,
             -2.20358979e-02, -2.20358979e-02, -4.19417445e-03,
             -4.19417445e-03,  1.55850551e-03, -1.55850551e-03,
             -1.55850551e-03,  1.55850551e-03,  1.01021661e-03,
              1.01021661e-03, -1.01021661e-03, -1.01021661e-03,
              1.39092556e-01, -6.64925506e-03, -6.64925506e-03,
             -9.77990177e-03, -9.77990177e-03, -5.90378859e-03,
              5.90378859e-03, -4.64503318e-03,  4.64503318e-03,
             -4.64503318e-03,  4.64503318e-03,  9.97674368e-03,
             -9.97674368e-03,  9.97674368e-03, -9.97674368e-03,
              1.55663113e-01, -1.16164603e-02, -1.16164603e-02,
              6.04837460e-03,  6.04837460e-03,  1.88289818e-01,
              1.55850551e-03, -1.55850551e-03, -1.55850551e-03,
              1.55850551e-03,  1.01021661e-03, -1.01021661e-03,
              1.01021661e-03, -1.01021661e-03, -2.20358979e-02,
             -2.20358979e-02, -4.19417445e-03, -4.19417445e-03,
              2.20039773e-01,  1.00374481e-02, -1.00374481e-02,
             -1.00374481e-02,  1.00374481e-02,  1.48172960e-03,
             -1.48172960e-03,  1.48172960e-03, -1.48172960e-03,
              1.48172960e-03,  1.48172960e-03, -1.48172960e-03,
             -1.48172960e-03,  6.48432287e-03,  6.48432287e-03,
              6.48432287e-03,  6.48432287e-03,  1.48581833e-01,
             -2.18755560e-03,  2.18755560e-03,  1.59201872e-01,
              1.39092556e-01, -5.90378859e-03,  5.90378859e-03,
             -6.64925506e-03, -6.64925506e-03, -4.64503318e-03,
             -4.64503318e-03,  4.64503318e-03,  4.64503318e-03,
             -9.77990177e-03, -9.77990177e-03,  9.97674368e-03,
              9.97674368e-03, -9.97674368e-03, -9.97674368e-03,
              1.48581833e-01, -2.18755560e-03,  2.18755560e-03,
              1.57970819e-01, -1.24259614e-02,  1.24259614e-02,
             -1.24259614e-02,  1.24259614e-02,  2.14571566e-02,
             -2.14571566e-02, -2.14571566e-02,  2.14571566e-02,
              1.35412579e-01,  1.01805363e-02, -1.01805363e-02,
              1.55663113e-01, -1.16164603e-02, -1.16164603e-02,
              6.04837460e-03,  6.04837460e-03,  1.59201872e-01,
              1.35412579e-01,  1.01805363e-02, -1.01805363e-02,
              1.66727286e-01,  2.87373749e-01,  2.78584086e-01,
              1.54628390e-05, -1.54628390e-05,  2.78893217e-01,
              2.00119947e-03,  2.00119947e-03,  6.95224679e-05,
             -6.95224679e-05
       \end{itemize} 
   \item[\#20] 
       \begin{itemize}
            \item \textbf{Geometry:}\\
            O 0.0 0.0 0.0\\
            H 0.0 0.0 0.8\\
            H 0.0 0.9463000876874145 -0.32328956686350335
            \item \textbf{Pauli strings:}\\
            'IIIIII', 'IIIIIZ', 'IIIZXX', 'IIIIYY', 'IIIXXX', 'IIIYXY', 'IIIIZZ', 'IIIZZI', 'IIIXZI', 'IIIXII', 'IIIZII', 'IIZIII', 'ZXXIII', 'IYYIII', 'XXXIII', 'YXYIII', 'IZZIII', 'ZZIIII', 'XZIIII', 'XIIIII', 'ZIIIII', 'IIIIZI', 'IIIZZZ', 'IIIXZZ', 'IIIXIZ', 'IIIZIZ', 'IIIZYY', 'IIIIXX', 'IIIXYY', 'IIIYYX', 'IIZIIZ', 'IIZZXX', 'IIZIYY', 'IIZXXX', 'IIZYXY', 'IZXIZX', 'IIXIZX', 'IZXIIX', 'IIXIIX', 'ZXXIIZ', 'IYYIIZ', 'ZXXZXX', 'IYYZXX', 'ZXXIYY', 'IYYIYY', 'ZXXXXX', 'IYYXXX', 'ZXXYXY', 'IYYYXY', 'XXXIIZ', 'YXYIIZ', 'XXXZXX', 'YXYZXX', 'XXXIYY', 'YXYIYY', 'XXXXXX', 'YXYXXX', 'XXXYXY', 'YXYYXY', 'IZZIIZ', 'IZZZXX', 'IZZIYY', 'IZZXXX', 'IZZYXY', 'ZXZIZX', 'IXIIZX', 'ZXZIIX', 'IXIIIX', 'XXZIZX', 'YYIIZX', 'XXZIIX', 'YYIIIX', 'ZZIIIZ', 'ZZIZXX', 'ZZIIYY', 'ZZIXXX', 'ZZIYXY', 'XZIIIZ', 'XIIIIZ', 'XZIZXX', 'XIIZXX', 'XZIIYY', 'XIIIYY', 'XZIXXX', 'XIIXXX', 'XZIYXY', 'XIIYXY', 'ZIIIIZ', 'ZIIZXX', 'ZIIIYY', 'ZIIXXX', 'ZIIYXY', 'IIZIZZ', 'IZXZXZ', 'IIXZXZ', 'IZXIXI', 'IIXIXI', 'IZXXXZ', 'IIXXXZ', 'IZXYYI', 'IIXYYI', 'ZXXIZZ', 'IYYIZZ', 'XXXIZZ', 'YXYIZZ', 'IZZIZZ', 'ZXZZXZ', 'IXIZXZ', 'ZXZIXI', 'IXIIXI', 'ZXZXXZ', 'IXIXXZ', 'ZXZYYI', 'IXIYYI', 'XXZZXZ', 'YYIZXZ', 'XXZIXI', 'YYIIXI', 'XXZXXZ', 'YYIXXZ', 'XXZYYI', 'YYIYYI', 'ZZIIZZ', 'XZIIZZ', 'XIIIZZ', 'ZIIIZZ', 'IIZZZI', 'IIZXZI', 'IIZXII', 'ZXXZZI', 'IYYZZI', 'ZXXXZI', 'IYYXZI', 'ZXXXII', 'IYYXII', 'XXXZZI', 'YXYZZI', 'XXXXZI', 'YXYXZI', 'XXXXII', 'YXYXII', 'IZZZZI', 'IZZXZI', 'IZZXII', 'ZZIZZI', 'ZZIXZI', 'ZZIXII', 'XZIZZI', 'XIIZZI', 'XZIXZI', 'XIIXZI', 'XZIXII', 'XIIXII', 'ZIIZZI', 'ZIIXZI', 'ZIIXII', 'IIZZII', 'ZXXZII', 'IYYZII', 'XXXZII', 'YXYZII', 'IZZZII', 'ZZIZII', 'XZIZII', 'XIIZII', 'ZIIZII', 'IZIIII', 'ZZZIII', 'XZZIII', 'XIZIII', 'ZIZIII', 'ZYYIII', 'IXXIII', 'XYYIII', 'YYXIII'
            \item \textbf{Coefficients:}\\
            -4.23118332e+00,  2.31350529e-01, -3.50615360e-02,
             -3.50615360e-02, -8.18077320e-03, -8.18077320e-03,
              1.79621986e-01, -1.53779836e-01, -1.80150757e-02,
              1.80150757e-02, -2.63527519e-01,  2.31350529e-01,
             -3.50615360e-02, -3.50615360e-02, -8.18077320e-03,
             -8.18077320e-03,  1.79621986e-01, -1.53779836e-01,
             -1.80150757e-02,  1.80150757e-02, -2.63527519e-01,
              2.83656064e-01,  2.74378270e-01, -8.92714251e-05,
              8.92714251e-05,  2.75303503e-01,  3.19636059e-03,
              3.19636059e-03,  1.48488475e-04, -1.48488475e-04,
              2.02719601e-01, -2.65917339e-02, -2.65917339e-02,
             -7.05558241e-03, -7.05558241e-03,  1.42365082e-02,
             -1.42365082e-02, -1.42365082e-02,  1.42365082e-02,
             -2.65917339e-02, -2.65917339e-02,  1.74142609e-02,
              1.74142609e-02,  1.74142609e-02,  1.74142609e-02,
             -2.95409903e-03, -2.95409903e-03, -2.95409903e-03,
             -2.95409903e-03, -7.05558241e-03, -7.05558241e-03,
             -2.95409903e-03, -2.95409903e-03, -2.95409903e-03,
             -2.95409903e-03,  1.37254917e-02,  1.37254917e-02,
              1.37254917e-02,  1.37254917e-02,  1.85220492e-01,
             -2.53578825e-02, -2.53578825e-02, -5.99237799e-03,
             -5.99237799e-03,  5.22121772e-04, -5.22121772e-04,
             -5.22121772e-04,  5.22121772e-04,  1.45890109e-03,
              1.45890109e-03, -1.45890109e-03, -1.45890109e-03,
              1.41070328e-01, -2.97671836e-03, -2.97671836e-03,
             -1.24494435e-02, -1.24494435e-02, -7.77346815e-03,
              7.77346815e-03, -5.14665288e-03,  5.14665288e-03,
             -5.14665288e-03,  5.14665288e-03,  5.57597409e-03,
             -5.57597409e-03,  5.57597409e-03, -5.57597409e-03,
              1.50657899e-01, -1.71076696e-02, -1.71076696e-02,
              7.43375953e-03,  7.43375953e-03,  1.85220492e-01,
              5.22121772e-04, -5.22121772e-04, -5.22121772e-04,
              5.22121772e-04,  1.45890109e-03, -1.45890109e-03,
              1.45890109e-03, -1.45890109e-03, -2.53578825e-02,
             -2.53578825e-02, -5.99237799e-03, -5.99237799e-03,
              2.20039773e-01,  9.02767765e-03, -9.02767765e-03,
             -9.02767765e-03,  9.02767765e-03,  1.87433664e-03,
             -1.87433664e-03,  1.87433664e-03, -1.87433664e-03,
              1.87433664e-03,  1.87433664e-03, -1.87433664e-03,
             -1.87433664e-03,  7.21358602e-03,  7.21358602e-03,
              7.21358602e-03,  7.21358602e-03,  1.47398773e-01,
             -2.85576106e-03,  2.85576106e-03,  1.57935788e-01,
              1.41070328e-01, -7.77346815e-03,  7.77346815e-03,
             -2.97671836e-03, -2.97671836e-03, -5.14665288e-03,
             -5.14665288e-03,  5.14665288e-03,  5.14665288e-03,
             -1.24494435e-02, -1.24494435e-02,  5.57597409e-03,
              5.57597409e-03, -5.57597409e-03, -5.57597409e-03,
              1.47398773e-01, -2.85576106e-03,  2.85576106e-03,
              1.64670911e-01, -1.24015263e-02,  1.24015263e-02,
             -1.24015263e-02,  1.24015263e-02,  1.34803342e-02,
             -1.34803342e-02, -1.34803342e-02,  1.34803342e-02,
              1.26152414e-01,  1.02379077e-02, -1.02379077e-02,
              1.50657899e-01, -1.71076696e-02, -1.71076696e-02,
              7.43375953e-03,  7.43375953e-03,  1.57935788e-01,
              1.26152414e-01,  1.02379077e-02, -1.02379077e-02,
              1.72875653e-01,  2.83656064e-01,  2.74378270e-01,
             -8.92714251e-05,  8.92714251e-05,  2.75303503e-01,
              3.19636059e-03,  3.19636059e-03,  1.48488475e-04,
             -1.48488475e-04
       \end{itemize}
    \item[\#21] 
       \begin{itemize}
            \item \textbf{Geometry:}\\
            O 0.0 0.0 0.0\\
            H 0.0 0.0 0.9\\
            H 0.0 0.7570400701499316 -0.2586316534908027
            \item \textbf{Pauli strings:}\\
            'IIIIII', 'IIIIIZ', 'IIIZXX', 'IIIIYY', 'IIIXXX', 'IIIYXY', 'IIIIZZ', 'IIIZZI', 'IIIXZI', 'IIIXII', 'IIIZII', 'IIZIII', 'ZXXIII', 'IYYIII', 'XXXIII', 'YXYIII', 'IZZIII', 'ZZIIII', 'XZIIII', 'XIIIII', 'ZIIIII', 'IIIIZI', 'IIIZZZ', 'IIIXZZ', 'IIIXIZ', 'IIIZIZ', 'IIIZYY', 'IIIIXX', 'IIIXYY', 'IIIYYX', 'IIZIIZ', 'IIZZXX', 'IIZIYY', 'IIZXXX', 'IIZYXY', 'IZXIZX', 'IIXIZX', 'IZXIIX', 'IIXIIX', 'ZXXIIZ', 'IYYIIZ', 'ZXXZXX', 'IYYZXX', 'ZXXIYY', 'IYYIYY', 'ZXXXXX', 'IYYXXX', 'ZXXYXY', 'IYYYXY', 'XXXIIZ', 'YXYIIZ', 'XXXZXX', 'YXYZXX', 'XXXIYY', 'YXYIYY', 'XXXXXX', 'YXYXXX', 'XXXYXY', 'YXYYXY', 'IZZIIZ', 'IZZZXX', 'IZZIYY', 'IZZXXX', 'IZZYXY', 'ZXZIZX', 'IXIIZX', 'ZXZIIX', 'IXIIIX', 'XXZIZX', 'YYIIZX', 'XXZIIX', 'YYIIIX', 'ZZIIIZ', 'ZZIZXX', 'ZZIIYY', 'ZZIXXX', 'ZZIYXY', 'XZIIIZ', 'XIIIIZ', 'XZIZXX', 'XIIZXX', 'XZIIYY', 'XIIIYY', 'XZIXXX', 'XIIXXX', 'XZIYXY', 'XIIYXY', 'ZIIIIZ', 'ZIIZXX', 'ZIIIYY', 'ZIIXXX', 'ZIIYXY', 'IIZIZZ', 'IZXZXZ', 'IIXZXZ', 'IZXIXI', 'IIXIXI', 'IZXXXZ', 'IIXXXZ', 'IZXYYI', 'IIXYYI', 'ZXXIZZ', 'IYYIZZ', 'XXXIZZ', 'YXYIZZ', 'IZZIZZ', 'ZXZZXZ', 'IXIZXZ', 'ZXZIXI', 'IXIIXI', 'ZXZXXZ', 'IXIXXZ', 'ZXZYYI', 'IXIYYI', 'XXZZXZ', 'YYIZXZ', 'XXZIXI', 'YYIIXI', 'XXZXXZ', 'YYIXXZ', 'XXZYYI', 'YYIYYI', 'ZZIIZZ', 'XZIIZZ', 'XIIIZZ', 'ZIIIZZ', 'IIZZZI', 'IIZXZI', 'IIZXII', 'ZXXZZI', 'IYYZZI', 'ZXXXZI', 'IYYXZI', 'ZXXXII', 'IYYXII', 'XXXZZI', 'YXYZZI', 'XXXXZI', 'YXYXZI', 'XXXXII', 'YXYXII', 'IZZZZI', 'IZZXZI', 'IZZXII', 'ZZIZZI', 'ZZIXZI', 'ZZIXII', 'XZIZZI', 'XIIZZI', 'XZIXZI', 'XIIXZI', 'XZIXII', 'XIIXII', 'ZIIZZI', 'ZIIXZI', 'ZIIXII', 'IIZZII', 'ZXXZII', 'IYYZII', 'XXXZII', 'YXYZII', 'IZZZII', 'ZZIZII', 'XZIZII', 'XIIZII', 'ZIIZII', 'IZIIII', 'ZZZIII', 'XZZIII', 'XIZIII', 'ZIZIII', 'ZYYIII', 'IXXIII', 'XYYIII', 'YYXIII'
            \item \textbf{Coefficients:}\\
            -4.20545766e+00,  2.39457798e-01, -3.03870183e-02,
             -3.03870183e-02, -5.50469270e-03, -5.50469270e-03,
              1.82978910e-01, -2.01800682e-01, -1.31690228e-02,
              1.31690228e-02, -2.66476129e-01,  2.39457798e-01,
             -3.03870183e-02, -3.03870183e-02, -5.50469270e-03,
             -5.50469270e-03,  1.82978910e-01, -2.01800682e-01,
             -1.31690228e-02,  1.31690228e-02, -2.66476129e-01,
              2.87373749e-01,  2.78584086e-01,  1.54628390e-05,
             -1.54628390e-05,  2.78893217e-01,  2.00119947e-03,
              2.00119947e-03,  6.95224680e-05, -6.95224680e-05,
              2.10214340e-01, -2.46156495e-02, -2.46156495e-02,
             -4.97251725e-03, -4.97251725e-03,  1.48714915e-02,
             -1.48714915e-02, -1.48714915e-02,  1.48714915e-02,
             -2.46156495e-02, -2.46156495e-02,  1.32260186e-02,
              1.32260186e-02,  1.32260186e-02,  1.32260186e-02,
             -2.24996623e-03, -2.24996623e-03, -2.24996623e-03,
             -2.24996623e-03, -4.97251725e-03, -4.97251725e-03,
             -2.24996623e-03, -2.24996623e-03, -2.24996623e-03,
             -2.24996623e-03,  1.53142811e-02,  1.53142811e-02,
              1.53142811e-02,  1.53142811e-02,  1.88289818e-01,
             -2.20358979e-02, -2.20358979e-02, -4.19417445e-03,
             -4.19417445e-03,  1.55850551e-03, -1.55850551e-03,
             -1.55850551e-03,  1.55850551e-03,  1.01021661e-03,
              1.01021661e-03, -1.01021661e-03, -1.01021661e-03,
              1.39092556e-01, -6.64925506e-03, -6.64925506e-03,
             -9.77990177e-03, -9.77990177e-03, -5.90378859e-03,
              5.90378859e-03, -4.64503318e-03,  4.64503318e-03,
             -4.64503318e-03,  4.64503318e-03,  9.97674368e-03,
             -9.97674368e-03,  9.97674368e-03, -9.97674368e-03,
              1.55663113e-01, -1.16164603e-02, -1.16164603e-02,
              6.04837460e-03,  6.04837460e-03,  1.88289818e-01,
              1.55850551e-03, -1.55850551e-03, -1.55850551e-03,
              1.55850551e-03,  1.01021661e-03, -1.01021661e-03,
              1.01021661e-03, -1.01021661e-03, -2.20358979e-02,
             -2.20358979e-02, -4.19417445e-03, -4.19417445e-03,
              2.20039773e-01,  1.00374481e-02, -1.00374481e-02,
             -1.00374481e-02,  1.00374481e-02,  1.48172960e-03,
             -1.48172960e-03,  1.48172960e-03, -1.48172960e-03,
              1.48172960e-03,  1.48172960e-03, -1.48172960e-03,
             -1.48172960e-03,  6.48432287e-03,  6.48432287e-03,
              6.48432287e-03,  6.48432287e-03,  1.48581833e-01,
             -2.18755560e-03,  2.18755560e-03,  1.59201872e-01,
              1.39092556e-01, -5.90378859e-03,  5.90378859e-03,
             -6.64925506e-03, -6.64925506e-03, -4.64503318e-03,
             -4.64503318e-03,  4.64503318e-03,  4.64503318e-03,
             -9.77990177e-03, -9.77990177e-03,  9.97674368e-03,
              9.97674368e-03, -9.97674368e-03, -9.97674368e-03,
              1.48581833e-01, -2.18755560e-03,  2.18755560e-03,
              1.57970819e-01, -1.24259614e-02,  1.24259614e-02,
             -1.24259614e-02,  1.24259614e-02,  2.14571566e-02,
             -2.14571566e-02, -2.14571566e-02,  2.14571566e-02,
              1.35412579e-01,  1.01805363e-02, -1.01805363e-02,
              1.55663113e-01, -1.16164603e-02, -1.16164603e-02,
              6.04837460e-03,  6.04837460e-03,  1.59201872e-01,
              1.35412579e-01,  1.01805363e-02, -1.01805363e-02,
              1.66727286e-01,  2.87373749e-01,  2.78584086e-01,
              1.54628390e-05, -1.54628390e-05,  2.78893217e-01,
              2.00119947e-03,  2.00119947e-03,  6.95224680e-05,
             -6.95224680e-05
       \end{itemize} 
   \item[\#22] 
       \begin{itemize}
            \item \textbf{Geometry:}\\
            O 0.0 0.0 0.0\\
            H 0.0 0.0 0.9\\
            H 0.0 0.8516700789186731 -0.290960610177153
            \item \textbf{Pauli strings:}\\
            'IIIIII', 'IIIIIZ', 'IIIZXX', 'IIIIYY', 'IIIIZZ', 'IIIZZI', 'IIIZII', 'IIZIII', 'ZXXIII', 'IYYIII', 'IZZIII', 'ZZIIII', 'ZIIIII', 'IIIIZI', 'IIIZZZ', 'IIIZIZ', 'IIIZYY', 'IIIIXX', 'IIZIIZ', 'IIZZXX', 'IIZIYY', 'IZXIZX', 'IIXIZX', 'IZXIIX', 'IIXIIX', 'ZXXIIZ', 'IYYIIZ', 'ZXXZXX', 'IYYZXX', 'ZXXIYY', 'IYYIYY', 'XXXXXX', 'YXYXXX', 'XXXYXY', 'YXYYXY', 'IZZIIZ', 'IZZZXX', 'IZZIYY', 'ZXZIZX', 'IXIIZX', 'ZXZIIX', 'IXIIIX', 'ZZIIIZ', 'ZZIZXX', 'ZZIIYY', 'XZIXXX', 'XIIXXX', 'XZIYXY', 'XIIYXY', 'ZIIIIZ', 'ZIIZXX', 'ZIIIYY', 'IIZIZZ', 'IZXZXZ', 'IIXZXZ', 'IZXIXI', 'IIXIXI', 'ZXXIZZ', 'IYYIZZ', 'IZZIZZ', 'ZXZZXZ', 'IXIZXZ', 'ZXZIXI', 'IXIIXI', 'XXZXXZ', 'YYIXXZ', 'XXZYYI', 'YYIYYI', 'ZZIIZZ', 'ZIIIZZ', 'IIZZZI', 'ZXXZZI', 'IYYZZI', 'XXXXZI', 'YXYXZI', 'XXXXII', 'YXYXII', 'IZZZZI', 'ZZIZZI', 'XZIXZI', 'XIIXZI', 'XZIXII', 'XIIXII', 'ZIIZZI', 'IIZZII', 'ZXXZII', 'IYYZII', 'IZZZII', 'ZZIZII', 'ZIIZII', 'IZIIII', 'ZZZIII', 'ZIZIII', 'ZYYIII', 'IXXIII'
            \item \textbf{Coefficients:}\\
            -4.25949713e+00,  2.32662956e-01,  3.54772125e-02,
              3.54772125e-02,  1.78805752e-01, -1.88569776e-01,
             -2.21558015e-01,  2.32662956e-01,  3.54772125e-02,
              3.54772125e-02,  1.78805752e-01, -1.88569776e-01,
             -2.21558015e-01,  2.84598573e-01,  2.75238365e-01,
              2.76163412e-01, -2.89285716e-03, -2.89285716e-03,
              2.04504482e-01,  2.73690858e-02,  2.73690858e-02,
              1.44503920e-02, -1.44503920e-02, -1.44503920e-02,
              1.44503920e-02,  2.73690858e-02,  2.73690858e-02,
              1.40099687e-02,  1.40099687e-02,  1.40099687e-02,
              1.40099687e-02,  1.65636905e-02,  1.65636905e-02,
              1.65636905e-02,  1.65636905e-02,  1.86046121e-01,
              2.51032370e-02,  2.51032370e-02, -1.18673870e-03,
              1.18673870e-03,  1.18673870e-03, -1.18673870e-03,
              1.36969992e-01,  1.04908618e-02,  1.04908618e-02,
             -1.40000129e-02,  1.40000129e-02, -1.40000129e-02,
              1.40000129e-02,  1.55458020e-01,  9.39710564e-03,
              9.39710564e-03,  1.86046121e-01, -1.18673870e-03,
              1.18673870e-03,  1.18673870e-03, -1.18673870e-03,
              2.51032370e-02,  2.51032370e-02,  2.20039773e-01,
              1.01521961e-02, -1.01521961e-02, -1.01521961e-02,
              1.01521961e-02,  6.05803110e-03,  6.05803110e-03,
              6.05803110e-03,  6.05803110e-03,  1.47421279e-01,
              1.58336373e-01,  1.36969992e-01,  1.04908618e-02,
              1.04908618e-02, -1.40000129e-02, -1.40000129e-02,
              1.40000129e-02,  1.40000129e-02,  1.47421279e-01,
              1.49231271e-01,  2.95202003e-02, -2.95202003e-02,
             -2.95202003e-02,  2.95202003e-02,  1.42523045e-01,
              1.55458020e-01,  9.39710564e-03,  9.39710564e-03,
              1.58336373e-01,  1.42523045e-01,  1.57529409e-01,
              2.84598573e-01,  2.75238365e-01,  2.76163412e-01,
             -2.89285716e-03, -2.89285716e-03
       \end{itemize}
   \item[\#23] 
       \begin{itemize}
            \item \textbf{Geometry:}\\
            O 0.0 0.0 0.0\\
            H 0.0 0.0 0.9\\
            H 0.0 0.9463000876874145 -0.32328956686350335
            \item \textbf{Pauli strings:}\\
            'IIIIII', 'IIIIIZ', 'IIIZXX', 'IIIIYY', 'IIIXXX', 'IIIYXY', 'IIIIZZ', 'IIIZZI', 'IIIXZI', 'IIIXII', 'IIIZII', 'IIZIII', 'ZXXIII', 'IYYIII', 'XXXIII', 'YXYIII', 'IZZIII', 'ZZIIII', 'XZIIII', 'XIIIII', 'ZIIIII', 'IIIIZI', 'IIIZZZ', 'IIIXZZ', 'IIIXIZ', 'IIIZIZ', 'IIIZYY', 'IIIIXX', 'IIIXYY', 'IIIYYX', 'IIZIIZ', 'IIZZXX', 'IIZIYY', 'IIZXXX', 'IIZYXY', 'IZXIZX', 'IIXIZX', 'IZXIIX', 'IIXIIX', 'ZXXIIZ', 'IYYIIZ', 'ZXXZXX', 'IYYZXX', 'ZXXIYY', 'IYYIYY', 'ZXXXXX', 'IYYXXX', 'ZXXYXY', 'IYYYXY', 'XXXIIZ', 'YXYIIZ', 'XXXZXX', 'YXYZXX', 'XXXIYY', 'YXYIYY', 'XXXXXX', 'YXYXXX', 'XXXYXY', 'YXYYXY', 'IZZIIZ', 'IZZZXX', 'IZZIYY', 'IZZXXX', 'IZZYXY', 'ZXZIZX', 'IXIIZX', 'ZXZIIX', 'IXIIIX', 'XXZIZX', 'YYIIZX', 'XXZIIX', 'YYIIIX', 'ZZIIIZ', 'ZZIZXX', 'ZZIIYY', 'ZZIXXX', 'ZZIYXY', 'XZIIIZ', 'XIIIIZ', 'XZIZXX', 'XIIZXX', 'XZIIYY', 'XIIIYY', 'XZIXXX', 'XIIXXX', 'XZIYXY', 'XIIYXY', 'ZIIIIZ', 'ZIIZXX', 'ZIIIYY', 'ZIIXXX', 'ZIIYXY', 'IIZIZZ', 'IZXZXZ', 'IIXZXZ', 'IZXIXI', 'IIXIXI', 'IZXXXZ', 'IIXXXZ', 'IZXYYI', 'IIXYYI', 'ZXXIZZ', 'IYYIZZ', 'XXXIZZ', 'YXYIZZ', 'IZZIZZ', 'ZXZZXZ', 'IXIZXZ', 'ZXZIXI', 'IXIIXI', 'ZXZXXZ', 'IXIXXZ', 'ZXZYYI', 'IXIYYI', 'XXZZXZ', 'YYIZXZ', 'XXZIXI', 'YYIIXI', 'XXZXXZ', 'YYIXXZ', 'XXZYYI', 'YYIYYI', 'ZZIIZZ', 'XZIIZZ', 'XIIIZZ', 'ZIIIZZ', 'IIZZZI', 'IIZXZI', 'IIZXII', 'ZXXZZI', 'IYYZZI', 'ZXXXZI', 'IYYXZI', 'ZXXXII', 'IYYXII', 'XXXZZI', 'YXYZZI', 'XXXXZI', 'YXYXZI', 'XXXXII', 'YXYXII', 'IZZZZI', 'IZZXZI', 'IZZXII', 'ZZIZZI', 'ZZIXZI', 'ZZIXII', 'XZIZZI', 'XIIZZI', 'XZIXZI', 'XIIXZI', 'XZIXII', 'XIIXII', 'ZIIZZI', 'ZIIXZI', 'ZIIXII', 'IIZZII', 'ZXXZII', 'IYYZII', 'XXXZII', 'YXYZII', 'IZZZII', 'ZZIZII', 'XZIZII', 'XIIZII', 'ZIIZII', 'IZIIII', 'ZZZIII', 'XZZIII', 'XIZIII', 'ZIZIII', 'ZYYIII', 'IXXIII', 'XYYIII', 'YYXIII'
            \item \textbf{Coefficients:}\\
            -4.29058773e+00,  2.25353366e-01, -3.93477423e-02,
             -3.93477423e-02, -7.10344802e-03, -7.10344802e-03,
              1.75838484e-01, -1.52491891e-01, -1.26797524e-02,
              1.26797524e-02, -2.06525872e-01,  2.25353366e-01,
             -3.93477423e-02, -3.93477423e-02, -7.10344802e-03,
             -7.10344802e-03,  1.75838484e-01, -1.52491891e-01,
             -1.26797524e-02,  1.26797524e-02, -2.06525872e-01,
              2.81072427e-01,  2.71213417e-01, -2.44604575e-04,
              2.44604575e-04,  2.72785490e-01,  3.98480953e-03,
              3.98480953e-03,  3.75599377e-04, -3.75599377e-04,
              1.97569819e-01, -2.87650179e-02, -2.87650179e-02,
             -5.56927627e-03, -5.56927627e-03,  1.38649938e-02,
             -1.38649938e-02, -1.38649938e-02,  1.38649938e-02,
             -2.87650179e-02, -2.87650179e-02,  1.74577416e-02,
              1.74577416e-02,  1.74577416e-02,  1.74577416e-02,
             -1.92386275e-03, -1.92386275e-03, -1.92386275e-03,
             -1.92386275e-03, -5.56927627e-03, -5.56927627e-03,
             -1.92386275e-03, -1.92386275e-03, -1.92386275e-03,
             -1.92386275e-03,  1.56326941e-02,  1.56326941e-02,
              1.56326941e-02,  1.56326941e-02,  1.83110898e-01,
             -2.78632350e-02, -2.78632350e-02, -4.79163160e-03,
             -4.79163160e-03,  3.49778896e-04, -3.49778896e-04,
             -3.49778896e-04,  3.49778896e-04,  7.21738086e-04,
              7.21738086e-04, -7.21738086e-04, -7.21738086e-04,
              1.38389308e-01, -7.51510351e-03, -7.51510351e-03,
             -9.88391410e-03, -9.88391410e-03, -5.25406264e-03,
              5.25406264e-03, -4.74614379e-03,  4.74614379e-03,
             -4.74614379e-03,  4.74614379e-03,  1.04780072e-02,
             -1.04780072e-02,  1.04780072e-02, -1.04780072e-02,
              1.51183602e-01, -1.37501971e-02, -1.37501971e-02,
              6.25079560e-03,  6.25079560e-03,  1.83110898e-01,
              3.49778896e-04, -3.49778896e-04, -3.49778896e-04,
              3.49778896e-04,  7.21738086e-04, -7.21738086e-04,
              7.21738086e-04, -7.21738086e-04, -2.78632350e-02,
             -2.78632350e-02, -4.79163160e-03, -4.79163160e-03,
              2.20039773e-01,  9.33937934e-03, -9.33937934e-03,
             -9.33937934e-03,  9.33937934e-03,  1.15641491e-03,
             -1.15641491e-03,  1.15641491e-03, -1.15641491e-03,
              1.15641491e-03,  1.15641491e-03, -1.15641491e-03,
             -1.15641491e-03,  6.56710735e-03,  6.56710735e-03,
              6.56710735e-03,  6.56710735e-03,  1.46573961e-01,
             -1.92918040e-03,  1.92918040e-03,  1.56848958e-01,
              1.38389308e-01, -5.25406264e-03,  5.25406264e-03,
             -7.51510351e-03, -7.51510351e-03, -4.74614379e-03,
             -4.74614379e-03,  4.74614379e-03,  4.74614379e-03,
             -9.88391410e-03, -9.88391410e-03,  1.04780072e-02,
              1.04780072e-02, -1.04780072e-02, -1.04780072e-02,
              1.46573961e-01, -1.92918040e-03,  1.92918040e-03,
              1.55931197e-01, -1.17034614e-02,  1.17034614e-02,
             -1.17034614e-02,  1.17034614e-02,  2.16044257e-02,
             -2.16044257e-02, -2.16044257e-02,  2.16044257e-02,
              1.33430948e-01,  1.07841780e-02, -1.07841780e-02,
              1.51183602e-01, -1.37501971e-02, -1.37501971e-02,
              6.25079560e-03,  6.25079560e-03,  1.56848958e-01,
              1.33430948e-01,  1.07841780e-02, -1.07841780e-02,
              1.63480744e-01,  2.81072427e-01,  2.71213417e-01,
             -2.44604575e-04,  2.44604575e-04,  2.72785490e-01,
              3.98480953e-03,  3.98480953e-03,  3.75599377e-04,
             -3.75599377e-04
       \end{itemize}
   \item[\#24] 
       \begin{itemize}
            \item \textbf{Geometry:}\\
            O 0.0 0.0 0.0\\
            H 0.0 0.0 1.0\\
            H 0.0 0.7570400701499316 -0.2586316534908027
            \item \textbf{Pauli strings:}\\
            'IIIIII', 'IIIIIZ', 'IIIZXX', 'IIIIYY', 'IIIXXX', 'IIIYXY', 'IIIIZZ', 'IIIZZI', 'IIIXZI', 'IIIXII', 'IIIZII', 'IIZIII', 'ZXXIII', 'IYYIII', 'XXXIII', 'YXYIII', 'IZZIII', 'ZZIIII', 'XZIIII', 'XIIIII', 'ZIIIII', 'IIIIZI', 'IIIZZZ', 'IIIXZZ', 'IIIXIZ', 'IIIZIZ', 'IIIZYY', 'IIIIXX', 'IIIXYY', 'IIIYYX', 'IIZIIZ', 'IIZZXX', 'IIZIYY', 'IIZXXX', 'IIZYXY', 'IZXIZX', 'IIXIZX', 'IZXIIX', 'IIXIIX', 'ZXXIIZ', 'IYYIIZ', 'ZXXZXX', 'IYYZXX', 'ZXXIYY', 'IYYIYY', 'ZXXXXX', 'IYYXXX', 'ZXXYXY', 'IYYYXY', 'XXXIIZ', 'YXYIIZ', 'XXXZXX', 'YXYZXX', 'XXXIYY', 'YXYIYY', 'XXXXXX', 'YXYXXX', 'XXXYXY', 'YXYYXY', 'IZZIIZ', 'IZZZXX', 'IZZIYY', 'IZZXXX', 'IZZYXY', 'ZXZIZX', 'IXIIZX', 'ZXZIIX', 'IXIIIX', 'XXZIZX', 'YYIIZX', 'XXZIIX', 'YYIIIX', 'ZZIIIZ', 'ZZIZXX', 'ZZIIYY', 'ZZIXXX', 'ZZIYXY', 'XZIIIZ', 'XIIIIZ', 'XZIZXX', 'XIIZXX', 'XZIIYY', 'XIIIYY', 'XZIXXX', 'XIIXXX', 'XZIYXY', 'XIIYXY', 'ZIIIIZ', 'ZIIZXX', 'ZIIIYY', 'ZIIXXX', 'ZIIYXY', 'IIZIZZ', 'IZXZXZ', 'IIXZXZ', 'IZXIXI', 'IIXIXI', 'IZXXXZ', 'IIXXXZ', 'IZXYYI', 'IIXYYI', 'ZXXIZZ', 'IYYIZZ', 'XXXIZZ', 'YXYIZZ', 'IZZIZZ', 'ZXZZXZ', 'IXIZXZ', 'ZXZIXI', 'IXIIXI', 'ZXZXXZ', 'IXIXXZ', 'ZXZYYI', 'IXIYYI', 'XXZZXZ', 'YYIZXZ', 'XXZIXI', 'YYIIXI', 'XXZXXZ', 'YYIXXZ', 'XXZYYI', 'YYIYYI', 'ZZIIZZ', 'XZIIZZ', 'XIIIZZ', 'ZIIIZZ', 'IIZZZI', 'IIZXZI', 'IIZXII', 'ZXXZZI', 'IYYZZI', 'ZXXXZI', 'IYYXZI', 'ZXXXII', 'IYYXII', 'XXXZZI', 'YXYZZI', 'XXXXZI', 'YXYXZI', 'XXXXII', 'YXYXII', 'IZZZZI', 'IZZXZI', 'IZZXII', 'ZZIZZI', 'ZZIXZI', 'ZZIXII', 'XZIZZI', 'XIIZZI', 'XZIXZI', 'XIIXZI', 'XZIXII', 'XIIXII', 'ZIIZZI', 'ZIIXZI', 'ZIIXII', 'IIZZII', 'ZXXZII', 'IYYZII', 'XXXZII', 'YXYZII', 'IZZZII', 'ZZIZII', 'XZIZII', 'XIIZII', 'ZIIZII', 'IZIIII', 'ZZZIII', 'XZZIII', 'XIZIII', 'ZIZIII', 'ZYYIII', 'IXXIII', 'XYYIII', 'YYXIII'
            \item \textbf{Coefficients:}\\
            -4.23118332e+00,  2.31350529e-01, -3.50615360e-02,
             -3.50615360e-02, -8.18077320e-03, -8.18077320e-03,
              1.79621986e-01, -1.53779836e-01, -1.80150757e-02,
              1.80150757e-02, -2.63527519e-01,  2.31350529e-01,
             -3.50615360e-02, -3.50615360e-02, -8.18077320e-03,
             -8.18077320e-03,  1.79621986e-01, -1.53779836e-01,
             -1.80150757e-02,  1.80150757e-02, -2.63527519e-01,
              2.83656064e-01,  2.74378270e-01, -8.92714251e-05,
              8.92714251e-05,  2.75303503e-01,  3.19636059e-03,
              3.19636059e-03,  1.48488475e-04, -1.48488475e-04,
              2.02719601e-01, -2.65917339e-02, -2.65917339e-02,
             -7.05558241e-03, -7.05558241e-03,  1.42365082e-02,
             -1.42365082e-02, -1.42365082e-02,  1.42365082e-02,
             -2.65917339e-02, -2.65917339e-02,  1.74142609e-02,
              1.74142609e-02,  1.74142609e-02,  1.74142609e-02,
             -2.95409903e-03, -2.95409903e-03, -2.95409903e-03,
             -2.95409903e-03, -7.05558241e-03, -7.05558241e-03,
             -2.95409903e-03, -2.95409903e-03, -2.95409903e-03,
             -2.95409903e-03,  1.37254917e-02,  1.37254917e-02,
              1.37254917e-02,  1.37254917e-02,  1.85220492e-01,
             -2.53578825e-02, -2.53578825e-02, -5.99237799e-03,
             -5.99237799e-03,  5.22121772e-04, -5.22121772e-04,
             -5.22121772e-04,  5.22121772e-04,  1.45890109e-03,
              1.45890109e-03, -1.45890109e-03, -1.45890109e-03,
              1.41070328e-01, -2.97671836e-03, -2.97671836e-03,
             -1.24494435e-02, -1.24494435e-02, -7.77346815e-03,
              7.77346815e-03, -5.14665288e-03,  5.14665288e-03,
             -5.14665288e-03,  5.14665288e-03,  5.57597409e-03,
             -5.57597409e-03,  5.57597409e-03, -5.57597409e-03,
              1.50657899e-01, -1.71076696e-02, -1.71076696e-02,
              7.43375953e-03,  7.43375953e-03,  1.85220492e-01,
              5.22121772e-04, -5.22121772e-04, -5.22121772e-04,
              5.22121772e-04,  1.45890109e-03, -1.45890109e-03,
              1.45890109e-03, -1.45890109e-03, -2.53578825e-02,
             -2.53578825e-02, -5.99237799e-03, -5.99237799e-03,
              2.20039773e-01,  9.02767765e-03, -9.02767765e-03,
             -9.02767765e-03,  9.02767765e-03,  1.87433664e-03,
             -1.87433664e-03,  1.87433664e-03, -1.87433664e-03,
              1.87433664e-03,  1.87433664e-03, -1.87433664e-03,
             -1.87433664e-03,  7.21358602e-03,  7.21358602e-03,
              7.21358602e-03,  7.21358602e-03,  1.47398773e-01,
             -2.85576106e-03,  2.85576106e-03,  1.57935788e-01,
              1.41070328e-01, -7.77346815e-03,  7.77346815e-03,
             -2.97671836e-03, -2.97671836e-03, -5.14665288e-03,
             -5.14665288e-03,  5.14665288e-03,  5.14665288e-03,
             -1.24494435e-02, -1.24494435e-02,  5.57597409e-03,
              5.57597409e-03, -5.57597409e-03, -5.57597409e-03,
              1.47398773e-01, -2.85576106e-03,  2.85576106e-03,
              1.64670911e-01, -1.24015263e-02,  1.24015263e-02,
             -1.24015263e-02,  1.24015263e-02,  1.34803342e-02,
             -1.34803342e-02, -1.34803342e-02,  1.34803342e-02,
              1.26152414e-01,  1.02379077e-02, -1.02379077e-02,
              1.50657899e-01, -1.71076696e-02, -1.71076696e-02,
              7.43375953e-03,  7.43375953e-03,  1.57935788e-01,
              1.26152414e-01,  1.02379077e-02, -1.02379077e-02,
              1.72875653e-01,  2.83656064e-01,  2.74378270e-01,
             -8.92714251e-05,  8.92714251e-05,  2.75303503e-01,
              3.19636059e-03,  3.19636059e-03,  1.48488475e-04,
             -1.48488475e-04
       \end{itemize}
   \item[\#25] 
       \begin{itemize}
            \item \textbf{Geometry:}\\
            O 0.0 0.0 0.0\\
            H 0.0 0.0 1.0\\
            H 0.0 0.8516700789186731 -0.290960610177153
            \item \textbf{Pauli strings:}\\
            'IIIIII', 'IIIIIZ', 'IIIZXX', 'IIIIYY', 'IIIXXX', 'IIIYXY', 'IIIIZZ', 'IIIZZI', 'IIIXZI', 'IIIXII', 'IIIZII', 'IIZIII', 'ZXXIII', 'IYYIII', 'XXXIII', 'YXYIII', 'IZZIII', 'ZZIIII', 'XZIIII', 'XIIIII', 'ZIIIII', 'IIIIZI', 'IIIZZZ', 'IIIXZZ', 'IIIXIZ', 'IIIZIZ', 'IIIZYY', 'IIIIXX', 'IIIXYY', 'IIIYYX', 'IIZIIZ', 'IIZZXX', 'IIZIYY', 'IIZXXX', 'IIZYXY', 'IZXIZX', 'IIXIZX', 'IZXIIX', 'IIXIIX', 'ZXXIIZ', 'IYYIIZ', 'ZXXZXX', 'IYYZXX', 'ZXXIYY', 'IYYIYY', 'ZXXXXX', 'IYYXXX', 'ZXXYXY', 'IYYYXY', 'XXXIIZ', 'YXYIIZ', 'XXXZXX', 'YXYZXX', 'XXXIYY', 'YXYIYY', 'XXXXXX', 'YXYXXX', 'XXXYXY', 'YXYYXY', 'IZZIIZ', 'IZZZXX', 'IZZIYY', 'IZZXXX', 'IZZYXY', 'ZXZIZX', 'IXIIZX', 'ZXZIIX', 'IXIIIX', 'XXZIZX', 'YYIIZX', 'XXZIIX', 'YYIIIX', 'ZZIIIZ', 'ZZIZXX', 'ZZIIYY', 'ZZIXXX', 'ZZIYXY', 'XZIIIZ', 'XIIIIZ', 'XZIZXX', 'XIIZXX', 'XZIIYY', 'XIIIYY', 'XZIXXX', 'XIIXXX', 'XZIYXY', 'XIIYXY', 'ZIIIIZ', 'ZIIZXX', 'ZIIIYY', 'ZIIXXX', 'ZIIYXY', 'IIZIZZ', 'IZXZXZ', 'IIXZXZ', 'IZXIXI', 'IIXIXI', 'IZXXXZ', 'IIXXXZ', 'IZXYYI', 'IIXYYI', 'ZXXIZZ', 'IYYIZZ', 'XXXIZZ', 'YXYIZZ', 'IZZIZZ', 'ZXZZXZ', 'IXIZXZ', 'ZXZIXI', 'IXIIXI', 'ZXZXXZ', 'IXIXXZ', 'ZXZYYI', 'IXIYYI', 'XXZZXZ', 'YYIZXZ', 'XXZIXI', 'YYIIXI', 'XXZXXZ', 'YYIXXZ', 'XXZYYI', 'YYIYYI', 'ZZIIZZ', 'XZIIZZ', 'XIIIZZ', 'ZIIIZZ', 'IIZZZI', 'IIZXZI', 'IIZXII', 'ZXXZZI', 'IYYZZI', 'ZXXXZI', 'IYYXZI', 'ZXXXII', 'IYYXII', 'XXXZZI', 'YXYZZI', 'XXXXZI', 'YXYXZI', 'XXXXII', 'YXYXII', 'IZZZZI', 'IZZXZI', 'IZZXII', 'ZZIZZI', 'ZZIXZI', 'ZZIXII', 'XZIZZI', 'XIIZZI', 'XZIXZI', 'XIIXZI', 'XZIXII', 'XIIXII', 'ZIIZZI', 'ZIIXZI', 'ZIIXII', 'IIZZII', 'ZXXZII', 'IYYZII', 'XXXZII', 'YXYZII', 'IZZZII', 'ZZIZII', 'XZIZII', 'XIIZII', 'ZIIZII', 'IZIIII', 'ZZZIII', 'XZZIII', 'XIZIII', 'ZIZIII', 'ZYYIII', 'IXXIII', 'XYYIII', 'YYXIII'
            \item \textbf{Coefficients:}\\
            -4.29058773e+00,  2.25353366e-01, -3.93477423e-02,
             -3.93477423e-02, -7.10344802e-03, -7.10344802e-03,
              1.75838484e-01, -1.52491891e-01, -1.26797524e-02,
              1.26797524e-02, -2.06525872e-01,  2.25353366e-01,
             -3.93477423e-02, -3.93477423e-02, -7.10344802e-03,
             -7.10344802e-03,  1.75838484e-01, -1.52491891e-01,
             -1.26797524e-02,  1.26797524e-02, -2.06525872e-01,
              2.81072427e-01,  2.71213417e-01, -2.44604575e-04,
              2.44604575e-04,  2.72785490e-01,  3.98480953e-03,
              3.98480953e-03,  3.75599377e-04, -3.75599377e-04,
              1.97569819e-01, -2.87650179e-02, -2.87650179e-02,
             -5.56927627e-03, -5.56927627e-03,  1.38649938e-02,
             -1.38649938e-02, -1.38649938e-02,  1.38649938e-02,
             -2.87650179e-02, -2.87650179e-02,  1.74577416e-02,
              1.74577416e-02,  1.74577416e-02,  1.74577416e-02,
             -1.92386275e-03, -1.92386275e-03, -1.92386275e-03,
             -1.92386275e-03, -5.56927627e-03, -5.56927627e-03,
             -1.92386275e-03, -1.92386275e-03, -1.92386275e-03,
             -1.92386275e-03,  1.56326941e-02,  1.56326941e-02,
              1.56326941e-02,  1.56326941e-02,  1.83110898e-01,
             -2.78632350e-02, -2.78632350e-02, -4.79163160e-03,
             -4.79163160e-03,  3.49778896e-04, -3.49778896e-04,
             -3.49778896e-04,  3.49778896e-04,  7.21738086e-04,
              7.21738086e-04, -7.21738086e-04, -7.21738086e-04,
              1.38389308e-01, -7.51510351e-03, -7.51510351e-03,
             -9.88391410e-03, -9.88391410e-03, -5.25406264e-03,
              5.25406264e-03, -4.74614379e-03,  4.74614379e-03,
             -4.74614379e-03,  4.74614379e-03,  1.04780072e-02,
             -1.04780072e-02,  1.04780072e-02, -1.04780072e-02,
              1.51183602e-01, -1.37501971e-02, -1.37501971e-02,
              6.25079560e-03,  6.25079560e-03,  1.83110898e-01,
              3.49778896e-04, -3.49778896e-04, -3.49778896e-04,
              3.49778896e-04,  7.21738086e-04, -7.21738086e-04,
              7.21738086e-04, -7.21738086e-04, -2.78632350e-02,
             -2.78632350e-02, -4.79163160e-03, -4.79163160e-03,
              2.20039773e-01,  9.33937934e-03, -9.33937934e-03,
             -9.33937934e-03,  9.33937934e-03,  1.15641491e-03,
             -1.15641491e-03,  1.15641491e-03, -1.15641491e-03,
              1.15641491e-03,  1.15641491e-03, -1.15641491e-03,
             -1.15641491e-03,  6.56710735e-03,  6.56710735e-03,
              6.56710735e-03,  6.56710735e-03,  1.46573961e-01,
             -1.92918040e-03,  1.92918040e-03,  1.56848958e-01,
              1.38389308e-01, -5.25406264e-03,  5.25406264e-03,
             -7.51510351e-03, -7.51510351e-03, -4.74614379e-03,
             -4.74614379e-03,  4.74614379e-03,  4.74614379e-03,
             -9.88391410e-03, -9.88391410e-03,  1.04780072e-02,
              1.04780072e-02, -1.04780072e-02, -1.04780072e-02,
              1.46573961e-01, -1.92918040e-03,  1.92918040e-03,
              1.55931197e-01, -1.17034614e-02,  1.17034614e-02,
             -1.17034614e-02,  1.17034614e-02,  2.16044257e-02,
             -2.16044257e-02, -2.16044257e-02,  2.16044257e-02,
              1.33430948e-01,  1.07841780e-02, -1.07841780e-02,
              1.51183602e-01, -1.37501971e-02, -1.37501971e-02,
              6.25079560e-03,  6.25079560e-03,  1.56848958e-01,
              1.33430948e-01,  1.07841780e-02, -1.07841780e-02,
              1.63480744e-01,  2.81072427e-01,  2.71213417e-01,
             -2.44604575e-04,  2.44604575e-04,  2.72785490e-01,
              3.98480953e-03,  3.98480953e-03,  3.75599377e-04,
             -3.75599377e-04
       \end{itemize}
   \item[\#26] 
       \begin{itemize}
            \item \textbf{Geometry:}\\
            O 0.0 0.0 0.0\\
            H 0.0 0.0 1.0\\
            H 0.0 0.9463000876874145 -0.32328956686350335
            \item \textbf{Pauli strings:}\\
            'IIIIII', 'IIIIIZ', 'IIIZXX', 'IIIIYY', 'IIIIZZ', 'IIIZZI', 'IIIZII', 'IIZIII', 'ZXXIII', 'IYYIII', 'IZZIII', 'ZZIIII', 'ZIIIII', 'IIIIZI', 'IIIZZZ', 'IIIZIZ', 'IIIZYY', 'IIIIXX', 'IIZIIZ', 'IIZZXX', 'IIZIYY', 'IZXIZX', 'IIXIZX', 'IZXIIX', 'IIXIIX', 'ZXXIIZ', 'IYYIIZ', 'ZXXZXX', 'IYYZXX', 'ZXXIYY', 'IYYIYY', 'XXXXXX', 'YXYXXX', 'XXXYXY', 'YXYYXY', 'IZZIIZ', 'IZZZXX', 'IZZIYY', 'ZXZIZX', 'IXIIZX', 'ZXZIIX', 'IXIIIX', 'ZZIIIZ', 'ZZIZXX', 'ZZIIYY', 'XZIXXX', 'XIIXXX', 'XZIYXY', 'XIIYXY', 'ZIIIIZ', 'ZIIZXX', 'ZIIIYY', 'IIZIZZ', 'IZXZXZ', 'IIXZXZ', 'IZXIXI', 'IIXIXI', 'ZXXIZZ', 'IYYIZZ', 'IZZIZZ', 'ZXZZXZ', 'IXIZXZ', 'ZXZIXI', 'IXIIXI', 'XXZXXZ', 'YYIXXZ', 'XXZYYI', 'YYIYYI', 'ZZIIZZ', 'ZIIIZZ', 'IIZZZI', 'ZXXZZI', 'IYYZZI', 'XXXXZI', 'YXYXZI', 'XXXXII', 'YXYXII', 'IZZZZI', 'ZZIZZI', 'XZIXZI', 'XIIXZI', 'XZIXII', 'XIIXII', 'ZIIZZI', 'IIZZII', 'ZXXZII', 'IYYZII', 'IZZZII', 'ZZIZII', 'ZIIZII', 'IZIIII', 'ZZZIII', 'ZIZIII', 'ZYYIII', 'IXXIII'
            \item \textbf{Coefficients:}\\
            -4.33138971e+00,  2.19444898e-01,  4.39315347e-02,
              4.39315347e-02,  1.73662918e-01, -1.40250575e-01,
             -1.67222158e-01,  2.19444898e-01,  4.39315347e-02,
              4.39315347e-02,  1.73662918e-01, -1.40250575e-01,
             -1.67222158e-01,  2.77904698e-01,  2.67362579e-01,
              2.69849309e-01, -4.98150555e-03, -4.98150555e-03,
              1.91653293e-01,  3.08611264e-02,  3.08611264e-02,
              1.33762486e-02, -1.33762486e-02, -1.33762486e-02,
              1.33762486e-02,  3.08611264e-02,  3.08611264e-02,
              1.83742032e-02,  1.83742032e-02,  1.83742032e-02,
              1.83742032e-02,  1.69611643e-02,  1.69611643e-02,
              1.69611643e-02,  1.69611643e-02,  1.80545878e-01,
              3.08071177e-02,  3.08071177e-02,  4.98031213e-05,
             -4.98031213e-05, -4.98031213e-05,  4.98031213e-05,
              1.36137150e-01,  1.14604904e-02,  1.14604904e-02,
             -1.45180846e-02,  1.45180846e-02, -1.45180846e-02,
              1.45180846e-02,  1.50815706e-01,  1.12577244e-02,
              1.12577244e-02,  1.80545878e-01,  4.98031213e-05,
             -4.98031213e-05, -4.98031213e-05,  4.98031213e-05,
              3.08071177e-02,  3.08071177e-02,  2.20039773e-01,
              9.35069230e-03, -9.35069230e-03, -9.35069230e-03,
              9.35069230e-03,  6.23558540e-03,  6.23558540e-03,
              6.23558540e-03,  6.23558540e-03,  1.45345459e-01,
              1.55835217e-01,  1.36137150e-01,  1.14604904e-02,
              1.14604904e-02, -1.45180846e-02, -1.45180846e-02,
              1.45180846e-02,  1.45180846e-02,  1.45345459e-01,
              1.47203463e-01,  2.92065229e-02, -2.92065229e-02,
             -2.92065229e-02,  2.92065229e-02,  1.39941592e-01,
              1.50815706e-01,  1.12577244e-02,  1.12577244e-02,
              1.55835217e-01,  1.39941592e-01,  1.53747842e-01,
              2.77904698e-01,  2.67362579e-01,  2.69849309e-01,
             -4.98150555e-03, -4.98150555e-03
       \end{itemize}
\end{description}

\section*{Convergence graphs of the real device runs}
\begin{figure}[h!]
\centering
    \includegraphics[trim=20 0 0 0,clip,width=0.9\linewidth]{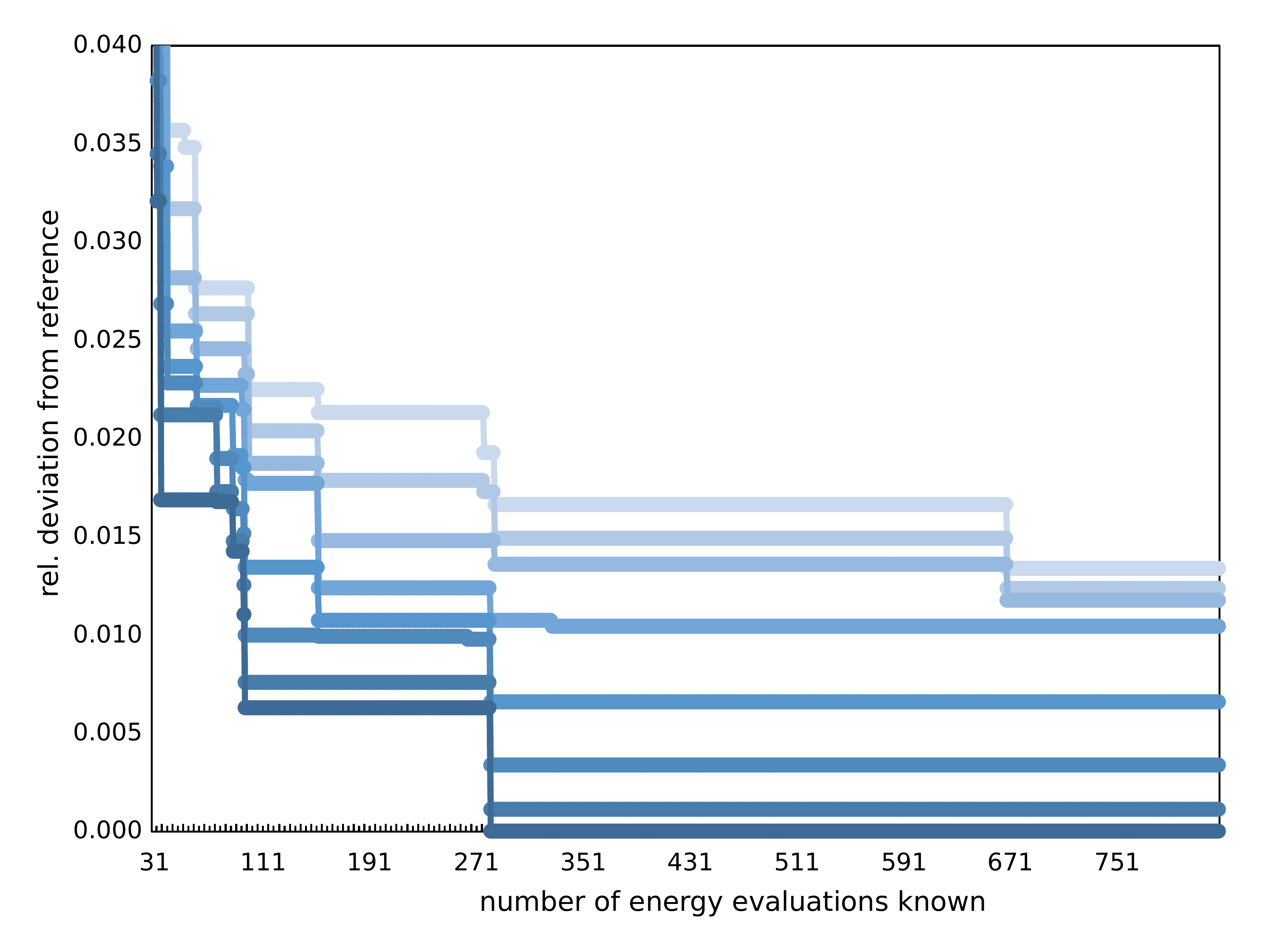}
    \caption{Convergence of the relative deviation of the best seen energy so far from the reference on the IBM Q System One ("Repetition 1" initial points.}
    \label{fig:ehnigen-conv-rep1}
\end{figure}
\begin{figure}
\centering
    \includegraphics[trim=20 0 0 0,clip,width=0.9\linewidth]{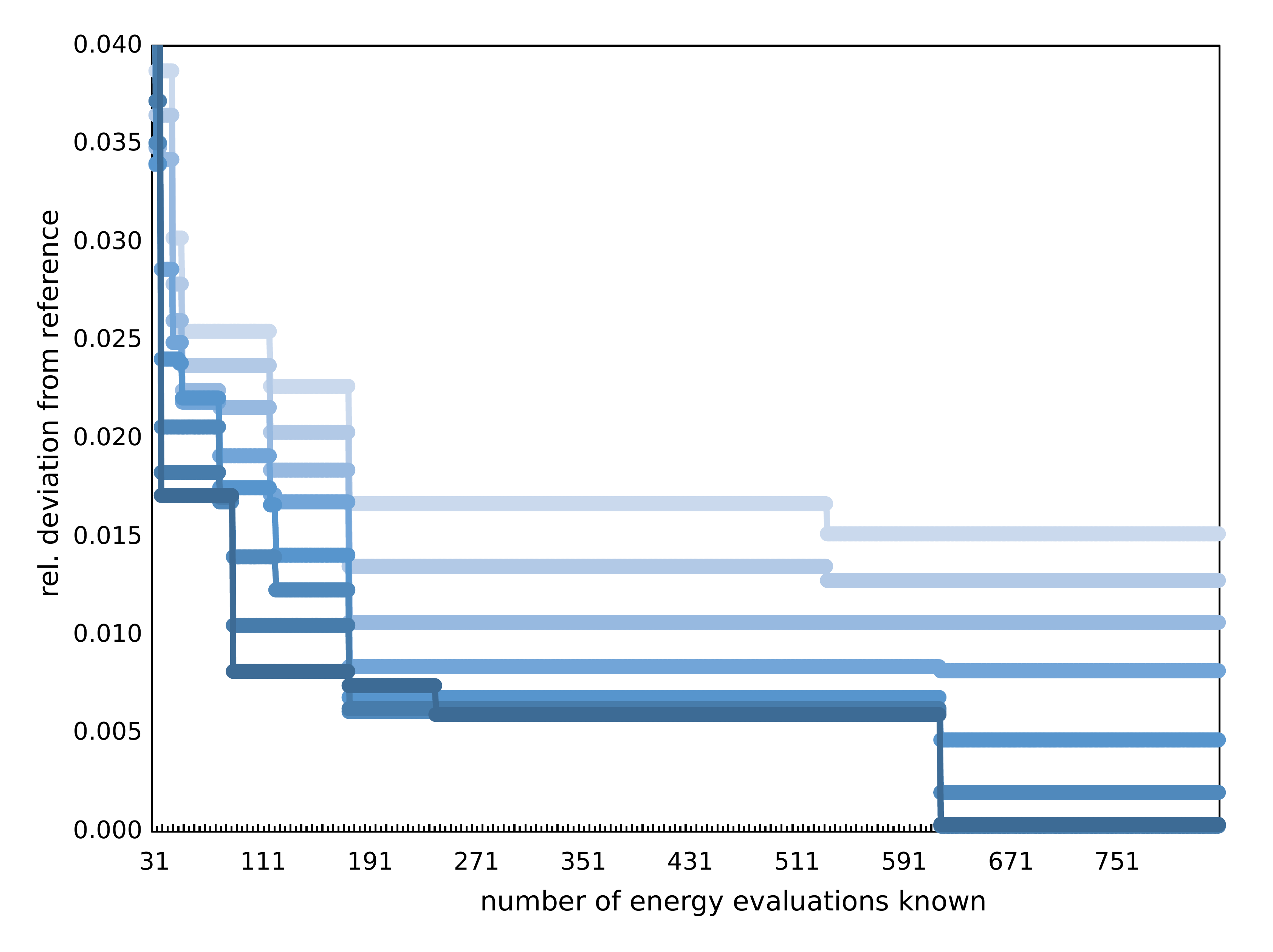}
    \caption{Convergence of the relative deviation of the best seen energy so far from the reference on the IBM Q System One ("Repetition 2" initial points.}
    \label{fig:ehnigen-conv-rep2}
\end{figure}
\begin{figure}
\centering
    \includegraphics[trim=20 0 0 0,clip,width=0.9\linewidth]{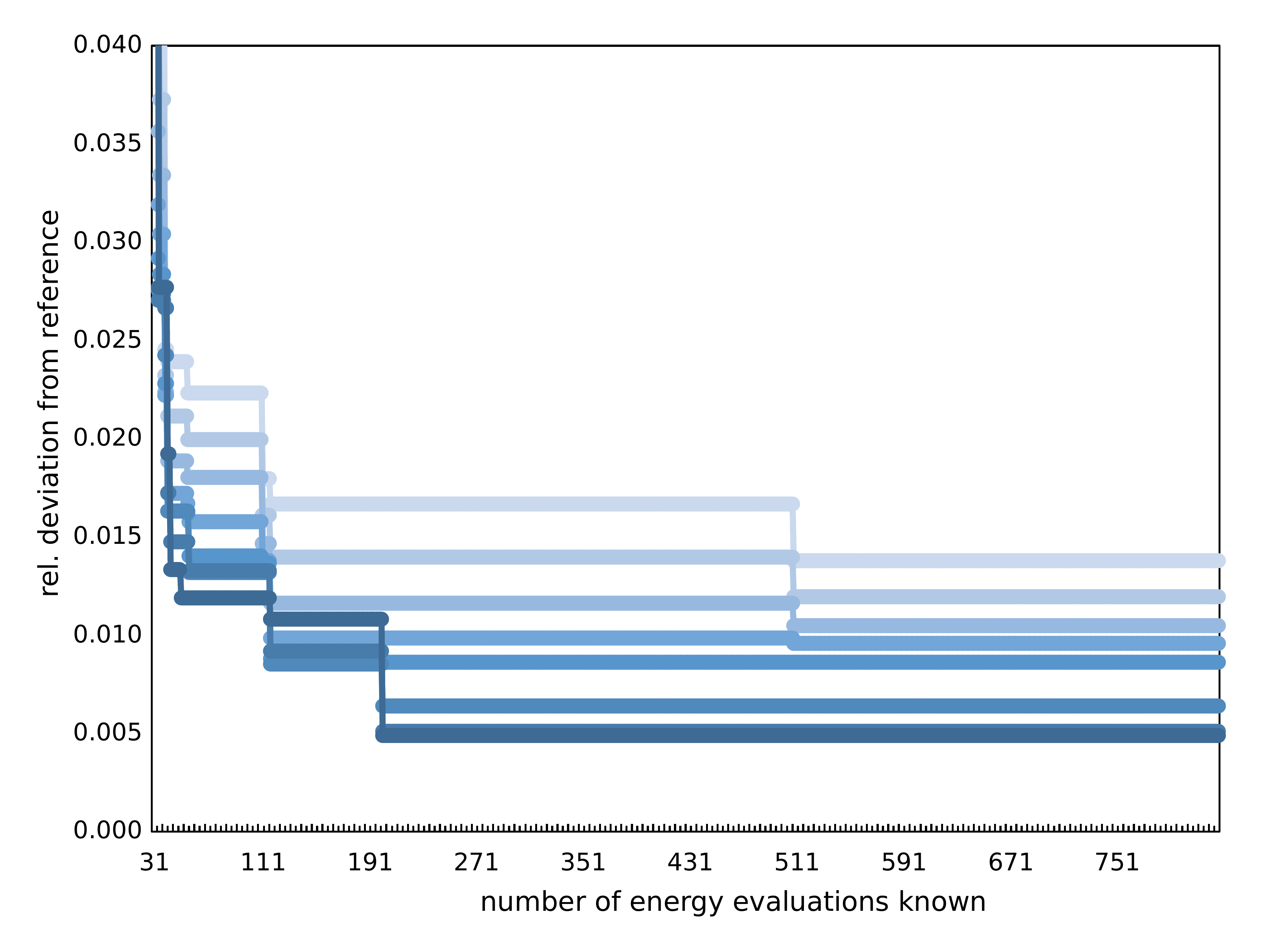}
    \caption{Convergence of the relative deviation of the best seen energy so far from the reference on the IBM Q System One ("Repetition 4" initial points.}
    \label{fig:ehnigen-conv-rep4}
\end{figure}
\begin{figure}
\centering
    \includegraphics[trim=20 0 0 0,clip,width=0.9\linewidth]{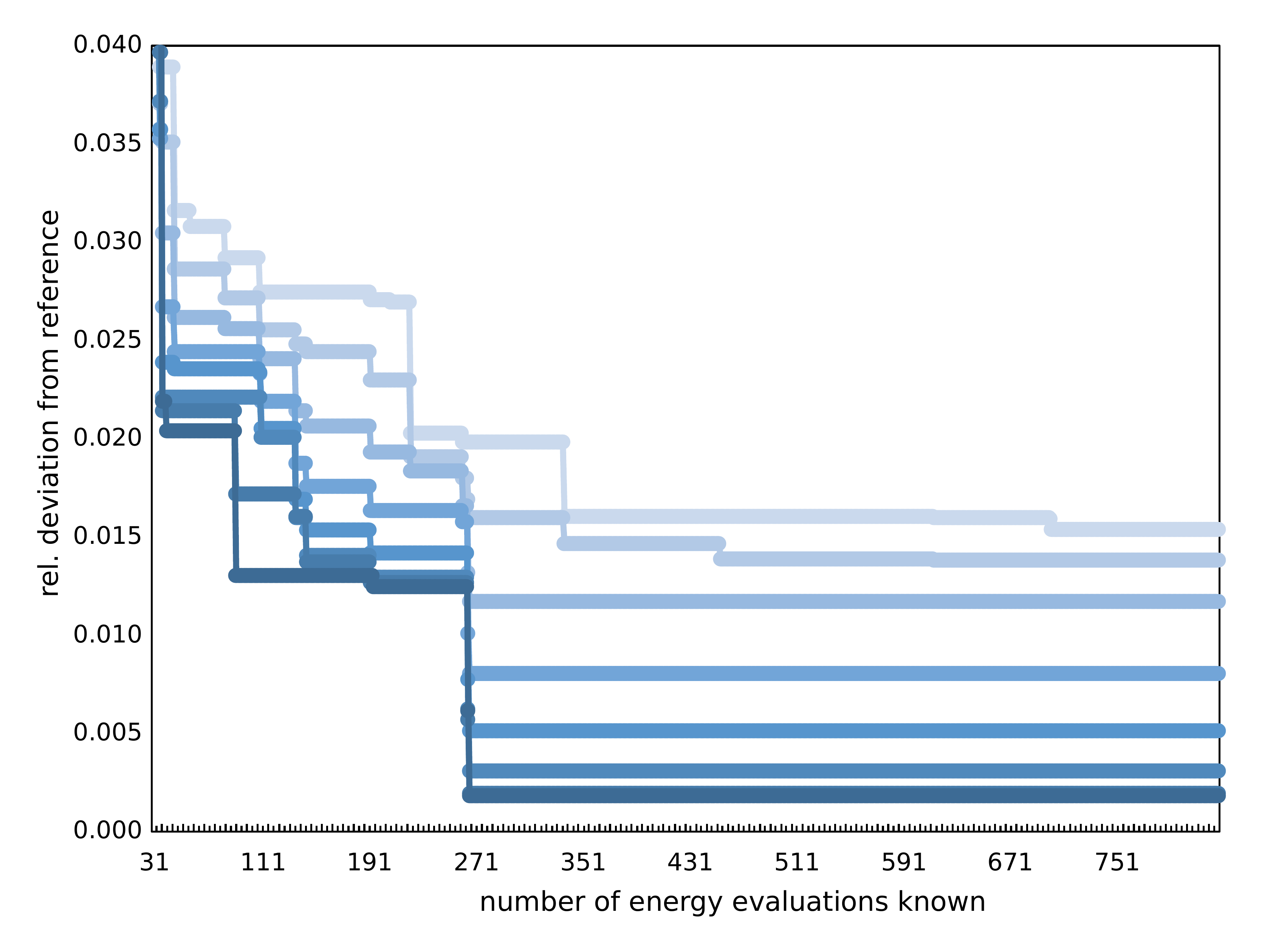}
    \caption{Convergence of the relative deviation of the best seen energy so far from the reference on the IBM Q System One ("Repetition 5" initial points.}
    \label{fig:ehnigen-conv-rep5}
\end{figure}